\documentclass[a4paper,11pt]{article}
\usepackage{jheppub} 
\usepackage{lineno}
\usepackage{nicefrac}
\usepackage[english]{babel}

\usepackage{soul}

\usepackage[bitstream-charter]{mathdesign}
\usepackage[amsmath]{empheq}
\numberwithin{equation}{section}
\DeclareSymbolFont{usualmathcal}{OMS}{cmsy}{m}{n}
\DeclareSymbolFontAlphabet{\mathcal}{usualmathcal}

\usepackage{comment}
\usepackage{braket}
\usepackage{bbm}
\usepackage{bm}
\usepackage{tikz}
\usepackage{color}
\usepackage{graphicx}

\usepackage{subcaption}

\usepackage{enumitem}

\newcommand{\be}{\begin{equation}}
	\newcommand{\ee}{\end{equation}}
\newcommand{\ba}{\begin{aligned}}
	\newcommand{\ea}{\end{aligned}}

\newcommand{\bs}[1]{\boldsymbol{\mathbf{#1}}}
\newcommand{\F}[4][{}]{\mathcal F_f^{#1}\left[\begin{smallmatrix}
#2\\
#3
\end{smallmatrix}
\right]
(#4)
}
\newcommand{\tr}{\mathrm{tr}}

\newcommand{\ETA}{\eta}

\title{\boldmath
Entanglement entropy of two disjoint intervals\\and spin structures in interacting chains\\in and out of equilibrium
}

\author[a]{Vanja Mari\'c,}
\author[b]{Saverio Bocini,}
\author[a]{Maurizio Fagotti}
\affiliation[a]{Universit\'e Paris-Saclay, CNRS, LPTMS, 91405, Orsay, France}
\affiliation[b]{Univ Lyon, ENS de Lyon, CNRS, Laboratoire de Physique, F-69342 Lyon, France}

\abstract{
We take the paradigm of interacting spin chains, the Heisenberg spin-$\frac{1}{2}$ XXZ model, as a reference system and consider interacting models that are related to it
by Jordan-Wigner transformations and restrictions to sub-chains. 
An example is the fermionic analogue of the gapless XXZ Hamiltonian, which, in a continuum scaling limit, is described by the massless Thirring model. 
We work out the R\'enyi-$\alpha$ entropies of disjoint blocks in the ground state and extract the universal scaling functions describing the R\'enyi-$\alpha$ tripartite information in the limit of infinite lengths.
We consider also the von Neumann entropy, but only in the limit of large distance.
We show how to use the entropies of spin blocks to unveil the  spin structures of the underlying massless Thirring model. 
Finally, we speculate about the tripartite information after global quenches and conjecture its asymptotic behaviour in the limit of infinite time and small quench. The resulting conjecture for the ``residual tripartite information'', which corresponds to the limit in which the intervals' lengths are infinitely larger than their (large) distance, supports the claim of universality recently made studying noninteracting spin chains.
Our mild assumptions imply that the residual tripartite information after a small quench of the anisotropy in the gapless phase of XXZ is equal to $-\log 2$.
}

\begin{document}
\maketitle
\flushbottom

\section{Introduction}

\label{sec:intro}
The entanglement entropies of subsystems in critical  chains at low temperature have become standard tools to extract universal properties of the underlying conformal field theories~\cite{Amico2008,Calabrese2009Entanglement}. To the best of our knowledge, the first evidence of it dates back to 1994, when Ref.~\cite{Holzhey1994Geometric} has shown that the leading order of the asymptotic expansion of the von Neumann entropy of an interval in the limit of large length is completely characterised by the central charge $c$
\begin{equation}
S[A]=-\mathrm{tr}[\rho_A\log\rho_A]\sim \frac{c}{3}\log|A|\, .
\end{equation}
Here $\rho_A$ is the reduced density matrix of the interval $A$ and $|A|$ denotes $A$'s length. 
About ten years later that result was generalised~\cite{Calabrese2004Entanglement} to other settings, e.g. finite temperature and finite systems. A similar formula was also obtained for the R\'enyi entropies
\begin{equation}\label{Renyi entropy single block}
S_\alpha[A]=\frac{1}{1-\alpha}\log \mathrm{tr}[\rho_A^\alpha]\sim \frac{c}{6}\left(1+\frac{1}{\alpha}\right)\log|A|\, .
\end{equation}
The interest in the subject exploded. 
On the one hand, it was found that some corrections to the scaling are universal and determined by the conformal dimensions of the operators of the theory~\cite{Calabrese2010Parity, Cardy2010}.
On the other hand, it was realised that changing the topology of the subsystem uncovers additional universal properties~\cite{Caraglio2008Entanglement,Furukawa2009Mutual}. A huge effort was then directed to the study of the entanglement entropies of disjoint blocks~\cite{Casini2009Remarks,Rajabpour2012,Calabrese2009Entanglement1,Calabrese2011Entanglement,Coser2014OnRenyi,Ruggiero2018,Alba2010Entanglement,Blanco2011,Alba2011Entanglement,Fagotti2010disjoint,Fagotti2012New,Fries2019,Balasubramanian2011,Grava2021,Ares2021Crossing}. The problem was traced back to the asymptotic behaviour of partition functions on Riemann surfaces with arbitrarily high genus, and few explicit solutions have been obtained. We mention the Ising model~\cite{Calabrese2011Entanglement} and the free boson compactified on a circle~\cite{Calabrese2009Entanglement1} or on an orbifold~\cite{Alba2011Entanglement}. Restricting to two disjoint blocks, a key role is played by a linear combination of the entropies of subsystems, built up with the two blocks and the region in between, which is also known as tripartite information
\begin{equation}\label{tripartite information def}
I_{3}^{(\alpha)}(A,B,C)=S_\alpha[A]+S_\alpha[B]+S_\alpha[C]-S_\alpha[A\cup B]-S_\alpha[A\cup C]-S_\alpha[B\cup C]+S_\alpha[A\cup B\cup C]\, .
\end{equation}
By construction, that combination simplifies both contributions coming from the boundaries of the subsystems and those extensive with subsystems' lengths, and it turns out that it is even free from the log divergences (in the subsystems' lengths) that characterise the entanglement entropies at criticality~\cite{Casini2009Remarks,Hayden2013Holographic}. In the ground state of a conformal field theory, $I_{3}^{(\alpha)}(A,B,C)$ turns out to be a nontrivial function of the cross ratio 
\begin{equation}\label{eq:x4}
x=\frac{|A||C|}{(|A|+|B|)(|B|+|C|)}
\end{equation}
and exhibits the symmetry $x\leftrightarrow 1-x$.

An important aspect that distinguishes the entropy of disjoint blocks from that of an interval is its sensitivity to duality mappings, in that such transformations generally preserve the locality of only a subset of operators (including the Hamiltonian density), breaking it instead into semilocality for the rest of them. This affects the meaning itself of subsystem, especially when it is not simply connected. 
A consequence of it on the behaviour of the entanglement entropies was firstly observed in the ground state of XY spin chains~\cite{Igloi2010,Fagotti2010disjoint}. At the critical point the tripartite information has a nontrivial scaling limit despite vanishing in the corresponding free-fermion model that emerges after a Jordan-Wigner transformation~\cite{Casini2005,Casini2009reduced_density,Casini2009Entanglement}.

Since the first step towards an analytical investigation into the entanglement entropies of spin blocks  is often a mapping to quasiparticles, it is important to understand how such mappings affect the entropies and how to ultimately fix the lack of equivalence.
In that respect, Ref.~\cite{Fagotti2010disjoint} proposed an expansion of the tripartite information based on the decomposition of the reduced density matrix into four pseudo reduced density matrices (pseudo-RDMs) that unfold the spin-flip symmetry. 
On the other hand, the critical spin-chain models studied in Ref.~\cite{Fagotti2010disjoint} admit a quantum field theory description of the low-energy and large-distance limit (the Ising model is described by a massless Majorana field theory, while the XX model by a Dirac one); as mentioned above, the R\'enyi entropy $S_\alpha$ of two 
disjoint intervals can then be written in terms of partition functions on  particular Riemann surfaces.
Ref.~\cite{Coser2016Spin} examined what 
the lattice procedure of Ref.~\cite{Fagotti2010disjoint} translates into in the quantum field theory description 
and established a one-to-one correspondence between the partition functions with given spin structures  and the moments of the pseudo-RDMs introduced in Ref.~\cite{Fagotti2010disjoint}.
It was  argued, in particular, that the entropy of disjoint blocks of spins is obtained by summing over all spin structures of the fermion model on the underlying Riemann surface, i.e., over all combinations of antiperiodic, also called Neveu-Schwarz (NS), and periodic, also called Ramond (R), boundary conditions on the cycles of the surface. The entropy of Jordan-Wigner fermions, instead, corresponds to antiperiodic boundary conditions over all cycles (see also Ref.~\cite{Camargo2021}).
As discussed in Ref.~\cite{Headrick2010} in detail, indeed, the discrepancy between the entanglement entropy of two disjoint intervals for the free Dirac fermion~\cite{Casini2005,Casini2009reduced_density,Casini2009Entanglement} and for the free boson compactified on a circle with the corresponding radius~\cite{Calabrese2009Entanglement1}, in our notations $R=1$, can be traced back to the spin structure. The result for the boson matches, instead, that for the modular invariant Dirac fermion~\cite{Calabrese2011Entanglement,Coser2016Spin}, which is obtained by summing over all boundary conditions. 
We mention that similar problems related to Bose-Fermi duality and modular invariance for the Dirac fermion on the torus arise when computing the entanglement of a single interval at finite temperature
~\cite{Lokhande2015,Mukhi2018,Arias2020}.

The tripartite information in the XXZ model can be deduced from the results of Refs~\cite{Calabrese2009Entanglement1,Furukawa2009Mutual} on the free compactified boson---see also Ref.~\cite{Alba2011Entanglement}. In light of Refs~\cite{Fagotti2010disjoint,Coser2016Spin}, we argue that it is possible to approach this problem also in the fermionic picture by considering the massless Thirring model. Concerning Ref.~\cite{Coser2016Spin}, we note that interacting spin chains described by the free compactified boson with radius $R\neq 1$ have been left out of the discussion. An analogous comment holds true for Ref.~\cite{Fagotti2010disjoint}, which was focussed on noninteracting spin chains. Both the construction of Ref.~\cite{Fagotti2010disjoint} and the arguments exhibited in Ref.~\cite{Coser2016Spin} about the correspondence between partition functions and moments of the pseudo-RDMs do not seem, however, to rely on the absence of interactions ($R=1$). Thus we argue that the main observations of Refs~\cite{Fagotti2010disjoint,Coser2016Spin} apply also to the Heisenberg XXZ chain, whose low-energy properties are captured by the massless Thirring model, and, in turn, by the free boson compactified on a circle. The problem of computing the partition function of the massless Thirring model on the higher genus Riemann surface with specified spin structure has been addressed in the works by Freedman, Pilch~\cite{Freedman1988Thirring,Freedman1989Thirring}, and Wu~\cite{Wu1989Determinants}, by direct manipulations of fermionic path integrals. We warn the reader that those results are not totally satisfactory: they display inconsistencies about factors and signs,
which were commented and partially resolved in Ref.~\cite{Sachs1996}. The tripartite information, however, can be computed without knowing all the details of the partition functions and we will fix the remaining ambiguities by imposing the known results for the XXZ chain. These observations are the starting point of our investigations.

The first part of this paper is mainly derivative of Refs~\cite{Freedman1989Thirring,Calabrese2009Entanglement1,Fagotti2012New}. On the one hand,  we consider the tripartite information in the XXZ fermion chain, which is described by the massless Thirring model with antiperiodic boundary conditions on all cycles: A direct application of Ref.~\cite{Freedman1989Thirring} allows us to generalize  the analysis of Ref.~\cite{Casini2005}, focussed on the free Dirac fermion, to the interacting case, where we report a contrastingly nonzero tripartite information. On the other hand, we  
reproduce the result for the XXZ spin chain/free compactified boson obtained in Ref.~\cite{Calabrese2009Entanglement1} by summing over all the periodic and anti-periodic spin structures computed in Ref.~\cite{Freedman1989Thirring}. 
We also consider a family of interacting spin-chain models (with local translationally invariant Hamiltonians) with higher central charge, that are related to XXZ by Jordan-Wigner transformations and restrictions to sub-chains. We call the latter models Jordan-Wigner deformations of XXZ. They are interacting generalizations of some free models introduced in Ref.~\cite{Suzuki1971The}, whose tripartite information was studied in Ref.~\cite{Fagotti2012New}. Their low energy description is a sum of independent Thirring models with different species of fermions and correlated spin structures. Knowing the contribution of each spin structure, we can easily work out their tripartite information.

In the second part of the paper we set up the inverse problem of computing the contribution of each spin structure from the knowledge of the tripartite information in a family of spin chains. The simplest and most accessible case is the torus, for which we can express the partition functions of the massless Thirring model with given boundary conditions in terms of the \mbox{R\'enyi-$2$} tripartite information of four JW deformations of XXZ. 
This yields a procedure to check the QFT predictions against numerical simulations (e.g., with tensor network algorithms). 
While effective on paper, this method is affected by a problem that makes the numerical investigations challenging: both the range of the interactions and the central charge of the models to consider increase by $1$ for each additional deformation. That is to say, even restricting to the torus, the numerical algorithm should be good enough to allow for the analysis of the entropies of large spin blocks also in a model with range $4$ and central charge $4$. The latter requirement is particularly challenging because the entropy is proportional to the central charge, so the numerical simulations become less and less effective~\cite{Pollmann2009Theory,Stojevic2015Conformal}.
In order to circumvent those issues, 
we propose an alternative procedure based on considering Hamiltonians obtained by ``interleaving'' XXZ critical Hamiltonians with noncritical ones. This breaks shift invariance from one-site into two-sites but provides a bound both on the range of the interactions, which can be as small as $2$ for every model to consider, and on their central charge, which remains the same as in the original system (in the specific case of XXZ, it remains equal to $1$). Remarkably, even if we focus on the torus and on the XXZ model, the procedure that we propose can be applied to Riemann surfaces with higher genus (though, it could become trickier) and to other spin-chain models with spin-flip symmetry.

In the third part of the paper we go out of equilibrium and speculate about the behaviour of the tripartite information at late times after having prepared the system in a state that is  close to but still different from the ground state.  Using the terminology of Ref.~\cite{Calabrese2012Quantum}, that would correspond to a ``small global quench'' from another critical Hamiltonian. We have recently investigated the tripartite information after global quenches in quantum spin chains that can be mapped to free fermions~\cite{Maric2022Universality,Maric2023Universality,Maric2023universality2}. In critical systems, in particular, we showed that it can remain nonzero also in the limit of infinite time and infinite lengths (taking the latter limit after the former one). Remarkably, the tripartite information remains a function of the cross ratio~\eqref{eq:x4}, but the symmetry $x\leftrightarrow 1-x$, which is always present in the ground state, doesn't survive the limits. Refs~\cite{Maric2022Universality,Maric2023Universality,Maric2023universality2} also showed that the limit $x\rightarrow 1^-$ (taken after the limits of infinite time and lengths), which was dubbed ``residual tripartite information'', can be nonzero. This can be surprising because it corresponds to the configuration in which the distance between the blocks is negligible with respect to the blocks' lengths, therefore it could be naively identified with the degenerate case in which the blocks are adjacent, for which the tripartite information is identically zero. 
Clearly the limits do not commute, but there is something more: in all the (noninteracting) cases investigated in Refs~\cite{Maric2022Universality,Maric2023Universality,Maric2023universality2}, the residual tripartite information was either $0$ or $-\log 2$. 
Here we provide evidence that this is not just a special feature of noninteracting models, we find indeed that also the residual tripartite information after quenches of the anistropy in the gapless phase of XXZ should be expected to be $-\log 2$.

\section{Summary of results}

\begin{figure}
\centering

\begin{tikzpicture}[thick,scale=0.9, every node/.style={scale=1.0}]

\begin{scope}[yshift=3.5 cm]
\node[inner sep=0pt] (img1) at (0.,0.) {\includegraphics[width= 4cm]{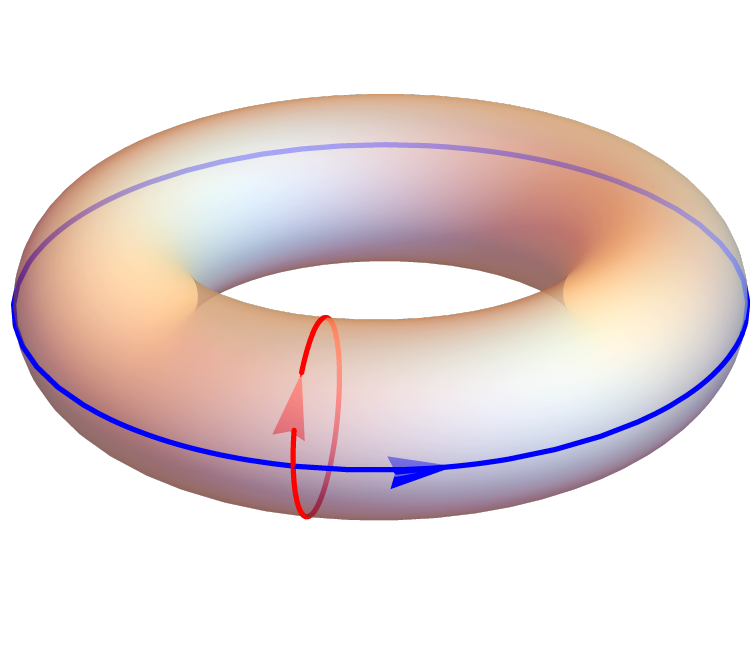}};
\draw (-0.8,-0.2) node{\textcolor{red}{\Large $a$}};
\draw (0.3,-0.5) node{\textcolor{blue}{\Large $b$}};

\draw (7,1) node[text width=5 cm, align=left]{Torus with modular parameter $\tau=i\frac{K(1-x)}{K(x)}$};

\draw (7.5,-1) node[text width=6 cm, align=left]{$Z_{\varepsilon,\delta}^{\mathrm{QFT}}\equiv $ partition function of fermionic QFT on the torus, with boundary conditions $\varepsilon,\delta$ on cycle $a,b$ respectively};

\end{scope}

\draw[<->] (2.2,0.1)   -- (5.8,0.1);
\draw (4.7,0.1) node[anchor=center, text width = 3 cm]{\small Jordan-Wigner transformation};

\draw (0,0.1) node[anchor=center]{\large Fermionic XXZ};

\draw (0,-1) node[anchor=center]{Ground state};

\draw (0,-2) node[anchor=center]{ $I_3^{(\alpha=2)}= \log \frac{Z_{\mathrm{NS},\mathrm{NS}}^\mathrm{Thirring}}{Z_{\mathrm{NS},\mathrm{NS}}^\mathrm{Dirac}} $};

\def\u{-3.}

\draw[->] (0,\u+0.2)   -- (0,\u-2.);

\draw (1.2,\u-0.75) node[text width=1.5 cm, anchor=center, align=left]{Small quantum quench};
\draw (0,\u-2.5) node[anchor=center]{Stationary state};

\draw (0,\u-3.5) node[anchor=center]{ $I_3^{(\alpha=2)}= \log \frac{Z_{\mathrm{NS},\mathrm{NS}}^\mathrm{Thirring}}{Z_{\mathrm{NS},\mathrm{NS}}^\mathrm{Dirac}} $};

\draw[very thick] (-3.1,-0.5) rectangle ++(6.2,-7);

\begin{scope}[xshift=8 cm]
\draw (0,0.1) node[anchor=center]{\large XXZ};

\draw (0,-1) node[anchor=center]{Ground state};

\draw (0,-2) node[anchor=center]{ $I_3^{(\alpha=2)}= \log \sum\limits_{\varepsilon,\delta=\mathrm{NS},\mathrm{R}}\frac{(-1)^{4\epsilon\delta}Z_{\varepsilon,\delta}^\mathrm{Thirring}}{2Z_{\mathrm{NS},\mathrm{NS}}^\mathrm{Dirac}} $};

\draw[->] (0,\u+0.2)   -- (0,\u-2.);

\draw (1.2,\u-0.75) node[text width=1.5 cm, anchor=center, align=left]{Small quantum quench};
\draw (0,\u-2.5) node[anchor=center]{Stationary state};

\draw (0,\u-3.5) node[anchor=center]{ $I_3^{(\alpha=2)}= \log \sum\limits_{\delta=\mathrm{NS},\mathrm{R}}\frac{Z_{\mathrm{NS},\delta}^\mathrm{Thirring}}{2Z_{\mathrm{NS},\mathrm{NS}}^\mathrm{Dirac}} $};

\draw[very thick] (-3.1,-0.5) rectangle ++(6.2,-7);

\end{scope}

\end{tikzpicture}

\caption{Summary of the relation between the tripartite information $I_3^{(\alpha)}(x)$ and QFT partition functions for the R\'enyi index $\alpha=2$. The Thirring coupling constant is fixed by the model parameters.}
\label{fig:torus summary}

\end{figure}
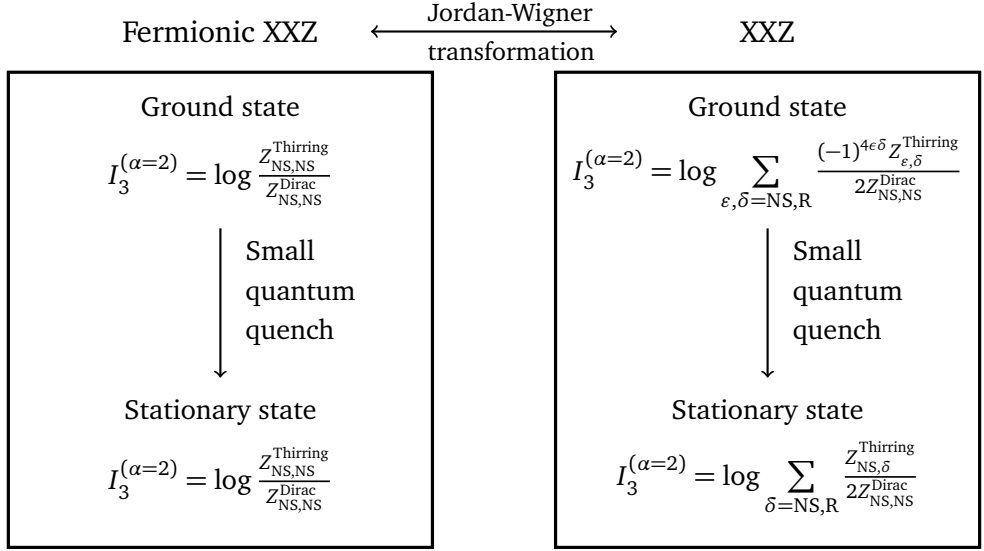

We consider the XXZ spin-$\frac{1}{2}$ chain described by the Hamiltonian
\begin{equation}\label{XXZ spin chain Hamiltonian}
\bs H_{\mathrm{XXZ}}=\sum_{\ell}\left(\bs\sigma_\ell^x\bs\sigma_{\ell+1}^x+\bs\sigma_\ell^y\bs\sigma_{\ell+1}^y+\Delta \bs\sigma_\ell^z\bs\sigma_{\ell+1}^z-2h\bs\sigma_\ell^z\right) \, ,
\end{equation}
where $h$ represents the coupling with an external magnetic field in the $z$ direction and $\Delta$ is usually referred to as ``anisotropy''. In a strip of the parameter space including $h=0\wedge|\Delta|<1$, the low-energy properties of the model are described by a conformal field theory with central charge $c=1$. 
By exploiting the correspondence between the XXZ spin-$\frac{1}{2}$ chain and the massless Thirring model, we point out the spin structures of the former. This observation allows us to compute the tripartite information of several related translationally invariant models. We report, in particular, the results for the following models (see Figure~\ref{fig:torus summary} for a summary of a relation between the tripartite information and Thirring partition functions in the case $\alpha=2$)
\begin{itemize}
\item \ul{XXZ fermion chain}. This is a model of spinless interacting fermions, dual to the XXZ spin chain by the Jordan-Wigner transformation $\bs c^\dag_\ell=\prod_{j<\ell}\bs \sigma_j^z\, \frac{\bs \sigma_\ell^x+i\bs \sigma_\ell^y}{2}$, where $\bs c^\dag$ are creation operators of spinless fermions satisfying $\{\bs c^\dag_\ell,\bs c_n\}=\delta_{\ell n}\bs 1$ and $\{\bs c_\ell,\bs c_n\}=0$. Its Hamiltonian reads
\begin{equation}\label{XXZ fermionic chain Hamiltonian}
\bs H_{\mathrm{fXXZ}}=2\sum_\ell \left( \bs c_{\ell+1} \bs c^\dag_\ell +\bs c_{\ell}\bs c^\dag_{\ell+1} -2(\Delta+h) \bs c^\dag_{\ell} \bs c_{\ell}+2\Delta \bs c^\dag_\ell \bs c_{\ell}\bs c^\dag_{\ell+1} \bs c_{\ell+1} \right) \, .
\end{equation}

\item \ul{Jordan-Wigner $\nu$-deformation of the XXZ spin chain}. This is dual to $\nu$ uncoupled XXZ fermion sub-chains interleaved with one another
\begin{equation}\label{eq:JWdef}
\begin{split}
\bs H_{\mathrm{XXZ}}^{(\nu)}=&\sum_\ell \Big[\bs \sigma_\ell^x (\bs\sigma_{\ell+1}^z\bs\sigma_{\ell+2}^z\cdots\bs \sigma_{\ell+\nu-1}^z)\bs \sigma_{\ell+\nu}^x+\bs \sigma_\ell^y(\bs\sigma_{\ell+1}^z\bs\sigma_{\ell+2}^z\cdots\bs \sigma_{\ell+\nu-1}^z)\bs\sigma_{\ell+\nu}^y\\ &+\Delta \bs \sigma_\ell^z\bs \sigma_{\ell+\nu}^z-2h\bs\sigma_\ell^z \Big]\, .
\end{split}
\end{equation}
\end{itemize}

\begin{figure}
    \centering
    \includegraphics[width=0.8\textwidth]{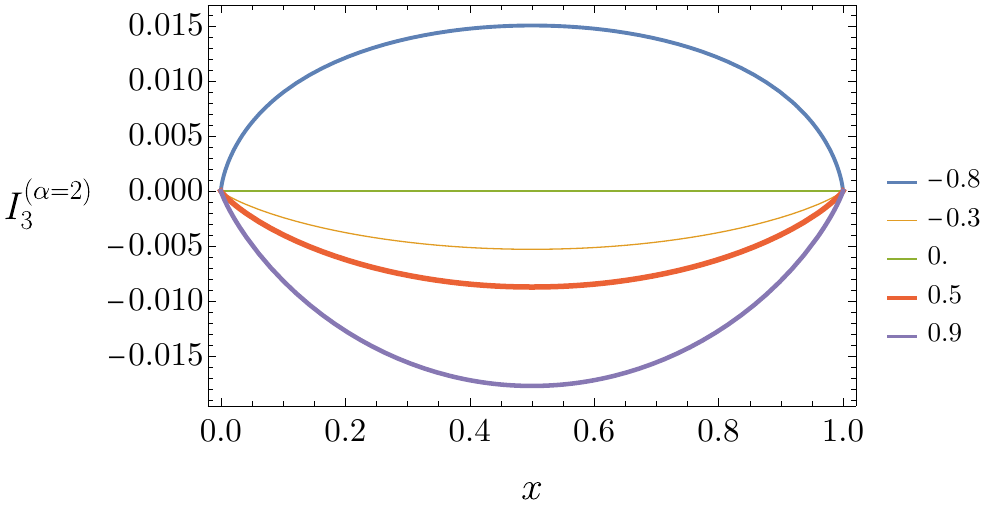}
    \caption{The R\'enyi-$2$ tripartite information $I_3^{(\alpha=2)}$ as a function of the cross ratio $x$ in the ground state of the fermionic XXZ chain for $h=0$ and different values of $\Delta$, reported in the legend. Depending on the model parameters, the tripartite information can be positive, negative or zero (the free case is $\Delta=0$).}
    \label{fig:fermionic}
\end{figure}

\begin{description}
\item[XXZ fermion chain:] 
Contrary to free-fermion systems, such as the Kitaev model, the R\'enyi-$\alpha$ tripartite information in the XXZ fermion chain is not identically zero. Specifically, we find
\begin{equation}\label{tripartite info fermionic XXZ}
     I_{3}^{(\alpha)}(x)=\frac{1}{\alpha-1}\log \left(  \sum_{\vec \varepsilon\in \{0,\frac{1}{2}\}^{\alpha-1}}\frac{\Theta\left[
    \begin{smallmatrix}
  \vec \varepsilon \\ \vec 0    
    \end{smallmatrix}\right]\left(\vec 0|\frac{1}{\ETA} \Omega\right)\Theta\left[
    \begin{smallmatrix}
  \vec \varepsilon \\ \vec 0    
    \end{smallmatrix}\right]\left(\vec 0|4\ETA \Omega\right)}{\left|\Theta\left(\vec 0 | \Omega\right)\right| ^2}\right)\, .
\end{equation}
Here $\ETA$ is the universal critical exponent proportional to the square of the compactification radius of the underlying bosonic theory, which for $h=0$ reads $\ETA=\frac{\arccos(-\Delta)}{\pi}$; $\Omega$ is the $(\alpha-1)\times (\alpha-1)$ period matrix of the Riemann surface $\mathcal R_\alpha$ with elements
\begin{equation}\label{period matrix def}
[\Omega]_{\ell n}=\frac{2i}{\alpha}\sum_{k=1}^{\alpha-1}\sin(\tfrac{\pi k}{\alpha})\cos(\tfrac{2\pi k(\ell-n)}{\alpha}) \beta_{\frac{k}{\alpha}}(x), \quad \beta_{\frac{k}{\alpha}}(x)\equiv \tfrac{P_{-k/\alpha}(2x-1)}{P_{-k/\alpha}(1-2x)}\, ,
\end{equation}
where $P_\mu(z)$ denotes the Legendre functions;  $
\Theta[\vec e](\vec z|M)
$ is the Riemann-Siegel theta function with characteristic, i.e., 
\begin{equation}\label{theta function with characteristic def}
\Theta[\vec e](\vec z|M)=\sum_{\vec m\in \mathbb Z^{\alpha-1}}e^{i\pi  (\vec m+\vec \varepsilon)^t M (\vec m+\vec \varepsilon)+2\pi i (\vec m+\vec\varepsilon)\cdot (\vec z+\vec\delta)}\qquad \vec e=\begin{pmatrix}
\vec \varepsilon\\
\vec \delta
\end{pmatrix}\, .
\end{equation}
Note that it is customary to omit the characteristic when it is zero, i.e., $\Theta\equiv \Theta[\Vec{0}]$. The behavior of the tripartite information for $\alpha=2$ is shown in Fig.~\ref{fig:fermionic}. The curves for higher $\alpha$ are barely distinguishable from those for $\alpha=2$ so we do not present them.

We also obtain the von Neumann tripartite information in the limit of small $x$ 
\begin{equation}\label{expansion tripartite information fermionic vN}
\begin{split}
      I_{3}^{(\mathrm{vN})}(x) =&-s_{\mathrm{vN}}(1)\frac{x}{2}+s_{\mathrm{vN}}(\tfrac{1}{\ETA})\left(\frac{x}{4}\right)^{\tfrac{1}{\ETA}}+s_{\mathrm{vN}}(4\ETA)\left(\frac{x}{4}\right)^{4\ETA}\\ & + s_{\mathrm{vN}}(\tfrac{1}{4\ETA}+\ETA)\left(\frac{x}{4}\right)^{\tfrac{1}{4\ETA}+\ETA}   +o\big(x^{\min(1,\tfrac{1}{\ETA},4\ETA)}\big), \qquad \textrm{as } x\to 0,
\end{split}
\end{equation}
where $s_{\mathrm{vN}}$ is defined as~\cite{Calabrese2011Entanglement}
\begin{equation}\label{s vN}
      s_{\mathrm{vN}}(\mathcal{A})=\frac{\sqrt{\pi}\Gamma(\mathcal{A}+1)}{2\Gamma(\mathcal{A}+\frac{3}{2})} ,\quad \mathcal{A}>0.
\end{equation}

We find, in particular, that the tripartite information is  negative for $\frac{1}{4}\leq\ETA\leq1$ (i.e., $\Delta\geq-\frac{\sqrt{2}}{2}$ when $h=0$), while in the XXZ spin chain it is always non-negative. For the R\'enyi-$\alpha$ tripartite information with $\alpha=2,3$ we report an unusual behavior: for certain values of $\Delta$, tripartite information changes the sign as a function of $x$ (see Fig.~\ref{fig:fermionic sign change}).

\begin{figure}
    \centering
    \includegraphics[width=0.75\textwidth]{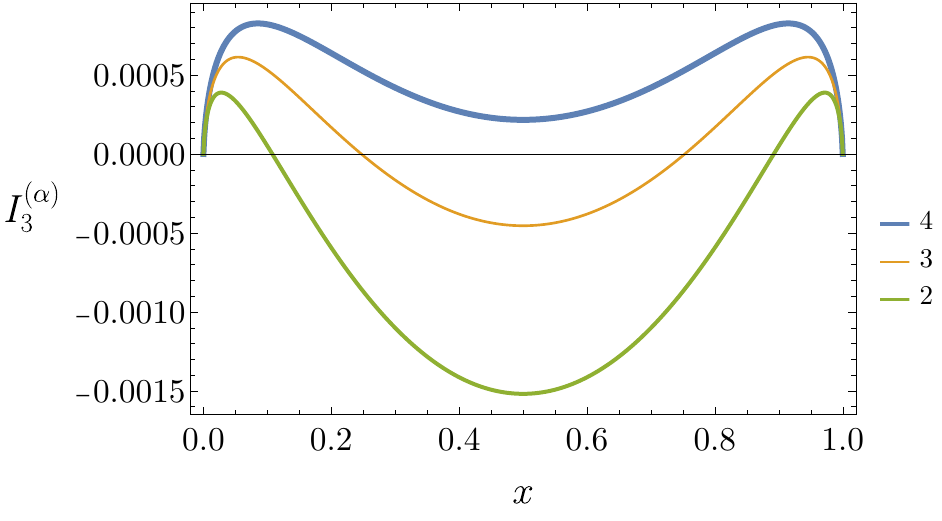}
    \caption{The R\'enyi-$\alpha$ tripartite information $I_3^{(\alpha)}$ as a function of the cross ratio $x$ in the ground state of the fermionic XXZ chain for $\Delta=-0.76$ and $h=0$. The values of the R\'enyi index $\alpha$ are reported in the legend. For $\alpha=2,3$ tripartite information changes sign as a function of $x$.}
    \label{fig:fermionic sign change}
\end{figure}

\begin{figure}
    \centering
	\begin{minipage}{0.85\textwidth}
		\centering
		\includegraphics[width=1\textwidth]{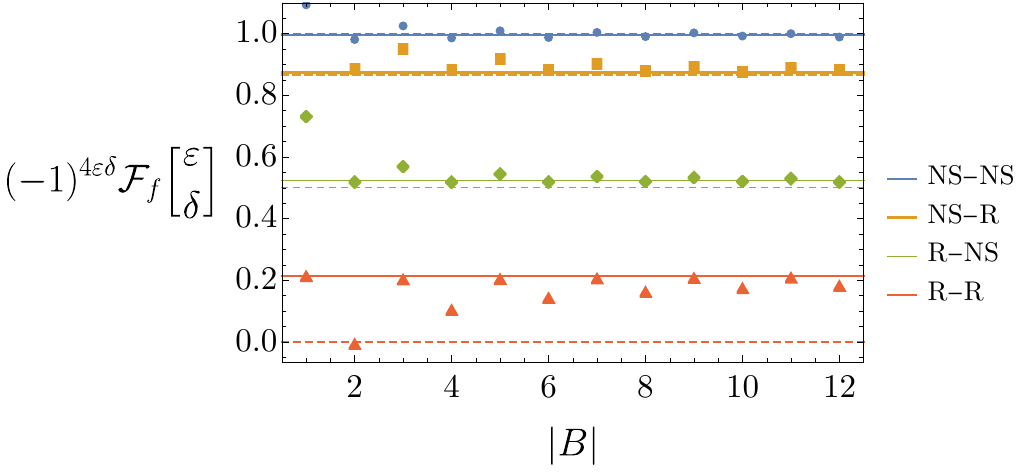}
		\subcaption{Contribution from different spin structures for $\alpha=2$ and $(\Delta,h)=(-0.3, 0) $. The three subsystems have equal lengths $|A|=|B|=|C|$, so the cross ratio is $x=1/4$. The dashed lines show the free case ($\Delta=0$) for comparison.}
	\end{minipage}
\\ \vspace{0.8 cm}
	\begin{minipage}{0.85\textwidth}
		\centering
		\includegraphics[width=1\textwidth]{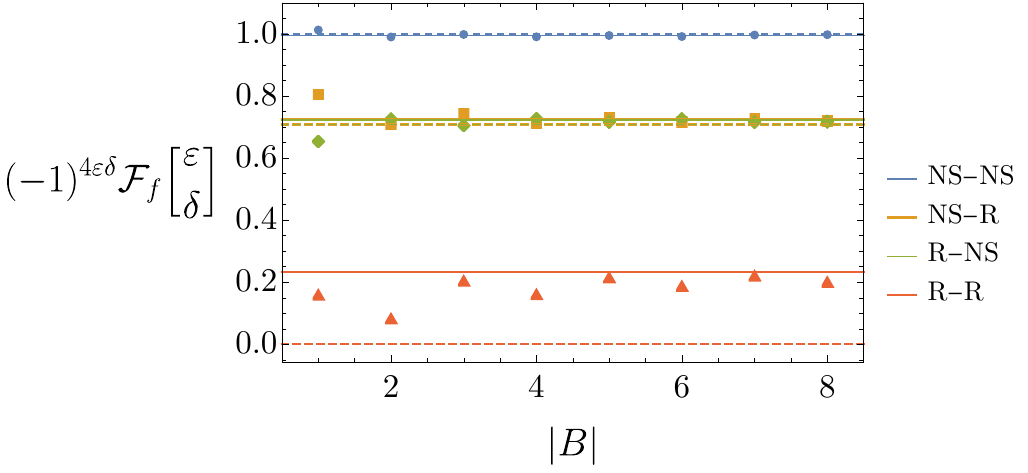}
		\subcaption{Contribution from different spin structures for $\alpha=2$ and $(\Delta,h)=(-0.3, 0) $. The subsystem lengths are set to be $(|A|,|B|,|C|)=(2\ell,\ell,3\ell)$ for $\ell=1,2,3,\ldots$, so the cross ratio is $x=1/2$. The lines R-NS and NS-R overlap. The dashed lines show the free case for comparison.}
	\end{minipage}
     
        \caption{Contributions of different spin structures $(\varepsilon,\delta)$ to the R\'enyi-$2$ tripartite information of three finite adjacent subsystems $A,B,C$ in the XXZ spin chain on infinite lattice.  Namely, function $\F{\varepsilon}{\delta}{A,B,C}$ (see Eq.~\eqref{eq:I3spinstr} and Eq.~\eqref{eq:spinstruct}) is shown for different subsystem lengths and all four torus spin structures. In each plot the lengths of the three subsystems are chosen so that the cross ratio $x=|A||C|/[(|A|+|B|)(|B|+|C|)]$ is fixed. The numerical data (represented by points) is obtained by considering the method of fusion of models, discussed in the text, and by incorporating tensor networks simulations. As the subsystems lengths get bigger, the data can be seen to converge towards the prediction (solid lines), given by Eq.~\eqref{eq:FfXXZ}, obtained from the massless Thirring QFT considerations. }
    \label{fig:fixed_x}
\end{figure}

\item[Expansion in the spin structure.]
The tripartite information in the ground state of a spin-flip invariant spin-$\frac{1}{2}$ chain Hamiltonian can be expanded as follows
\begin{equation}\label{eq:I3expansion}
I_3^{(\alpha)}(A,B,C)=\frac{1}{\alpha-1}\log\left(\frac{1}{2^{\alpha-1}}\sum_{\vec\varepsilon,\vec\delta\in\{0,\frac{1}{2}\}^{\alpha-1}}(-1)^{4\vec \varepsilon\cdot\vec\delta}\F{\vec\varepsilon}{\vec\delta}{A,B,C}\right)
\end{equation} 
where $\F{\vec\varepsilon}{\vec\delta}{A,B,C}$ describes the contribution from the underlying fermionic system with given spin structure---see Sec.~\ref{sec:fermions vs spins}. 
\item[XXZ spin chain.] 
In the Heisenberg XXZ model we find
\begin{equation}\label{eq:FfXXZ}
\F{\vec\varepsilon}{\vec\delta}{A,B,C} = 
   \sum_{\vec \mu\in \{0,\frac{1}{2}\}^{\alpha-1}}(-1)^{4\vec \mu\cdot\vec\delta}\frac{\Theta\left[
    \begin{smallmatrix}
  \vec\epsilon+\vec \mu \\ \vec 0    
    \end{smallmatrix}\right]\left(\vec 0|\frac{1}{\ETA} \Omega\right)\Theta\left[
    \begin{smallmatrix}
  \vec \mu \\ \vec 0    
    \end{smallmatrix}\right]\left(\vec 0|4\ETA \Omega\right)
    }{\left[\Theta(\vec 0 |\Omega
      )\right]^2} \, .
\end{equation} 
Note, in particular, that the fermionic tripartite information---\eqref{tripartite info fermionic XXZ}---is obtained by taking only the term with $\vec\varepsilon=\vec\delta=0$. While the sum over all spin structures has the well-known invariance under $\ETA\rightarrow\frac{1}{\ETA}$ (this is shown explicitly in Appendix~\ref{appendix reproducing CCT}), each individual term exhibits another symmetry:
\begin{equation}
\F{\vec\varepsilon}{\vec\delta}{A,B,C}\xrightarrow{\ETA\rightarrow\frac{1}{4\ETA}} (-1)^{4\vec\varepsilon\cdot\vec\delta}\F{\vec\varepsilon}{\vec\delta}{A,B,C}\, .
\end{equation}
We point out that the transformation $\ETA\rightarrow\frac{1}{4\ETA}$ distinguishes the even spin structures from the odd ones. Here we call a spin structure even or odd depending on whether $(-1)^{4\vec\varepsilon\cdot \vec\delta}=1$ or $(-1)^{4\vec\varepsilon\cdot \vec\delta}=-1$ respectively.
For $h=0$ this transformation connects two XXZ chains as long as $\Delta\geq -\frac{\sqrt{2}}{2}$. For $h=0$ and  $\Delta< -\frac{\sqrt{2}}{2}$ a similar effect can be achieved by applying a double transformation
$
\ETA\rightarrow\frac{1}{4\ETA}\rightarrow 4\ETA
$,
which doesn't preserve the terms separately but, effectively, acts as it reversed the sign of the odd spin structures.

\item[Jordan-Wigner deformations:] We work out the tripartite information in the Jordan-Wigner deformations---\eqref{eq:JWdef}---of the XXZ spin chain. In agreement with the discussion of Ref.~\cite{Fagotti2012New}, we find that the tripartite information is given by
\begin{equation}\label{eq:JW def tripartite explicitly}
I_3^{(\alpha;\nu)}(x)=\frac{1}{\alpha-1}\log\left(\frac{1}{2^{\alpha-1}}\sum_{\vec\varepsilon,\vec\delta\in\{0,\frac{1}{2}\}^{\alpha-1}}(-1)^{4\vec \varepsilon\cdot\vec\delta}\F[(\nu)]{\vec\varepsilon}{\vec\delta}{x}\right)
\end{equation}
with
\begin{equation}
\F[(\nu)]{\vec\varepsilon}{\vec\delta}{x}=\Bigl(\F{\vec\varepsilon}{\vec\delta}{x}\Bigr)^\nu\, ,
\end{equation} 
where $\F{\vec\varepsilon}{\vec\delta}{x}$ is given by Eq.~\eqref{eq:FfXXZ}. For $\nu=2$ the spin tripartite information matches the fermionic one, which, in turn, is equal to twice the tripartite information in the fermionic XXZ chain
\begin{equation}\label{tripartite info JW 2}
    I_{3}^{(\alpha;\nu=2)}(x)= 2\ I_{3}^{(\alpha;\mathrm{fXXZ})}(x) \; ,
\end{equation}
where $I_{3}^{(\alpha;\mathrm{fXXZ})}$ is given by Eq.~\eqref{tripartite info fermionic XXZ}. For small $x$ and any $\nu$ we have computed the von Neumann tripatite information. The results are given in Sec.~\ref{sec JW deformations}. The tripartite information can be positive, negative or zero, depending on the model parameters.

\begin{figure}
    \centering
    \includegraphics[width=0.75\textwidth]{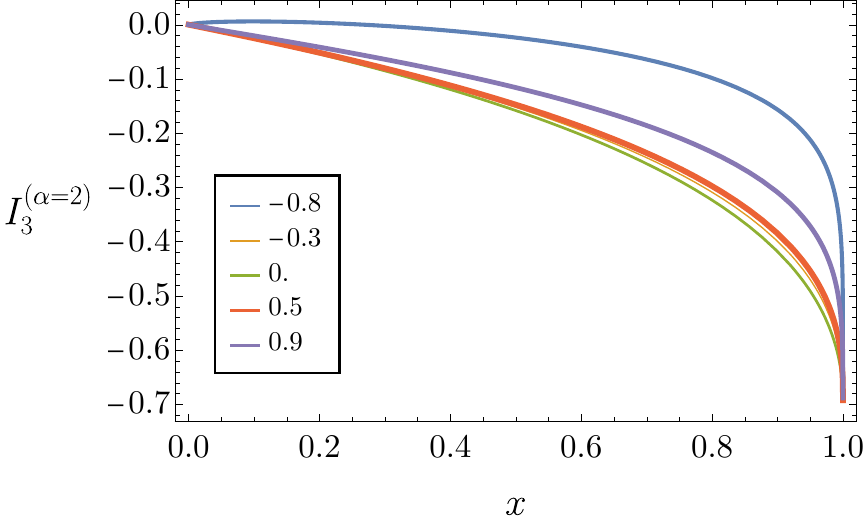}
    \caption{Prediction for the R\'enyi-$2$ tripartite information $I_3^{(\alpha=2)}$ as a function of the cross ratio $x$ in the stationary state after the small quantum quench in the XXZ spin chain for $h=0$ and different values of $\Delta$, reported in the legend. The tripartite information depends on the model parameters, but, irrespectively of them, it reaches the value $-\log 2$ as $x\to 1^-$.}
    \label{fig:quench}
\end{figure}

\item[From tripartite information to spin structures:]
We propose two different procedures for computing the contribution of each spin structure of the massless Thirring model on a Riemann surface to the R\'enyi-$\alpha$ tripartite information in the XXZ chain, with the main focus on the torus, i.e. $\alpha=2$.

\item \ul{Jordan-Wigner $\nu$-deformation of the XXZ spin chain}
We express the partition function with given spin structure in terms of the R\'enyi-$2$ tripartite information of four JW deformations of XXZ (including the XXZ model itself). 
For example, the term corresponding to  Ramond boundary conditions on both cycles, which we have indicated by
$\F{\nicefrac{1}{2}}{\nicefrac{1}{2}}{x}$, can be written as follows
\begin{equation}\label{from tripartite info in JW def to spin structure}
\F{\nicefrac{1}{2}}{\nicefrac{1}{2}}{x}=\tfrac{4e^{4 I_3^{(2;1)}(x)}-12 e^{2I_3^{(2;1)}(x)}e^{I_3^{(2;2)}(x)}+3e^{2I_3^{(2;2)}(x)}+8e^{ I_3^{(2;1)}(x)}e^{ I_3^{(2;3)}(x)}-3e^{ I_3^{(2;4)}(x)}}{12e^{ I_3^{(2;1)}(x)}e^{ I_3^{(2;2)}(x)}-8e^{3I_3^{(2;1)}(x)}-4 e^{ I_3^{(2;3)}(x)}}\, ,
\end{equation}
where $I_3^{(2;\nu)}(x)$ denotes the R\'enyi-2 tripartite information of the JW $\nu$-deformation of XXZ. 
In fact, a posteriori we observe that this particular term can be obtained just studying the tripartite information in XXZ chains. By exploiting the symmetry mentioned before, we indeed find
\begin{equation}
\F[(\ETA)]{\nicefrac{1}{2}}{\nicefrac{1}{2}}{x}=e^{ I_3^{(2)}(x;\ETA)}-e^{ I_3^{(2)}(x;\min(\nicefrac{1}{(4\ETA)},4\ETA))}\, ,
\end{equation}
where the subscripts $\ETA$, $4\ETA$, and $\nicefrac{1}{(4\ETA)}$ identify the XXZ model. 

\item \ul{Interleaved models} The second procedure is based on the study of a class of two-site shift invariant Hamiltonians such as the following
\begin{equation}
\bs H=\bs H_e+\bs H_o
\end{equation}
where $\bs H_e$ and $\bs H_o$ are commuting Hamiltonians, $[\bs H_e,\bs H_o]=0$, with
\begin{align}
\bs H_e=&\sum_\ell\left( \bs\sigma_{2\ell}^x\bs\sigma_{2\ell+1}^z\bs\sigma_{2\ell+1}^x+\bs\sigma_{2\ell}^y\bs\sigma_{2\ell+1}^z\bs\sigma_{2\ell+1}^y+\Delta\bs\sigma_{2\ell}^z\bs\sigma_{2\ell+2}^z -2h\bs\sigma_{2\ell}^z \right)\, \label{interleaved XXZ XY the XXZ part}\\
\bs H_o=&-\sum_\ell\left[(1+\gamma)\bs\sigma_{2\ell-1}^x\bs\sigma_{2\ell}^z\bs\sigma_{2\ell+1}^x+(1-\gamma)\bs\sigma_{2\ell-1}^y\bs\sigma_{2\ell}^z\bs\sigma_{2\ell+1}^y-2h'\bs \sigma_{2\ell-1}^z\right]\label{interleaved XXZ XY the XY part} \, .
\end{align}
This describes an XXZ model interleaved with an XY model, and the ground state is the tensor product of two Fock states associated with independent sets of Jordan-Wigner fermions.

We point out
\begin{equation}
\F{\vec\varepsilon}{\vec\delta}{A,B,C}=\F[(\mathrm{XXZ}_{\Delta,h})]{\vec\varepsilon}{\vec\delta}{A_e,B_e,C_e}\F[(\mathrm{XY}_{\gamma,h'})]{\vec\varepsilon}{\vec\delta}{A_o,B_o,C_o}\, ,
\end{equation}
where we used the notation $X_{e(o)}$ to indicate the set of even (odd) sites of subsystem $X$ and the elements on the right hand side correspond to the underlying XXZ---\eqref{eq:FfXXZ}---and XY chains--- Section~\ref{s:noncrit}. 
As a result, the ground state tripartite information can still be written as a linear combination of Thirring partition functions with given spin structure, the coefficients of the combination depending on $\gamma$ and $h'$. 
We exploit this connection with the Thirring partition functions to perform a numerical check of the analytical predictions for given spin structure. Some results are presented in Fig.~\ref{fig:fixed_x}. To the best of our knowledge, this is the first numerical analysis of its kind in an interacting spin chain (we note, in particular, that the numerical analysis of Refs~\cite{Fagotti2012New,Coser2016Spin} had the same goal but the technique that was used is specific to noninteracting spin chains).

\item[Global quenches from critical states:] 
Refs~\cite{Maric2022Universality,Maric2023Universality} studied the stationary value of the tripartite information long after having prepared the system in a critical state (a collection of Fermi seas). It was in particular observed that the main effect of the quench is to turn some algebraically decaying correlations into exponentially decaying ones. The  correlations that become exponentially small are those associated with the vertex operator in the underlying free boson compactified on a circle. Assuming this to remain true also in the interacting case, we speculate about the behaviour of the tripartite information after global quenches. This corresponds to dropping all spin structures with $\vec\varepsilon\neq \vec 0$ and we conjecture, in particular, a formula for the tripartite information in the generalised Gibbs ensemble emerging after a quench of the anisotropy $\Delta_0\rightarrow \Delta$ in the XXZ spin-$\frac{1}{2}$ chain in the limit of small quench
\begin{equation}\label{eq:I3quench}
\lim_{\Delta_0\rightarrow\Delta}\lim_{t\rightarrow\infty}I_3^{(\alpha)}(x)=\frac{1}{\alpha-1}\log \frac{\Theta(\vec 0|\frac{1}{\ETA} \Omega)\Theta(\vec 0|4\ETA \Omega)}{[\Theta(\vec 0 | \Omega)] ^2}\, .
\end{equation}
The behavior for $\alpha=2$ is presented in Fig.~\ref{fig:quench}. We have not been able to compare this conjecture with data from numerical simulations, but the interest aroused by nonequilibrium time evolution in lattice systems is triggering rapid development of efficient algorithms that could make a numerical verification possible  before long.  

From Eq.~\eqref{eq:I3quench} we have been able to obtain the large $x$ expansion of the von Neumann tripartite information,
\begin{equation}\label{tripartite expansion quench vN}
\begin{split}
   & \lim_{\Delta_0\rightarrow\Delta}\lim_{t\rightarrow\infty}I_3^{(\mathrm{vN})}(x)\\&\xrightarrow{1-x\ll 1}-\log 2 \  + s_{\mathrm{vN}}(\ETA)\left(\frac{1-x}{4}\right)^{\ETA}+s_{\mathrm{vN}}(\tfrac{1}{4\ETA})\left(\frac{1-x}{4}\right)^{\frac{1}{4\ETA}}+o\left((1-x)^{\min(\ETA, \tfrac{1}{4\ETA})}\right), 
\end{split}
\end{equation}
where the function $s_{\mathrm{vN}}$ is defined in Eq.~\eqref{s vN}.

Eqs~\eqref{eq:I3quench} and \eqref{tripartite expansion quench vN} allow us to compute the residual tripartite information, which is the limit $x\rightarrow1^-$. Remarkably, we find that it is independent of the anisotropy and equals $-\log 2$. Even if our analysis is based on a conjecture that deserves further investigation, our findings support the claim of universality made in Ref.~\cite{Maric2022Universality}, which was based on the study of a wide class of noninteracting spin-chain models. 
\end{description}

\section{Tripartite information}

\subsection{Fermions vs spins: spin structures}\label{sec:fermions vs spins}

In spin-lattice systems, the tensor product structure of the Hilbert space allows for a clear notion of separable states and entanglement. In composite fermionic systems the Fock space does not exhibit the same  structure, and the situation is not as simple. We refer the reader to Ref.~\cite{Banuls2007} for a discussion on this topic. In this section we explore the differences between the entanglement entropies of blocks of spins in spin-$\frac{1}{2}$ chains with spin-flip symmetry and the entropies of the corresponding spinless fermions defined through a Jordan-Wigner transformation.

\paragraph{Reduced density matrix.}In a spin-$\frac{1}{2}$ system, such as the XXZ chain, the sites of a subsystem $X$ are associated with the indices of the spin operators $\bs\sigma_\ell^{\gamma}$, $\ell\in X$, for $\gamma=x,y,z$.
If the whole system is in a state $\ket{\Psi}$, the reduced density matrix $\rho_X$ of a subsystem $X$ reads
\begin{equation}\label{density matrix sigma basis tensor}
        \rho_X=\frac{1}{2^{|X|}}\sum_{\gamma_{\ell}\in\{0,x,y,z\},\ell\in X} \bra{\Psi}\prod_{\ell\in X}\bs \sigma^{\gamma_\ell}_\ell \ket{\Psi}\bigotimes_{\ell\in X} \sigma^{\gamma_\ell} \;,
\end{equation}
where the sum is over an orthogonal basis of operators on $X$, given by all possible products of Pauli matrices (referred to in the following as ``strings''), and we use the standard convention $\sigma^0\equiv \mathbb{I}$. Note that we use bold fonts for operators defined in the Hilbert space of the whole lattice and unbold fonts for the ones defined in the Hilbert space associated to a subsystem or a single site.

In systems of spinless fermions one can associate a subsystem $X=\{\ell_1,\ell_2,\ldots, \ell_{|X|}\}$ with the indices of the fermionic operators $\bs c_\ell^\dagger, \bs c_\ell$ that create and annihilate, respectively, a fermion at site $\ell$. In general, such a notion of a fermionic subsystem is problematic: the operators constructed with fermions in the set $\{\bs c_\ell| \ell\in X\}$ do not  commute with every operator in the complementary set $\{\bs c_\ell| \ell\in \bar X\}$. However, we deal with states in which expectation values of products of an odd number of fermionic operators vanish and such problems do not arise.  By introducing the fermionic operators $ c_\ell^\dagger, c_\ell$, $\ell\in X$, that are defined on the $2^{|X|}$ dimensional Fock space asscociated to the subsystem (note the unbold notation), one can define the operator
\begin{equation}\label{reduced density matrix fermions smaller Hilbert space}
         \rho_{f,X}= \frac{1}{2^{|X|}}\sum_{F_X} \braket{\Psi|\bs F_X|\Psi} F_X^\dagger
      \, ,
\end{equation}
where we have denoted
\begin{equation}
    \bs F_X= (\bs c_{\ell_1}^\dagger)^{\gamma_{1}}(\bs c_{\ell_1})^{\delta_{1}} \ldots (\bs c_{\ell_{|X|}}^\dagger)^{\gamma_{{|X|}}}(\bs c_{\ell_{|X|}})^{\delta_{{|X|}}} \; , \quad   F_X= (c_{\ell_1}^\dagger)^{\gamma_{1}}(c_{\ell_1})^{\delta_{1}} \ldots ( c_{\ell_{|X|}}^\dagger)^{\gamma_{{|X|}}}(c_{\ell_{|X|}})^{\delta_{{|X|}}} \; ,
\end{equation}
with $\gamma_{\ell},\delta_{\ell}\in\{0,1\}, \ell\in X$ a generic fermionic operator on the subsystem. The sum in \eqref{reduced density matrix fermions smaller Hilbert space} is over all such operators, i.e. over all $\gamma_{\ell},\delta_{\ell}\in\{0,1\}, \ell\in X$. Since we deal with states where the expectation value of any product of an odd number of fermions vanishes, the sum can be further restricted with the condition $(-1)^{\sum_j \gamma_j}=(-1)^{\sum_j \delta_j}$. The operator~\eqref{reduced density matrix fermions smaller Hilbert space} is a fermionic reduced density matrix in the sense that it reproduces the expectation values of operators associated with the subsystem,
\begin{equation}
    \tr \left(\rho_{f,X} F_X \right)=\bra{\Psi}\bs F_X\ket{\Psi}.
\end{equation}
The entanglement entropies of a subsystem $X$ of fermions can then be computed in an analogous way to spins, through $S_\alpha(X)=\log[\tr(\rho_{f,X}^\alpha)]/(1-\alpha)$ and $S_1(X)=-\tr[\rho_{f,X}\log\rho_{f,X}]$. One can also take it as a definition of fermionic entanglement entropies.

\paragraph{Jordan-Wigner transformation.} In view of comparing the entropies of spins and fermions under the Jordan-Wigner transformation, due to its non-local properties it is more convenient to work with the Hilbert space associated to the whole lattice. Thus, instead of the reduced density matrices, we introduce the operators,
\begin{equation}\label{eq:Rspin}
\begin{split}
\bs R_X(\mathrm{spins}) &=\mathrm{tr}[\bs 1_X]\rho_X\otimes I_{\bar X}\\
& =\sum_{\gamma_{\ell}\in\{0,x,y,z\},\ell\in X} \braket{\Psi|\prod_{\ell\in X}\bs \sigma^{\gamma_\ell}_\ell |\Psi}\prod_{\ell\in X} \bs \sigma_\ell^{\gamma_\ell}
\end{split}
\end{equation}
for spins, and
\begin{equation}\label{reduced density matrix fermions}
\begin{split}
\bs R_X &= \sum_{\bs F_X} \braket{\Psi|\bs F_X|\Psi}\bs F_X^\dagger \\
& =\!\!\!\!\!\!\sum_{\gamma_{\ell},\delta_{\ell}\in\{0,1\}, \ell\in X\atop (-1)^{\sum_j \gamma_j}=(-1)^{\sum_j \delta_j}} \!\!\!\!\!\!\braket{\Psi|(\bs c_{\ell_1}^\dagger)^{\gamma_{1}}(\bs c_{\ell_1})^{\delta_{1}} \ldots (\bs c_{\ell_{|X|}}^\dagger)^{\gamma_{{|X|}}}(\bs c_{\ell_{|X|}})^{\delta_{{|X|}}}|\Psi}
        ( \bs c^\dag_{\ell_{|X|}})^{\delta_{{|X|}}}( \bs c_{\ell_{|X|}})^{\gamma_{{|X|}}}\cdots ( \bs c_{\ell_1}^\dag)^{\delta_{1}}   
        (\bs c_{\ell_1})^{\gamma_{1}}
\end{split}
\end{equation}
for fermions. Note that \eqref{density matrix sigma basis tensor} and \eqref{eq:Rspin} do not differ only in the normalisation, but also in the space where they are defined, as we emphasized by using or not using bold fonts, and similarly for \eqref{reduced density matrix fermions smaller Hilbert space} and \eqref{reduced density matrix fermions}. The entanglement entropies in terms of $\bs R_X$ read
\begin{equation}\label{entropies in terms of auxiliary R}
\begin{aligned}
S_\alpha(X)=&\frac{1}{1-\alpha}\log\frac{\mathrm{tr}[\bs R_X^\alpha]}{\mathrm{tr}[\bs 1]}+\log\mathrm{tr}[\bs 1_X]\equiv \frac{1}{1-\alpha}\log\braket{\Psi|\bs R_X^{\alpha-1}|\Psi}+\log\mathrm{tr}[\bs 1_X]\\
S_{1}(X)=&-\frac{\mathrm{tr}[\bs R_X\log \bs R_X]}{\mathrm{tr}[\bs 1]}+\log\mathrm{tr}[\bs 1_X]\equiv -\braket{\Psi|\log\bs R_X|\Psi}+\log\mathrm{tr}[\bs 1_X]\, .
\end{aligned}
\end{equation}

The Jordan-Wigner transformation $\bs \sigma_\ell^+=\prod_{j<\ell}(2\bs c^\dag_j \bs c_j-1) \bs c^\dag_\ell$ maps the spins to spinless fermions and vice versa. The relation between the entropy of the subsystem $X$ of spins and the corresponding one of the Jordan-Wigner fermions depends on whether $X$ is connected or not.
\begin{itemize}
\item If $X$ is a connected set of sites, the auxiliary operators for spins and fermions are equal~\cite{Vidal2003,Peschel2003,Peschel2009,Latorre2004}, i.e. $\bs R_X(\mathrm{spins})=\bs R_X$. 
The R\'enyi entropies of a single interval of spins and those  of the corresponding interval of Jordan-Wigner fermions
are, in turn, equal. Note that if we use the Jordan-Wigner transformation in the $2^{|X|}$ dimensional Hilbert space associated to the subsystem $X$, we obtain
$$\big(\bigotimes_{j=1}^{\ell-1}\sigma^z\big)\otimes\sigma^+ \otimes \big(\bigotimes_{j=\ell+1}^{|X|}\mathbb{I}\big)=\prod_{j<\ell}(2 c^\dag_j  c_j-1) c^\dag_\ell \, ,$$
and hence we can adopt the point of view that the reduced density matrix of spins is equal to the fermionic one, given by Eq.~\eqref{reduced density matrix fermions smaller Hilbert space}. 

\item When $X$ is not connected,
the non-locality of the Jordan-Wigner transformation starts having effects. In particular, it has been observed in Ref.~\cite{Fagotti2010disjoint} that the Renyi-$\alpha$ entanglement entropy of two disjoint intervals $X=A\cup C$ for a spin system in a state $\ket{\Psi}$ is equal to the entropy of the corresponding Jordan-Wigner fermions in the state
\begin{equation}\label{spin fermion state relation disjoint}
    \ket{\Psi'}=\left[\frac{1+(-1)^{\bs F_B} }{2}+\frac{1-(-1)^{\bs F_B} }{2}(-1)^{\bs F_C}\right]\ket{\Psi},
\end{equation}
    where $\bs F_B=\sum_{\ell\in B} \bs c^\dagger_\ell \bs c_\ell $ and $\bs F_C=\sum_{\ell\in C} \bs c^\dagger_\ell \bs c_\ell $ is the fermion number on interval $B$ and $C$ respectively and $(-1)^{\bs F}\equiv \exp(i\pi \bs F)$. 
That is to say, the operator $\bs R_X$ for spins, defined in~\eqref{eq:Rspin}, can be expressed as a sum of four fermionic operators,
\begin{equation}\label{step auxiliary R disjoint}
    \bs R_X (\mathrm{spins})= \frac{\bs R^{++}+\bs R^{+-}}{2}+(-1)^{\bs F_B} \frac{\bs R^{-+}-\bs R^{--}}{2},
\end{equation}
where
\begin{equation}
 \bs R^{++}=\bs R_{AC}, \qquad  \bs R^{-+}=\sum_{F_X} \braket{\Psi|(-1)^{\bs F_B}\bs F_X|\Psi} F_X^\dagger, \qquad\bs R^{s-}=(-1)^{\bs F_{C}} \bs R^{s+} (-1)^{\bs F_{C}}, \, s=\pm 1 ,
\end{equation}
and $\bs R_{AC}$ is defined in~\eqref{reduced density matrix fermions}. Furthermore, the string operator $(-1)^{\bs F_B}$ in~\eqref{step auxiliary R disjoint} can be removed without any effect on the entanglement entropies. Namely, $\bs R_X$ can be replaced by the operator 
\begin{equation}
    \bs R'_X(\mathrm{spins})= \frac{\bs R^{++}+\bs R^{+-}+\bs R^{-+}-\bs R^{--}}{2}
\end{equation}
in Eq.~\eqref{entropies in terms of auxiliary R}, indeed $\tr[ (\bs R_{X}'(\mathrm{spins}))^\alpha]=\tr[ (\bs R_X(\mathrm{spins})^\alpha]$. This allows us to express the entanglement entropies in a less pedantic way than through the auxiliary operators $\bs R_X$, with the help of the Jordan-Wigner transformation in the $2^{|A|+|C|}$ dimensional Hilbert space associated with the subsystem $A\cup C$. The exact relations are reported in the forthcoming Eq.~\eqref{eq:pseudoRDM}.
\end{itemize}

\paragraph{Example.} We illustrate the difference between the entanglement of spins and fermions, related through the Jordan-Wigner transformation, with a simple example. We consider a system of four qubits, dual to a system of four fermions. We denote the four adjacent sites by $A, B, C, D$ and consider the bipartition of the system into subsystems $A\cup C$ and $B\cup D$, thus each consisting of two disjoint sites.

The state
\begin{equation}
    \ket{\Psi_1}=\frac{1}{\sqrt{2}}\left(\ket{\downarrow\downarrow}+ \ket{\uparrow\uparrow}\right)_{AC}\otimes \frac{1}{\sqrt{2}}\left(\ket{\downarrow\downarrow}+ \ket{\uparrow\uparrow}\right)_{BD}
\end{equation}
is separable, i.e., it is  a product state with respect to the bipartition into $A\cup C$ and $B\cup D$ ($S_\alpha(A\cup C)=0$). In terms of fermions, instead, the state reads
\begin{equation}\label{example state in terms of fermions}
    \ket{\Psi_1}=\frac{1}{2}(1+\bs c_A^\dagger\bs  c_C^\dagger+\bs c_B^\dagger\bs  c_D^\dagger+\bs c_A^\dagger\bs  c_B^\dagger\bs c_C^\dagger\bs  c_D^\dagger)\ket{0}.
\end{equation}
Roughly speaking, the state is not a fermionic product state, as witnessed by the inequality $\braket{(\bs c_A\bs c_C)(\bs c_B \bs c_D)}_{\Psi_1}\neq \braket{\bs c_A\bs c_C}_{\Psi_1}\braket{\bs c_B \bs c_D}_{\Psi_1}$, in which only the left hand side is non-zero. The fermionic entropy $S_{\alpha,f}(A\cup C)$ is indeed nonzero and equals $\log 2$.

On the other hand, the state
\begin{equation}
    \ket{\Psi_2}=\frac{1}{2}(1+\bs c_A^\dagger\bs  c_C^\dagger)(1+\bs c_B^\dagger\bs  c_D^\dagger)\ket{0}
\end{equation}
can be interpreted as a fermionic product state and has zero fermionic entanglement entropy $S_{\alpha,f}(A\cup C)=0$. In terms of spins, it reads
\begin{equation}
    \ket{\Psi_2}=\frac{1}{\sqrt{2}}\left[\ket{\downarrow\downarrow}_{AC} \otimes \frac{1}{\sqrt{2}}( \ket{\downarrow\downarrow}+\ket{\uparrow\uparrow})_{BD}+\ket{\uparrow\uparrow}_{AC} \otimes \frac{1}{\sqrt{2}}( \ket{\downarrow\downarrow}-\ket{\uparrow\uparrow})_{BD}\right],
\end{equation}
which is instead an entangled state. The entropy of spins $S_\alpha(A\cup C)$ now equals $\log 2$. One could easily see that relation \eqref{spin fermion state relation disjoint} gives the correct entropy in both examples.

\paragraph{Spin structures: lattice.} The spin RDM $\rho_{AC}$ (defined in~\eqref{density matrix sigma basis tensor}) is equivalent to the sum of four pseudo-RDMs
\begin{equation}
    \rho_{AC}'= \frac{\rho^{++}_{AC}+\rho_{AC}^{+-}+\rho_{AC}^{-+}-\rho_{AC}^{--}}{2}\, ,
\end{equation}
in the sense that the two matrices have the same R\'enyi-$\alpha$ entanglement entropies ($\tr[ \rho_{AC}^\alpha]=\tr[ (\rho_{AC}')^\alpha]$). The four pseudo-RDMs are defined as follows
\begin{equation}\label{eq:pseudoRDM}
\begin{split}
& \rho_{AC}^{+s'}=\frac{1}{2^{|A|+|C|}} \left[\sum_{\mathrm{even}} \braket{\bs O_A \bs O_C} O_A\otimes O_C+s'\sum_{\mathrm{odd}}\braket{\bs O_A (-1)^{\bs F_B} \bs O_C } O_A\otimes O_C \right], \quad s'=\pm 1 \\
& \rho_{AC}^{-s'}=\frac{1}{2^{|A|+|C|}} \left[\sum_{\mathrm{even}} \braket{\bs O_A (-1)^{\bs F_B}\bs O_C} O_A\otimes O_C+s'\sum_{\mathrm{odd}}\braket{\bs O_A \bs O_C } O_A\otimes O_C \right] , \quad s'=\pm 1
\end{split}
\end{equation}
where $O_A=\bigotimes_{\ell\in A} \sigma^{\gamma_\ell}$ ($\bs O_A=\prod_{\ell\in A} \bs\sigma_\ell^{\gamma_\ell}$ ) denotes a generic string of Pauli operators on subsystem $A$, $O_C$ ($\bs O_C$) denotes an analogous string for subsystem $C$, $\sum_{\mathrm{even}}$ denotes the sum over all $O_A$ and $O_C$ which are even under $S^z_A\equiv \bigotimes_{\ell\in A}\sigma^z$ and $S^z_C\equiv\bigotimes_{\ell\in C}\sigma^z$ respectively ($[O_A,S_{A}^z]=0, [O_C,S_{C}^z]=0$), and $\sum_{\mathrm{odd}}$ denotes the sum over all $O_A$ and $O_C$ which are odd under $S^z_A$ and $S^z_C$ respectively ($\{O_A,S_{A}^z\}=0, \{O_C,S_{C}^z\}=0$).

Note that both $\rho_{AC}^{++}$ and $\rho_{AC}^{+-}$ are proper RDMs, they are indeed equivalent to the fermionic reduced density matrix~\eqref{reduced density matrix fermions smaller Hilbert space}. 
The matrices $\rho_{AC}^{\pm  s'}$, $s'=\pm 1$, on the other hand, cannot be interpreted as density matrices because they are not positive semidefinite. This is why, in general, we are referring to them as ``pseudo-RDMs'' rather than RDMs.

The R\'enyi-$\alpha$ tripartite information can be written as follows:
\begin{equation}\label{eq:I3spinstr}
I_3^{(\alpha)}(A,B,C)=\frac{1}{\alpha-1}\log\left(\frac{1}{2^{\alpha-1}}\sum_{\vec\varepsilon,\vec\delta\in\{0,\frac{1}{2}\}^{\alpha-1}}(-1)^{4\vec \varepsilon\cdot\vec\delta}\F{\vec\varepsilon}{\vec\delta}{A,B,C}\right)
\end{equation}
where
\begin{equation}\label{eq:spinstruct}
\F{\vec\varepsilon}{\vec\delta}{A,B,C}=\frac{\mathrm{tr}[\rho_{AB}^\alpha]\mathrm{tr}[\rho_{BC}^\alpha]}{\mathrm{tr}[\rho_A^\alpha]\mathrm{tr}[\rho_B^\alpha]\mathrm{tr}[\rho_C^\alpha]\mathrm{tr}[\rho_{ABC}^\alpha]}\mathrm{tr}\left[\prod\limits_{k=1}^\alpha\rho_{AC}^{s_k[\vec\varepsilon],s_k'[\vec \delta]}\right]
\end{equation}
with
\begin{equation}
\begin{aligned}
s_k[\vec\varepsilon]=&\begin{cases}(1-4\varepsilon_1) \, , &k=1\\
(1-4\varepsilon_k)(1-4\varepsilon_{k-1})\, , & 1<k< \alpha \\
(1-4\varepsilon_{\alpha-1}) \, , &k=\alpha
\end{cases}\\
s'_k[\vec\delta]=&\begin{cases}1 \, , &k=1\\
\prod\limits_{j=1}^{k-1}(1-4\delta_j)\, ,  & 1<k\leq \alpha
\end{cases}\, .
\end{aligned}
\end{equation}
Note that, due to more or less explicit symmetries, there are alternative equivalent identifications (e.g., we could change the sign of all $s_k'$). Incidentally, we isolated the sign $(-1)^{4\vec \varepsilon\cdot\vec\delta}$ both because it naturally emerges in the expansion in pseudo-RDMs and because it brings some advantages that will be discussed later.

Eq.~\eqref{eq:I3spinstr} is exact on the lattice. Ref.~\cite{Fagotti2010disjoint} exploited this structure to compute the R\'enyi entropies in spin systems that are mapped to free fermions by the Jordan-Wigner transformation. Indeed, in that case, $\rho^{s,s'}_{AC}$ turns out to be Gaussian  for each pair $(s,s')$, whereas the full reduced density matrix is not. When specialised to the ground state of critical chains, these notations make it manifest the correspondence pointed out in Ref.~\cite{Coser2016Spin}, according to which $\F{\vec\varepsilon}{\vec\delta}{A,B,C}$ is the partition function of  the  underlying fermionic quantum field theory on the Riemann surface associated with the R\'enyi-$\alpha$  entropy, with given boundary conditions along the cycles of the canonical homology basis. We now discuss this identification.

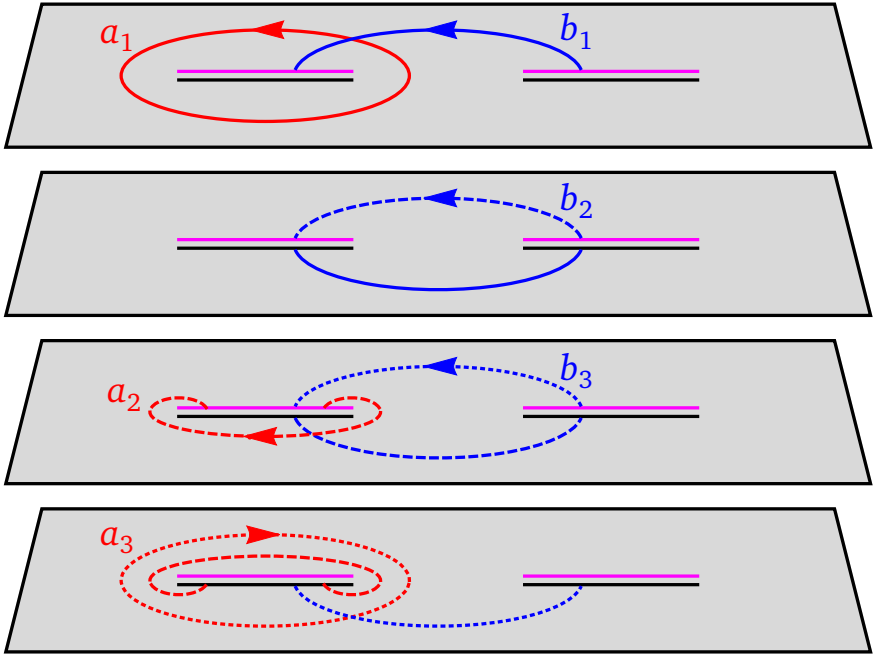
\begin{figure}
   
    \centering
    \begin{tikzpicture}[thick,scale=0.9, every node/.style={scale=1.0}]

\node[inner sep=0pt] (img1) at (0.,0.) {\includegraphics[width= \textwidth]{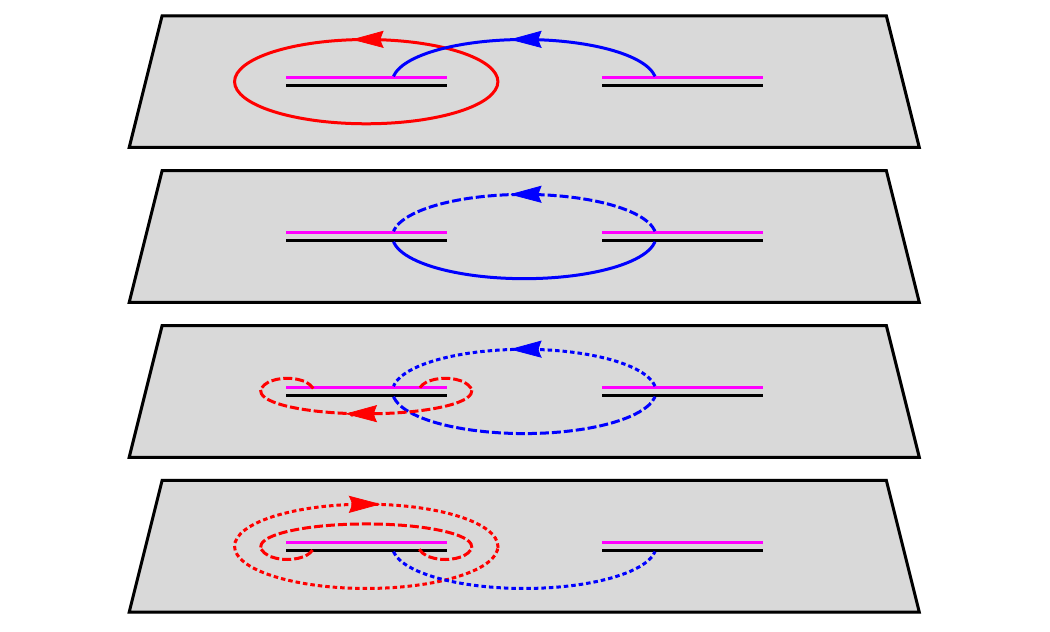}};

\draw (-4.7,4.3) node{\textcolor{red}{\Large $a_1$}};
\draw (2,4.4) node{\textcolor{blue}{\Large $b_1$}};
\draw (-4.6,-1) node{\textcolor{red}{\Large $a_2$}};
\draw (2,1.9) node{\textcolor{blue}{\Large $b_2$}};
\draw (-4.7,-3.1) node{\textcolor{red}{\Large $a_3$}};
\draw (2,-0.6) node{\textcolor{blue}{\Large $b_3$}};

\end{tikzpicture}

\caption{The Riemann surface arising in the computation of the R\'enyi-$4$ entropy of two disjoint blocks is represented through the cyclic joining of four sheets. The lower edge of a slit (black) should be identified with the upper edge of a slit above (magenta), in a cyclic way. The canonical homology basis used in this work is indicated. It is the same one as in Ref.~\cite{Coser2016Spin}.}
\label{fig:sheets}
\end{figure}

\paragraph{Spin structures: quantum field theory.} In QFT, the moments $\tr \rho_{A\cup C}^\alpha$ of the reduced density matrix of disjoint blocks can be interpreted in terms of partition functions on Riemann surfaces $\mathcal{R}_\alpha$ with genus $g=\alpha-1$. We refer the interested reader to Refs~\cite{Calabrese2009Entanglement,Calabrese2011Entanglement,Headrick2013Bose,Coser2014OnRenyi,Coser2016Spin,Calabrese2011Entanglement} for more detailed presentations of  such surfaces; here we mention just their main properties. The Riemann surface $\mathcal{R}_\alpha$ is given by the complex algebraic curve $(y,z)\in \mathbb{C}^2$, with $y^\alpha=(z-u_1)(z-u_2)[(z-v_1)(z-v_2)]^{\alpha-1}$. A key object characterising a Riemann surface is the so-called period matrix. Since it depends on the homology basis, we start by introducing such basis.
That is a set $\{a_j,b_j ; j=1,2,\ldots, g\}$ of closed oriented curves that cannot be contracted to a point, with the following intersection properties. Defining the intersection number $h\circ \tilde{h}$ of two oriented curves $h,\tilde{h}$ as their number of intersections, with the orientation taken into account, the cycles satisfy $a_i \circ a_j =b_i\circ b_j=0$ and $a_i\circ b_j=-b_i\circ a_j =\delta_{ij}$. In this work we use the same canonical cohomology basis as Ref.~\cite{Coser2016Spin} (and also Refs~\cite{Alvarez-Gaume1986,Verlinde1987}), shown, e.g., in Fig.~\ref{fig:sheets} for the case $\alpha=4$. More details can be found in Refs~\cite{Coser2016Spin,Coser2014OnRenyi}. The period matrix for this basis is given in Eq.~\eqref{period matrix def}. We remark that it is purely imaginary, a property that is lost with more than two disjoint blocks~\cite{Headrick2013Bose}. For $\alpha=2$ the Riemann surface is a torus and the period matrix reduces to the so-called modular parameter $\tau=\Omega_{11}=i\beta_{1/2}(x)$, where $\beta$ is defined in \eqref{period matrix def}. The modular parameter $\tau$ can also be expressed in terms of the complete elliptic integral $K$, as reported in Figure~\ref{fig:torus summary}.

For fermionic models on higher genus Riemann surfaces, the boundary conditions along the cycles of the canonical homology basis can be either periodic (Ramond)  or antiperiodic (Neveu-Schwarz). The set of boundary conditions over all cycles provides the \textit{spin structure} of the model on the underlying Riemann surface. We use  standard notations according to which $\vec{\varepsilon}$ denotes the vector of  the boundary conditions along $a$-cycles ($\varepsilon_i=0$ for antiperiodic and $\varepsilon_i=1/2$ for periodic boundary conditions), whereas  $\vec{\delta}$ is the vector of the boundary conditions along $b$-cycles ($\delta_i=0$ for antiperiodic and $\delta_i=1/2$ for periodic boundary conditions).
It is not accidental that the same symbols $\vec\varepsilon$ and $\vec \delta$ appear in \eqref{eq:I3spinstr}. Such a connection was established in Ref.~\cite{Coser2016Spin}, which showed that
the elements of the pseudo-RDMs introduced in~\eqref{eq:pseudoRDM} can be expressed in the fermionic coherent state path integral formalism based on Grassmann variables~\cite{Kleinert2009}. Specifically, two Grassmann variables $\zeta_\ell, \zeta_\ell^\dagger$ are associated with each lattice site $\ell$ and are then promoted to fields $\zeta(x), \zeta^\dagger(x)$ in the continuum limit. Let us assume, for the sake of simplicity, that the system is described by a single fermionic field $\bs \psi(x)$, satisfying the standard anticommutation relations $\{\bs \psi(x),\bs \psi^\dagger(y)\}=\bs{I}\delta(x-y)$. The coherent states $\ket{\zeta}$ then satisfy $\bs\psi(x)\ket{\zeta}=\zeta(x)\ket{\zeta}$, $\bra{\zeta}\bs\psi^\dagger(x)=\bra{\zeta}\zeta^\dagger(x)$. We also use the notation $\ket{\zeta}=\ket{\zeta_A,\zeta_B, \zeta_C, \zeta_{\overline{ABC}}}$ to specify different subsystems separately, i.e. $\zeta(x)=\zeta_A(x)$ for $x\in A$ etc. Then the elements of the pseudo RDMs can be written as
\begin{multline}
  \bra{\zeta_A,\zeta_C} \rho_{AC}^{s,s'} \ket{\eta_A,\eta_C} =e^{-\zeta_A^\dagger\eta_A-\zeta_C^\dagger\eta_C}\\ \int \mathcal{D}\zeta_{B}^\dagger\mathcal{D}\zeta_{B} \mathcal{D}\zeta_{\overline{ABC}}^\dagger\mathcal{D}\zeta_{\overline{ABC}} \braket{\zeta_A, -s \zeta_C, -s' \zeta_B, -\zeta_{\overline{ABC}}| 0}\braket{0| \eta_A, -s \eta_C, \zeta_B, \zeta_{\overline{ABC}} } \, ,
\end{multline}
where $\zeta^\dagger_A\eta_A=\sum_{\ell \in A}\zeta_\ell^\dagger\eta_\ell$.

In the continuum limit the overlaps $\braket{\zeta|0}$  and $\braket{0|\zeta'}$ can be represented as path integrals in the lower and upper half plane respectively. Ref.~\cite{Coser2016Spin} has proceeded along these lines and reinterpreted Eq.~\eqref{eq:spinstruct} in terms of partition functions. Namely,
\begin{equation}
\mathrm{tr}\left[\prod\limits_{k=1}^{\alpha-1}\rho_{AC}^{s_k[\vec\varepsilon],s_k'[\vec \delta]}\right]=Z_{\vec\varepsilon, \vec \delta}=\int_{\vec\varepsilon, \vec \delta} \mathcal{D} \psi^\dagger \mathcal{D}\psi \  e^{-S_E[\psi^\dagger,\psi]},
\end{equation}
where $S_E=\int \mathcal{L}[\psi^\dag,\psi] dx d\tau $ is the Euclidean action of the fermionic QFT description of the model, $\psi(x,\tau)$, $\psi^\dag(x,\tau)$ are Grassmann fields and the path integral is performed on the Riemann surface $\mathcal{R}_\alpha$, with spin structure $\vec\varepsilon$, $\vec \delta$ for the previously introduced canonical homology basis. When there are multiple fermionic fields, such as the ones constituting the Dirac spinor, the previous considerations generalize straightforwardly and one works then with multiple fields with correlated spin structures. Ref.~\cite{Coser2016Spin} has in this way identified the spin structures of the free Dirac and Majorana QFTs with the joint moments of the pseudo-RDMs defined in Ref.~\cite{Fagotti2010disjoint}. 
We remark that, while both Ref.~\cite{Coser2016Spin} and Ref.~\cite{Fagotti2010disjoint} focused on non-interacting spin chains, the presented arguments are in fact general. In this work, in turn, we will apply the same picture in the presence of interactions.

\subsection{Conformal critical systems}

In this section we briefly review the behavior of the entanglement entropies in critical spin chains described by a conformal field theory, and the corresponding formalism. We focus on ground states in the thermodynamic limit. More details about the formalism for computing the entanglement entropies in QFT can be found in reviews~\cite{Calabrese2009Entanglement,Rangamani2017}.

When the subsystem is a single interval $A=(u,v)$, the moments of the reduced density matrix for a criticial system described by a CFT with central charge $c$ are given by~\cite{Holzhey1994Geometric,Calabrese2004Entanglement}
\begin{equation}\label{trace rho alpha single block}
    \tr \rho_A^\alpha=c_\alpha \left(\frac{v-u}{a}\right)^{-\frac{c}{6}\left(\alpha-\frac{1}{\alpha}\right)},
\end{equation}
where $a$ is the lattice spacing and $c_\alpha$ is a non-universal constant. When the subsytem consists of two disjoint intervals, $A=(u_1,v_1)$ and $C=(u_2,v_2)$, Refs~\cite{Calabrese2009Entanglement1,Furukawa2009Mutual,Caraglio2008Entanglement} found
\begin{equation}\label{trace rho alpha CFT}
    \mathrm{tr} \rho_{A\cup C}^\alpha=c_\alpha^2 \left[\frac{a^2(u_2-u_1)(v_2-v_1)}{(v_1-u_1)(v_2-u_1)(v_2-u_2)(u_2-v_1)}\right]^{\frac{c}{6}\left(\alpha-\frac{1}{\alpha}\right)}\mathcal{F}(x),
\end{equation}
where $c_\alpha$ is the same non-universal constant as before, $x$ is the cross ratio~\eqref{eq:x4}, which can be written as
\begin{equation}\label{cross ratio x continuum}
    x=\frac{(v_1-u_1)(v_2-u_2)}{(u_2-u_1)(v_2-v_1)} \; ,
\end{equation}
and $\mathcal{F}(x)$ is a universal function of the cross ratio that is model dependent, in that it depends on the full operator content of the CFT. A very convenient way of expressing the result for disjoint blocks is through the tripartite information~\eqref{tripartite information def} of subsystems $A,B,C$, where $B$ is the interval between $A$ and $C$,
\begin{equation}
    I_3^{(\alpha)}(A,B,C)=I_3^{(\alpha)}(x)=\frac{1}{\alpha-1}\log \mathcal{F}_\alpha(x).
\end{equation}
Thus, the universal part of the entanglement entropy of disjoint blocks is determined by the tripartite information of three adjacent blocks.

The relation between the tripartite information of $A,B,C$ and the mutual information $I_2^{(\alpha)}(A,C)=S_\alpha[A]+S_\alpha[C]-S_\alpha[A\cup C]$ follows from Eqs~\eqref{Renyi entropy single block} and \eqref{tripartite information def}
\begin{equation}
    I_3^{(\alpha)}(A,B,C)=\frac{c}{6}\left(1+\frac{1}{\alpha}\right)\log(1-x) + I_2^{(\alpha)}(A,C) \; .
\end{equation}
In the limit $x\to 0$, corresponding to a large separation $|B|$ with respect to the lengths of $A$ and $C$, the mutual information is expected to vanish. Since the first term on the right hand side vanishes as well, we obtain $I_3^{(\alpha)}(x)=0$ for $x\to 0$. This accounts for the choice of normalization $\mathcal{F}_\alpha(0)=1$.

We mention that the result~\eqref{trace rho alpha single block} for the single block can be interpreted as a two-point function $\tr\rho_A^\alpha=\braket{\mathcal{T}_\alpha(u)\overline{\mathcal{T}}_\alpha(v)}$ of branch-point twist fields~\cite{Calabrese2004Entanglement,Cardy2008Form} acting at $u$ and $v$. The twist fields behave as spinless primary operators with scaling dimension $\Delta_\alpha=\tfrac{c}{12}(\alpha-\tfrac{1}{\alpha})$. Similarly, the result~\eqref{trace rho alpha CFT} can be interpreted as a four-point function $\tr \rho_{A\cup C}^\alpha=\braket{\mathcal{T}_\alpha(u_1)\overline{\mathcal{T}}_\alpha(v_1)\mathcal{T}_\alpha(u_2)\overline{\mathcal{T}}_\alpha(v_2)}$ and its form is dictated by global conformal invariance.

A general expectation is that the tripartite information $I_3^{(\alpha)}(x)$ (and the function $\mathcal{F}(x)$) exhibits the crossing symmetry $x\leftrightarrow 1-x$. One way to see this is to consider a finite periodic system of length $L$. There, the lengths $\ell$ entering the cross ratio are replaced by chord lengths $\tfrac{L}{\pi}\sin(\tfrac{\ell \pi}{L})$, so that the cross ratio becomes
\begin{equation}
    x=\frac{\sin\left(\pi\frac{v_1-u_1}{L}\right)\sin\left(\pi\frac{v_2-u_2}{L}\right)}{\sin\left(\pi\frac{u_2-u_1}{L}\right)\sin\left(\pi\frac{v_2-v_1}{L}\right)} \; .
\end{equation}
One has then
\begin{equation}
    I_3^{(\alpha)}(x)=I_3^{(\alpha)}(A,B,C)=I_3^{(\alpha)}(B,C,\overline{ABC})=I_3^{(\alpha)}(1-x)\, ,
\end{equation}
where the second equality is a consequence of the property that in pure states the entropy of a subsystem $X$ is equal to the entropy of its complement $\bar X$, and the third equality stems from the property that the cross ratio for disjoint intervals $B$ and $\overline{ABC}$ is equal to $1-x$.

\subsection{Noncritical systems}\label{s:noncrit}
\paragraph{Disordered phases.}
In a state with clustering properties and a finite correlation length, at the leading order in the lengths, the entanglement entropy of a subsystem of spins depends only on the number of subsystem's boundaries. This is known as ``area law''~\cite{Hastings2007} and is sufficient to conclude that the tripartite information is asymptotically zero. One could still wonder whether, in the presence of spin-flip symmetry, something nontrivial could be found by singling out the contribution coming from a given spin structure in the underlying fermionic theory.  We argue that the answer is no.  The spin structure in that case was briefly addressed in Ref.~\cite{Fagotti2010disjoint} through a numerical analysis in the paramagnetic phase of the  quantum XY model. It was inferred 
\begin{equation}\label{eq:noncrit}
\mathcal F_f\left[\begin{smallmatrix}
\vec\varepsilon\\
\vec\delta
\end{smallmatrix}
\right]=1\, .
\end{equation}
In Appendix~\ref{appendix spin structures and gapped spin chains} we lift this numerical observation to a prediction for any state satisfying the aforementioned properties. It follows that the tripartite information vanishes, $I_3^{(\alpha)}=0$.

\paragraph{Ordered phases.} Our proof of \eqref{eq:noncrit} relies on clustering and spin-flip symmetry; the only way to circumvent \eqref{eq:noncrit} in a gapped system is therefore by breaking one of the two. As we discussed before, spin flip  is however important, if not necessary, in order to give sense to the underlying fermionic subsystems, therefore we do not allow ourselves to give it up in the representation through Jordan-Wigner fermions. We thus examine symmetric ground states that break clustering.
We assume that the state is a cat state of the form
\begin{equation}
    \ket{\Psi}=\frac{1+(-1)^{\bs F}}{\sqrt{2}}\ket{\psi} \, ,
\end{equation}
where $\ket{\psi}$ has clustering properties and is macroscopically different from $(-1)^{\bs F}\ket{\psi}$. 
Ref.~\cite{Fagotti2010disjoint} considered also this case by numerically studying the ferromagnetic phase of the XY model; specifically, it was inferred 
\begin{equation}\label{eq:Fcat}
\mathcal F_f\left[\begin{smallmatrix}
\vec\varepsilon\\
\vec\delta
\end{smallmatrix}
\right]=(-1)^{4\vec \varepsilon\cdot\vec \delta}
\end{equation}
In Appendix~\ref{appendix spin structures and gapped spin chains} we provide arguments that this result should be expected in general.
It follows that the tripartite information in this case is $I_3^{(\alpha)}=\log 2$.

\paragraph{Scaling close to discontinuous phase transitions.} In the rest of the section we investigate a limit in which the second class of states approaches the former. Having in mind the physical interpretation of the two classes as ground states of systems in disordered and ordered phases, one could naively expect that, preserving the symmetry (spin flip), such a limit would correspond to a quantum phase transition in which the entanglement entropies exhibit logarithmic singularities  (the reader could, for example, think of the effect of changing the magnetic field in the quantum Ising model). This is however avoidable in a presence of a point with a discontinuous phase transition~\cite{Franchini2007Ellipses,Ercolessi2011Essential}, opening the door to the existence of states that satisfy the area law but that exhibit a nontrivial tripartite information. 
We consider, in particular, the following cat state
\begin{equation}
\ket{\bar \theta}=\tfrac{\ket{\theta}+\ket{-\theta}}{\sqrt{2}}\, ,
\end{equation}
where
$$
\ket{\theta}=e^{i\frac{\theta}{2}\sum_\ell \sigma_\ell^y}\ket{\Uparrow}
$$
and $\Uparrow$ denotes a string of all spin up. A reader familiar with the quantum XY model~\cite{Lieb1961,Barouch1971}, defined by the Hamiltonian
\begin{equation}\label{XY chain Hamiltonian}
    \bs H_{\mathrm{XY}}=\sum_{\ell} \left[(1+\gamma)\bs\sigma_\ell^x \bs\sigma_{\ell+1}^x +(1-\gamma)\bs\sigma_\ell^y \bs\sigma_{\ell+1}^y +2h\bs \sigma_\ell^z\right] \;, 
\end{equation}
might have recognised this as the symmetric ground state in a special region of the phase diagram within its (anti)ferromagnetic phase~\cite{Kurmann1982,Muller1985Implications,Franchini2017}, with $\cos^2(\theta)=(1-\gamma)/(1+\gamma)$. The limit $\theta\rightarrow 0$ brings the cat state into a symmetric product state, which can be interpreted as a ground state in a disordered phase (it is, indeed, the ground state in the paramagnetic phase of the isotropic XY model).

First we note that we are free to replace the coherent superposition by an incoherent one: the system is in the thermodynamic limit and $\ket{\theta}$ is macroscopically different from $\ket{-\theta}$ for every $\theta\neq 0$. 
The spin representation of the fermionic pseudo-RDM, \eqref{eq:pseudoRDM},  reads
\begin{equation}
\rho^{s,s'}=(-\cos\theta)^{\frac{1-s}{2}|B|}\rho^{(\theta;+)}_A\rho^{(\theta;+)}_C+s'(-\cos\theta)^{\frac{1+s}{2}|B|}\rho^{(\theta;-)}_A\rho^{(\theta;-)}_C
\end{equation}
where
\begin{equation}
\rho_A^{(\theta;s)}=\frac{e^{i\frac{\theta}{2}\sum_{\ell\in A} \sigma_\ell^y}\ket{\Uparrow_A}\bra{\Uparrow_A}e^{-i\frac{\theta}{2}\sum_{\ell\in A} \sigma_\ell^y}+s e^{-i\frac{\theta}{2}\sum_{\ell\in A} \sigma_\ell^y}\ket{\Uparrow_A}\bra{\Uparrow_A}e^{i\frac{\theta}{2}\sum_{\ell\in A} \sigma_\ell^y}}{2}
\end{equation}
and, in particular, $\rho_A^{(\theta;+)}$ is equivalent to the reduced density matrix of a spin block.
The spin structure can be easily computed, and we find
\begin{multline}\label{F XY circle exact}
\F{\varepsilon}{\delta}{A,B,C}=\tfrac{(1+\cos^{2(|A|+|B|)}(\theta))(1+\cos^{2(|B|+|C|)}(\theta))}{1+\cos^{2(|A|+|B|+|C|)}(\theta)}\\
\Bigl[\tfrac{1}{1+\cos^{2(-1)^{2\varepsilon}|B|}(\theta)}+\tfrac{(-1)^{2\delta}\tanh(|A|\log\cos\theta)\tanh(|C|\log\cos\theta)}{1+\cos^{-2(-1)^{2\varepsilon}|B|}(\theta)}\Bigr]
\end{multline}
Consistently with \eqref{eq:Fcat}, for any nonzero $\theta$ this approaches $(-1)^{4\epsilon\delta}$ in the limit of infinite lengths.
If however we take that limit while scaling also the angle so that $\theta=\frac{y}{\sqrt{|B|}}$ with $y$ finite, we get
\begin{multline}
\F{\varepsilon}{\delta}{A,B,C}\xrightarrow{|A|\sim|B|\sim|C|\sim\theta^{-2}\rightarrow\infty} \tfrac{(1+e^{-(\chi_1+1)y^2})(1+e^{-(\chi_2+1)y^2})}{1+e^{-(\chi_1+\chi_2+1)y^2}}\\
\Bigl[\tfrac{1}{1+e^{-(-1)^{2\varepsilon}y^2}}+\tfrac{(-1)^{2\delta}\tanh(\frac{\chi_1}{2}y^2 )\tanh(\frac{\chi_2}{2}y^2)}{1+e^{(-1)^{2\varepsilon}y^2}}\Bigr]
\end{multline}
where $\chi_1=\frac{|A|}{|B|}$ and $\chi_2=\frac{|C|}{|B|}$. This means that, for any arbitrarily large subsystems, we can find angles $\theta$ for which the spin structure is different from the basic values~\eqref{eq:noncrit} and \eqref{eq:Fcat}. This will be exploited in Section~\ref{sec Spin structure from the fusion of models} to uncover the spin structure of the Heisenberg XXZ model.

\section{Heisenberg XXZ model}

\subsection{Quantum field theory description}

The XXZ spin chain~\eqref{XXZ spin chain Hamiltonian} is critical in a range of parameters, e.g., in the absence of magnetic field $h$ the critical region is $\Delta\in[-1,1]$. Its low energy and large distance properties admit an effective Luttinger liquid description, developed in Refs~\cite{Luther1975,Haldane1980}. For a pedagogical introduction to the effective quantum field theory description in terms of Thirring and free compactified boson Lagrangian we refer the reader to Ref.~\cite{Affleck1988LesHouches}.

The identification of the QFT is based on the Jordan-Wigner transformation. It is first established perturbatively and is then lifted to a non-perturbative level by fixing the QFT parameters phenomenologically, through the comparison bewteen QFT predictions and exact lattice results. Specifically, the non-interacting model is identified with a free Dirac fermion and the interacting part, quartic in the Jordan-Wigner fermions, is treated as a small perturbation. This yields the massless Thirring model, whose Lagrangian in the Euclidean space reads
\begin{equation}
    \mathcal{L}=\Bar{\psi}\gamma^\mu\partial_{\mu}\psi-\frac{1}{2}\pi \lambda \ \Bar{\psi}\gamma^\mu\psi\Bar{\psi}\gamma_\mu\psi,
\end{equation}
where the spinors $\psi$ have two components and the Dirac matrices satisfy $\{\gamma_\mu,\gamma_\nu\}=\delta_{\mu,\nu}$ (e.g. $\gamma_0=\sigma^y, \ \gamma_1=\sigma^x$). Bosonization is then used to map the Thirring model to the free boson compactified on a circle, given by the Lagrangian
\begin{equation}
    \mathcal{L}=\frac{R^2}{8\pi}\partial_\mu\varphi\partial^\mu \varphi,
\end{equation}
where the bosonic fields are periodic in the target space with period $2\pi$, i.e. $\varphi\equiv \varphi +2\pi$, and the compactification radius $R$ is related to the Thirring coupling constant through $R^2=1+\lambda$. Note that throughout the literature different conventions for the compactification radius are used. In the one we use the T-duality reads $R\leftrightarrow 2/R$.

Next, symmetry arguments are used to conclude that there are no other relevant Lorentz-invariant interactions in a range of $R$, while the various non-renormalizible interactions can only lead to a renormalization of $R$. The compactification radius is finally determined phenomenologically, by comparing the lattice result for some quantity, typically obtained within Bethe ansatz, with the field theoretical one. Various quantities can be used to this purpose, such as some critical exponents~\cite{Luther1975}, finite-size spectrum~\cite{Haldane1980,Alcaraz1987,Lukyanov1998}, compressibility~\cite{Giamarchi2003book} and susceptibility~\cite{Eggert1994}. For zero magnetic field $h=0$ the identification is
\begin{equation}\label{XXZ identification parameters}
    \ETA=\frac{1}{2}R^2= \frac{1}{2}(1+\lambda)=\frac{1}{2K_L}= \frac{\arccos(-\Delta)}{\pi} \; ,
\end{equation}
where we have also indicated the relation with the Luttinger liquid parameter $K_L$ (see e.g.~\cite{Giamarchi2003book}).
The exponent $\ETA$ appears in the spin-correlation functions, whose large distance behavior for $h=0$ is given by~\cite{Luther1975}
\begin{equation}\label{correlation functions XXZ}
    \braket{\bs\sigma^x_0\bs\sigma^x_\ell}\sim (-1)^\ell \frac{1}{|\ell|^{\ETA}} ,\qquad \braket{\bs\sigma^z_0\bs\sigma^z_\ell}\sim \frac{1}{\ell^2}+(-1)^\ell\frac{1}{|\ell|^{1/\ETA}} \; .
\end{equation}
For a non-zero magnetic field the last equality in Eq.~\eqref{XXZ identification parameters} does not hold, nor Eq.~\eqref{correlation functions XXZ} for the spin correlation functions. In that case $\ETA$ is obtained numerically by solving  integral equations~\cite{Cabra1998} or through perturbative methods~\cite{Granet2019}.

\subsection{The free boson compactified on a circle}

The R\'enyi-$\alpha$ tripartite information for the free compactified boson was first computed for $\alpha=2$ in Ref.~\cite{Furukawa2009Mutual} and then was generalized to integer $\alpha\geq 2$ in Ref.~\cite{Calabrese2009Entanglement1}. The result reads
\begin{equation}\label{tripartite information compactified boson}
      I_3^{(\alpha)}(x)=\frac{1}{\alpha-1}\log \frac{\Theta(\vec 0|\ETA \Omega)\Theta(\vec 0 | \Omega /\ETA)}{\left[\Theta(\vec 0 |\Omega
      )\right]^2},
\end{equation}
where $\Omega$ is the period matrix~\cite{Calabrese2009Entanglement1} of the Riemann surface with genus $\alpha-1$, given in Eq.~\eqref{period matrix def}, and $\Theta\equiv\Theta[\vec 0]$ is the Riemann-Siegel theta function defined in Eq.~\eqref{theta function with characteristic def}. The result~\eqref{tripartite information compactified boson} was derived in Ref.~\cite{Calabrese2009Entanglement1} using older results about four-point correlation functions of twist fields~\cite{Dixon1987}. The result was rederived in Ref.~\cite{Coser2014OnRenyi}, where also the case of more than two intervals is discussed, using older results on partition functions on higher genus Riemann surfaces~\cite{Zamolodchikov1987,Alvarez-Gaume1986,Alvarez-Gaume1987,Verlinde1987,Dijkgraaf1988}. The formula~\eqref{tripartite information compactified boson} is manifestly invariant under $\ETA\to 1/\ETA$, i.e. $R\to 2/R$. Although not apparent from the expression above, the result also respects the crossing symmetry $x\leftrightarrow 1- x$.

The analytical continuation of \eqref{tripartite information compactified boson} and, accordingly, the limit $\alpha\rightarrow 1^+$ are not known in general. However, it is possible to perform the limit $\alpha\rightarrow 1^+$  for small $x$. The small $x$ expansion for $\ETA\neq 1$ reads~\cite{Calabrese2011Entanglement}
\begin{equation}\label{expansion XXZ small x}
     I_3^{(\alpha)}(x)=\frac{s\left(\alpha;\mathcal{A}_{\textrm{CB}}\right)}{\alpha-1}\left(\frac{x}{4\alpha^2}\right)^{\mathcal{A}_{\textrm{CB}}}+o\left(x^{\mathcal{A}_{\textrm{CB}}}\right), \qquad \ETA\neq 1  , \quad \textrm{as } x\to 0 ,
\end{equation}
where
\begin{equation}
    \mathcal{A}_{\textrm{CB}}=\mathrm{min}[\ETA,1/\ETA]
\end{equation}
and we have defined the function
\begin{equation}\label{s CCT}
  s(\alpha;\mathcal{A})=\alpha\sum_{j=1}^{\alpha-1}\left[\sin\left(\pi\frac{j}{\alpha}\right)\right]^{-2\mathcal{A}} \; , \qquad \alpha=2,3,4,\ldots
\end{equation}
Ref.~\cite{Calabrese2011Entanglement} also obtained the derivative at $\alpha=1$ of the analytical continuation of $s(\alpha;\mathcal{A})$ to complex $\alpha$,
\begin{equation}\label{s vN step}
s_{\mathrm{vN}}(\mathcal{A})\equiv \frac{d}{d\alpha}s(\alpha;\mathcal{A})\Big|_{\alpha=1}=\frac{\sqrt{\pi}\Gamma(\mathcal{A}+1)}{2\Gamma(\mathcal{A}+\frac{3}{2})},
\end{equation}
given also in Eq.~\eqref{s vN}.
This result can be used to obtain the small $x$ expansion of the von Neumann tripartite information,
\begin{equation}
     I_3^{(\mathrm{vN})}(x)=\frac{\sqrt{\pi}\Gamma(\mathcal{A}_{\textrm{CB}}+1)}{2\Gamma(\mathcal{A}_{\textrm{CB}}+\frac{3}{2})}\left(\frac{x}{4}\right)^{  \mathcal{A}_{\textrm{CB}}}+o\left(x^{\mathcal{A}_{\textrm{CB}}}\right), \quad \textrm{as } x\to 0
\end{equation}
At the self-dual point $\ETA=1$ the tripartite information \eqref{tripartite information compactified boson} is exactly zero for any $\alpha$, and therefore also for $\alpha=1$.

The formula~\eqref{tripartite information compactified boson} has been checked numerically in the XXZ chain for the lowest values of $\alpha$ using exact diagonalization~\cite{Furukawa2009Mutual} and tree tensor networks~\cite{Alba2011Entanglement}. We stress that it would not be correct to identify $\ETA$ with $1/\ETA$ in the chain. Namely, although the tripartite information in Eq.~\eqref{tripartite information compactified boson} is invariant under $\ETA$ and $1/\ETA$, some quantities, like the spin correlation functions in Eq.~\eqref{correlation functions XXZ}, are not. Note that the identification~\eqref{XXZ identification parameters} for the XXZ chain without magnetic field makes it evident that $\ETA\in(0,1]$ so the duality $\ETA\leftrightarrow 1/\ETA$ cannot be addressed there.

\subsection{Thirring model on a Riemann surface}
Here we point out that the tripartite information in the XXZ chain can also be computed within the fermionic approach, by computing \eqref{eq:spinstruct} for each $\vec{\varepsilon},\vec{\delta}$ separately and then summing them. As discussed in section \ref{sec:fermions vs spins}, the last term in \eqref{eq:spinstruct} in the fermionic field theory becomes a partition function on the higher genus Riemann surface with the specified spin structure. Although the arguments are general, the method has been applied so far only in non-interacting spin chains.

To derive the tripartite information in the XXZ chain in the fermionic picture we need to evaluate the partition functions of the massless Thirring model on a higher genus closed Riemann surface for different spin structures. Several papers~\cite{Freedman1988Thirring,Freedman1989Thirring,Wu1989Determinants,Sachs1996} have already tackled with this problem, by manipulating fermionic path integrals, and the results display inconsistencies about factors and the sign of the partition functions, as e.g. commented in Ref.~\cite{Sachs1996}. In this work we fix such ambiguities by requiring that  our results  reproduce the known predictions for the entanglement entropies of disjoint blocks in the bosonic picture~\cite{Calabrese2009Entanglement1}. In this way we fix some sign discrepancies but we do not investigate overall factors common to all spin structures, which are irrelevant to our purposes.

We conclude that, for purely imaginary period matrix $\Omega$, we have
\begin{equation}\label{partition function Thirring}
Z(\vec{\varepsilon},\vec{\delta};\lambda)=(-1)^{4\vec{\varepsilon}\cdot\vec{\delta}}Z(\vec{0},\vec{0};0)\ f_\alpha(\lambda)\ \frac{\Theta\left[
    \begin{matrix}
  \vec E \\ \vec F    
    \end{matrix}\right]\left(\vec 0|\tilde \Omega\right)}{\left[\Theta(\vec 0 |\Omega
      )\right]^2} \; ,
\end{equation}
where $Z(\vec{0},\vec{0};0)$ is the partition function for the free Dirac fermion with all antiperiodic boundary coniditions, $f_\alpha$ are unknown functions of $\lambda$ (neglected in Fig.~\ref{fig:torus summary}) such that $f_\alpha(0)=1$, $\Tilde{\Omega}$ is a $2(\alpha-1) \times 2(\alpha-1)$  matrix defined by
\begin{equation}
\tilde \Omega=\Bigl[\bigl(\tfrac{1}{4\ETA}+\ETA\bigr)\mathrm I+\bigl(\tfrac{1}{4\ETA}-\ETA\bigr)\sigma^x\Bigr]\otimes \Omega \; , \quad \ETA=\frac{1}{2}(1+\lambda),
\end{equation}
and the vectors
\begin{equation}
\vec E=\begin{pmatrix}
    1 \\ -1
\end{pmatrix}\otimes \vec \varepsilon\, ,\qquad \vec F=\begin{pmatrix}
    1 \\ 1
\end{pmatrix}\otimes\vec\delta
\end{equation}
carry the information about the spin structure. The formula~\eqref{partition function Thirring} differs from the result of Refs~\cite{Freedman1988Thirring,Freedman1989Thirring,Wu1989Determinants} in the sign $(-1)^{4\vec\varepsilon\cdot\vec\delta}$. In  the free case $\lambda=0$, studied in Ref.~\cite{Alvarez-Gaume1986}, the partition function of odd spin structures vanishes so that sign is irrelevant and formula~\eqref{partition function Thirring} captures  the relation between different spin structures
\begin{equation}
    \frac{Z(\vec\varepsilon,\vec\delta ; 0)}{Z(\vec 0,\vec 0 ; 0)}=\Bigg| \Theta\left[
    \begin{matrix}
  \vec \varepsilon \\ \vec \delta    
    \end{matrix}\right]\left(\vec 0|\Omega\right) /\Theta\left(\vec 0|\Omega\right) \Bigg|^2 \, .
\end{equation}

\subsection{Spin structures}

In this section we compute the function $\mathcal{F}_f$, defined in \eqref{eq:spinstruct}, for each spin structure in the XXZ chain.
We know that the partition function $Z(\vec{0},\vec{0};0)$ is equal, up to a non-universal multiplicative constant, to the trace in Eq.~\eqref{trace rho alpha CFT} with $c=1$ and $\mathcal{F}(x)=1$, where the latter is a result of Ref.~\cite{Casini2005,Casini2009reduced_density,Casini2009Entanglement}. With this insight, and using also \eqref{trace rho alpha single block}, we conclude
\begin{equation}\label{spin structure XXZ chain step}
\mathcal{F}_f  \left[\begin{matrix}
  \vec \varepsilon \\ \vec \delta    
    \end{matrix}\right] =(-1)^{4\vec\varepsilon\cdot \vec\delta} \mathcal{C}_\alpha \frac{\Theta\left[
    \begin{matrix}
  \vec E \\ \vec F    
    \end{matrix}\right]\left(\vec 0|\tilde \Omega\right)}{\left[\Theta(\vec 0 |\Omega
      )\right]^2} \; ,
\end{equation}
for some non-universal constant $\mathcal{C_\alpha}$, that is independent of the spin structure. The numerator in Eq.~\eqref{spin structure XXZ chain step} is expressed in terms of a $2(\alpha-1)\times2(\alpha-1)$ matrix $\tilde \Omega$. It is possible to express it in terms of $(\alpha-1)\times (\alpha-1)$ matrices, by resorting to the identity
\begin{equation}
\Theta\left[
    \begin{matrix}
  \vec E \\ \vec F    
    \end{matrix}\right]\left(\vec 0|\tilde \Omega\right)=\sum_{\vec \mu\in \{0,\frac{1}{2}\}^{\alpha-1}}(-1)^{4\vec \mu\cdot\vec\delta}\Theta\left[
    \begin{matrix}
  \vec \mu \\ \vec 0    
    \end{matrix}\right]\left(\vec 0|\tfrac{1}{\ETA} \Omega\right)\Theta\left[
    \begin{matrix}
  \vec \varepsilon+\vec \mu \\ \vec 0    
    \end{matrix}\right]\left(\vec 0|4\ETA \Omega\right)\, ,
\end{equation}
proven in Appendix \ref{appendix reproducing CCT}.

The tripartite information~\eqref{eq:I3spinstr} is obtained by summing over all spin structures. The constant $\mathcal{C}_\alpha$ can be fixed by requiring that the tripartite information vanishes at $x\to 0$, which is a consequence of clustering (an equivalent trick was used in Ref.~\cite{Calabrese2011Entanglement} for the Ising CFT). This fixes the constant to $\mathcal{C}_\alpha=1$. In this way we obtain the result in Eq.~\eqref{eq:FfXXZ}. In Appendix \ref{appendix reproducing CCT} we show that formula~\eqref{step tripartite info XXZ} reproduces the known result~\eqref{tripartite information compactified boson} obtained from studying the free compactified boson.

\paragraph{Small $x$ expansion}
The small $x$ expansion of \eqref{eq:FfXXZ} can be derived using the methods developed in Ref.~\cite{Calabrese2011Quantum} for the free compactified boson and Ising CFT. Details of the procedure are presented in Appendix~\ref{appendix small x expansion}. We are going to look separately at three different classes of spin structures.
\begin{enumerate}[label={\arabic*)}]
    \item Spin structures with $\vec{\varepsilon}=\vec{0}$\\
    These are the only spin structures resulting in a non-zero value in the limit $x\to 0$. The small $x$ expansion is
    \begin{equation}\label{expansion spin structure epsilon zero}
\begin{split}
      &\mathcal{F}_f  \left[\begin{matrix}
  \vec 0 \\ \vec \delta     
    \end{matrix}\right] =1+s(\alpha; 1/\ETA)\left(\frac{x}{4\alpha^2}\right)^{\tfrac{1}{\ETA}}+s(\alpha;4\ETA)\left(\frac{x}{4\alpha^2}\right)^{4\ETA}-s(\alpha;1)\frac{x}{2\alpha^2}\\ & +\sum\limits_{0\leq j_1<j_2 \leq \alpha-1} \frac{4\ (-1)^{4\sum_{k=j_1+1}^{j_2} \delta_{k}}  }{\left[\sin\left(\pi\frac{j_2-j_1}{\alpha}\right)\right]^{2\big(\tfrac{1}{4\ETA}+\ETA \big)}}\left(\frac{x}{4\alpha^2}\right)^{\tfrac{1}{4\ETA}+\ETA}   +o\big(x^{\min(\tfrac{1}{\ETA},4 \ETA,1,\tfrac{1}{4\ETA}+\ETA)}\big)
\end{split}
\end{equation}
where $s$ is defined in Eq.~\eqref{s CCT}.

\item Spin structures with two ``domain walls'' in $\vec{\varepsilon}$ (imagining $\varepsilon_{-1}=\varepsilon_\alpha=0$ at the boundaries), i.e.
\begin{equation}
    \vec{\varepsilon}^\mathrm{T}=\frac{1}{2}(\underbrace{0,0,\ldots,0}_{j_1}, \underbrace{1,1,\ldots,1}_{j_2-j_1},0,0,\ldots,0) \; , \quad 0\leq j_1< j_2\leq \alpha-1 \; .
\end{equation}
The small $x$ expansion is
\begin{equation}\label{expansion spin structure epsilon two domain walls}
\begin{split}
    &\mathcal{F}_f  \left[\begin{matrix}
  \vec \varepsilon \\ \vec \delta    
    \end{matrix}\right] \\ & =\frac{2}{\left[\sin\left(\pi\frac{j_2-j_1}{\alpha}\right)\right]^{\tfrac{1}{2\ETA}}}\left(\frac{x}{4\alpha^2}\right)^{\tfrac{1}{4\ETA}}+(-1)^{4\vec{\varepsilon}\cdot\vec\delta}\frac{2}{\left[\sin\left(\pi\frac{j_2-j_1}{\alpha}\right)\right]^{2\ETA}}\left(\frac{x}{4\alpha^2}\right)^{\ETA} +o\left(x^{\min(\ETA,\tfrac{1}{4\ETA})}\right)  
\end{split}
  \end{equation}
independently of $\vec{\delta}$.

\item Spin structures with four or more ``domain walls'' in $\vec{\varepsilon}$, i.e. all the remaining spin structures.\\
These spin structures exist only for $\alpha\geq 4$ and they are subleading with respect to the spin-structures of type $1$,
\begin{equation}\label{expansion spin structure epsilon more than two domain walls}
    \mathcal{F}_f  \left[\begin{matrix}
  \vec \varepsilon \\ \vec \delta    
    \end{matrix}\right]= O\left(x^{\min(2\ETA,\tfrac{1}{2\ETA})}\right) \; .
\end{equation}
\end{enumerate}
The small $x$-expansion for the tripartite information in the XXZ chain, given by Eq.~\eqref{expansion XXZ small x}, comes from the first term in \eqref{expansion spin structure epsilon zero} and the second term in \eqref{expansion spin structure epsilon two domain walls} (and taking the logarithm). Note that the second term in \eqref{expansion spin structure epsilon two domain walls} is subleading for $\ETA > 1/2$, but summing over spin structures kills the leading term.

\section{Fermionic XXZ and models with higher central charge}\label{sec fermionic XXZ and JW deformations}

\subsection{Fermionic XXZ chain}\label{sec fermionic XXZ}

\begin{figure}
        \centering
        \includegraphics[width=0.75\textwidth]{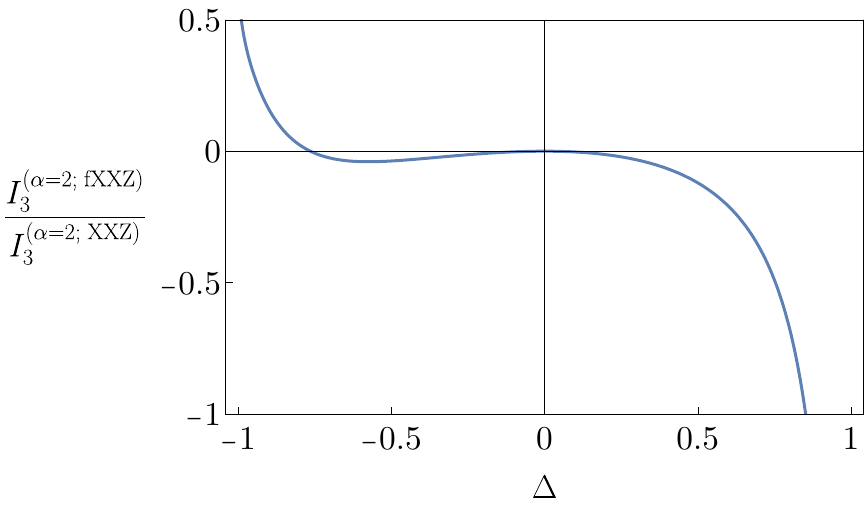}
    \caption{Ratio between the R\'enyi-$2$ tripartite information in the ground state of the XXZ fermionic  chain~\eqref{XXZ fermionic chain Hamiltonian} and that in the XXZ spin chain~\eqref{XXZ spin chain Hamiltonian} vs $\Delta$ for $x=1/2$ and $h=0$. For $\Delta=0$, corresponding to non-interacting chains, the tripartite information in the fermionic model is zero. In its vicinity the tripartite information in the fermionic chain is much smaller than the one in the corresponding spin chain. 
    }
    \label{fig:ratio tripartite spins fermions}
\end{figure}

In the previous section we have computed the contribution of each spin structure to the tripartite information of the XXZ chain. From the result for the spin structure $\vec{\varepsilon}=\vec{\delta}=\vec{0}$ (antiperiodic boundary conditions on all cycles), we can readily compute the tripartite information in the fermionic XXZ chain~\eqref{XXZ fermionic chain Hamiltonian}. In this way we obtain the result presented in Eq.~\eqref{tripartite info fermionic XXZ}. In Fig.~\ref{fig:ratio tripartite spins fermions} the result is compared to the one for the XXZ spin chain with $\alpha=2$. Prediction~\eqref{tripartite info fermionic XXZ} is manifestly invariant under $\ETA \leftrightarrow 1/(4\ETA)$, which is a symmetry around the free point $\ETA=1/2$. This symmetry should be contrasted to the symmetry $\ETA\leftrightarrow 1/\ETA$ of the free compactified boson in Eq.~\eqref{tripartite information compactified boson}. At the free point $\ETA=1/2$ we get $I_{3,\mathrm{f}}^{(\alpha)}(x)=0$, because of the identity~\cite{Headrick2013Bose}
\begin{equation}
|\Theta(\vec 0 | \Omega)|^2= \sum_{\vec \varepsilon\in \{0,\frac{1}{2}\}^{\alpha-1}}\Bigl|\Theta\left[
    \begin{smallmatrix}
  \vec \varepsilon \\ \vec 0    
    \end{smallmatrix}\right]\left(\vec 0|2 \Omega\right)\Bigr|^2\, .
\end{equation}
The vanishing of the tripartite information in the non-interacting case, i.e. in the Kitaev chain at its phase transition, is in agreement with the known result for the free Dirac fermion~\cite{Casini2005,Casini2009Entanglement,Casini2009reduced_density}.

\paragraph{Small $x$ expansion.} Putting $\vec{\delta}=\vec 0$ in \eqref{expansion spin structure epsilon zero} and taking the logarithm, we get the small $x$ expansion for the tripartite information in the fermionic XXZ chain,
  \begin{equation}\label{expansion tripartite information fermionic XXZ}
\begin{split}
      I_{3,\mathrm{f}}^{(\alpha)}(x) =&\frac{1}{\alpha-1}\bigg[-s(\alpha;1)\frac{x}{2\alpha^2}+s(\alpha;\tfrac{1}{\ETA})\left(\frac{x}{4\alpha^2}\right)^{\tfrac{1}{\ETA}}+s(\alpha;4\ETA)\left(\frac{x}{4\alpha^2}\right)^{4\ETA}\\ & + s(\alpha;\tfrac{1}{4\ETA}+\ETA)\left(\frac{x}{4\alpha^2}\right)^{\tfrac{1}{4\ETA}+\ETA}   +o\big(x^{\min(1,\tfrac{1}{\ETA},4\ETA)}\big)\bigg]
\end{split}
\end{equation}
Note that the term with power $\tfrac{1}{4\ETA}+\ETA$ contributes only for $\ETA=1/2$. It is instructive to single out the $\ETA$-dependent leading term,
\begin{equation}\label{expansion tripartite information fermionic XXZ cases}
     I_{3,\mathrm{f}}^{(\alpha)}(x)=\begin{cases}
    s(\alpha;4\ETA)\left(\frac{x}{4\alpha^2}\right)^{4\ETA} +o(x^{4\ETA}), & \ETA<\tfrac{1}{4}\\
    -s(\alpha;1)\frac{x}{4\alpha^2}+o(x), & \ETA =\tfrac{1}{4}\\
    -s(\alpha;1)\frac{x}{2\alpha^2} +o(x), & \tfrac{1}{4}<\ETA < 1, \ETA\neq \tfrac{1}{2}\\   
    -s(\alpha;1)\frac{x}{4\alpha^2} +o(x), & \ETA =1\\  
         s(\alpha;\tfrac{1}{\ETA})\left(\frac{x}{4\alpha^2}\right)^{\tfrac{1}{\ETA}} +o(x^{\tfrac{1}{\ETA}}), & \ETA >1\\
     \end{cases}
\end{equation}

It is now possible to compute the small $x$ expansion of the von Neumann tripartite information using formula~\eqref{s vN step}, derived in Ref.~\cite{Calabrese2011Entanglement} in the context of the spin chain. The result is given in Eq.~\eqref{expansion tripartite information fermionic vN}, or, singling out the leading term, by
\begin{equation}\label{expansion tripartite information fermionic vN cases}
     I_{3,\mathrm{f}}^{(\mathrm{vN})}(x)=\begin{cases}
    s_{\mathrm{vN}}(4\ETA)\left(\frac{x}{4}\right)^{4\ETA} +o(x^{4\ETA}), & \ETA<\tfrac{1}{4}\\
    -s_{\mathrm{vN}}(1)\frac{x}{4}+o(x), & \ETA =\tfrac{1}{4}\\
    -s_{\mathrm{vN}}(1)\frac{x}{2} +o(x), & \tfrac{1}{4}<\ETA < 1, \ETA\neq \tfrac{1}{2}\\   
    -s_{\mathrm{vN}}(1)\frac{x}{4} +o(x), & \ETA =1\\  
         s_{\mathrm{vN}}(\tfrac{1}{\ETA})\left(\frac{x}{4}\right)^{\tfrac{1}{\ETA}} +o(x^{\tfrac{1}{\ETA}}), & \ETA >1\\
     \end{cases}
\end{equation}
Interestingly, while the tripartite information in the XXZ chain is positive (or zero for $\ETA=1$), we can see clearly that for small $x$ the tripartite information in the fermionic XXZ chain is negative in the parameter region $\ETA\in[\tfrac{1}{4},\tfrac{1}{2})\cup (\tfrac{1}{2},1]$.

\subsection{Jordan-Wigner deformations}\label{sec JW deformations}
There is a straightforward way to construct spin-chain models with higher central charge starting from a  massless system with local interactions.
It is sufficient to move operators apart in such a way that their distance is multiplied by a given number $n$, e.g.,
$
\sigma_\ell^\alpha\sigma_{\ell+1}^\beta\sigma_{\ell+3}^\gamma\rightarrow\sigma_\ell^\alpha\sigma_{\ell+n}^\beta\sigma_{\ell+3n}^\gamma
$.
The resulting Hamiltonian is written as a sum of $\nu$ commuting extensive operators acting on distinct sublattices and, appropriately rescaling the lengths ($\ell\rightarrow\frac{\ell}{\nu}$), the low-energy physics is described by the sum of independent CFTs. Such a theory is arguably uninteresting. 

If the Hamiltonian is invariant under a spin flip, a Jordan-Wigner (JW) transformation allows one to reinterpret the spin-chain Hamiltonian as a fermion one with local interactions. This opens the door to applying the same procedure to the fermion Hamiltonian, which in the spin representation has the effect of moving the spins apart, but additional spin operators in the direction of the quantisation axis can appear. For example we have
$$
\sigma_\ell^x\sigma_{\ell+2}^x\xrightarrow{\nu} \sigma_\ell^x(\prod_{j=1\atop j\neq \nu}^{2\nu-1}\sigma_{\ell+j}^z)\sigma_{\ell+2\nu}^x\, ,
$$
which are also known as ``cluster interactions''~\cite{Raussendorf2001,Doherty2009,Son2011,Giampaolo2015}.  Because of the  spin structure, such a deformation has now a nontrivial effect. This leads us to Hamiltonian~\eqref{eq:JWdef}, that we call ``Jordan-Wigner-$\nu$ deformations of the XXZ chain''. We will see in Sec.~\ref{sec Spin structure from the fusion of models} that JW deformations characterise the spin structure almost completely.

Applying the JW transformation to the JW deformation of the XXZ chain, given by Hamiltonian~\eqref{eq:JWdef}, we get
\begin{equation}\label{eq:JWdef fermionic}
\bs H_{\mathrm{XXZ}}^{(\nu)}=2\sum_\ell \bs c_{\ell+\nu} \bs c^\dag_\ell +\bs c_{\ell}\bs c^\dag_{\ell+\nu} -2(\Delta+h) \bs c^\dag_{\ell} \bs c_{\ell}+2\Delta \bs c^\dag_\ell \bs c_{\ell}\bs c^\dag_{\ell+\nu} \bs c_{\ell+\nu}
\end{equation}
The model is a sum of fermionic XXZ chain Hamiltonians (see Eq.~\eqref{XXZ fermionic chain Hamiltonian}), on sublattices $\nu\mathbb Z+j=\{,\ldots,-\nu+j,j,\nu+j,2\nu+j\ldots\}$, for $j=0,1,\ldots,\nu-1$. The Hamiltonians are mutually commuting so they can be diagonalized simultaneously. In particular, all the fermionic correlation functions in the ground state factorize into those involving only fermions on the same sublattice. This also implies the factorization of the function in Eq.~\eqref{spin structure XXZ chain step},
\begin{equation}\label{lattice factorization spin structure JW deformation}
    \F{\vec\varepsilon}{\vec\delta}{A,B,C} =\prod_{j=1}^\nu \F[(\mathrm{XXZ})]{\vec\varepsilon}{\vec\delta}{A|_j,B|_j,C|_j},
\end{equation}
where we use the notation $A|_j=A\cap (\nu \mathbb Z+j)$ for the intersection between set $A$ and  sublattice $j$. The factorization~\eqref{lattice factorization spin structure JW deformation} is exact on the lattice. In the limit of large subsystems we have a dependence on the configuration only through the cross ratio, so
\begin{equation}
    \F{\vec\varepsilon}{\vec\delta}x =\left( \F[(\mathrm{XXZ})]{\vec\varepsilon}{\vec\delta}x\right)^\nu,
\end{equation}
Thus, the contribution of each spin structure to the tripartite information in the Jordan-Wigner deformations is trivially related to the corresponding contribution in  the XXZ chain. By summing over spin structures, we get the tripartite information in the Jordan-Wigner deformation of the XXZ chain,
\begin{equation}\label{eq:I3expansion JWdef}
I_3^{(\alpha;\nu)}(x)=\frac{1}{\alpha-1}\log\left(\frac{1}{2^{\alpha-1}}\sum_{\vec\varepsilon,\vec\delta\in\{0,\frac{1}{2}\}^{\alpha-1}}(-1)^{4\vec \varepsilon\cdot\vec\delta}\left(\F[(\mathrm{XXZ})]{\vec\varepsilon}{\vec\delta}x\right)^\nu\right),
\end{equation}
where $\mathcal{F}_f^{(\mathrm{XXZ})}$ is given by Eq.~\eqref{eq:FfXXZ}. In the free case $\ETA=1/2$ the result reduces to
\begin{equation}
I_3^{(\alpha;\nu)}(x)=\frac{1}{\alpha-1}\log\left(\frac{1}{2^{\alpha-1}}\sum_{\vec\varepsilon,\vec\delta\in\{0,\frac{1}{2}\}^{\alpha-1}}(-1)^{4{\vec \varepsilon}\cdot {\vec \delta}}\left|\frac{\Theta\left[
    \begin{smallmatrix}
  \vec\epsilon\\ \vec \delta   
    \end{smallmatrix}\right](0|\Omega)}{\Theta(0|\Omega)}\right|^{2\nu}\right)\, ,
\end{equation}
which could be deduced already from Refs~\cite{Fagotti2012New,Coser2016Spin}.

\paragraph{Jordan-Wigner 2-deformation} Let us focus on the particular case $\nu=2$ and derive a simplified expression. Writing explicitly the square as a product we have
\begin{equation}\label{step tripartite info JW 2 deformation}
\begin{split}
  & \frac{1}{2^{\alpha-1}}\sum_{\vec\varepsilon,\vec\delta\in\{0,\frac{1}{2}\}^{\alpha-1}}(-1)^{4\vec \varepsilon\cdot\vec\delta}\left(\F[(\mathrm{XXZ})]{\vec\varepsilon}{\vec\delta}x\right)^2=\\&  \frac{1}{2^{\alpha-1}}\sum_{\vec\varepsilon,\vec\delta\in\{0,\frac{1}{2}\}^{\alpha-1}}(-1)^{4\vec \varepsilon\cdot\vec\delta} \sum_{\vec{\mu}_1,\vec{\mu_2}\in\{0,\frac{1}{2}\}^{\alpha-1}} \prod_{j=1}^2 (-1)^{4\vec{\mu}_j\cdot \vec{\delta}}
  \frac{\Theta\left[
    \begin{smallmatrix}
  \vec\epsilon+\vec{\mu}_j \\ \vec 0    
    \end{smallmatrix}\right]\left(\vec 0|\frac{1}{\ETA} \Omega\right)\Theta\left[
    \begin{smallmatrix}
  \vec{\mu}_j \\ \vec 0    
    \end{smallmatrix}\right]\left(\vec 0|4\ETA \Omega\right)
    }{\left[\Theta(\vec 0 |\Omega
      )\right]^2}
\end{split}
\end{equation}
The sum over $\vec\delta$-dependent terms is
\begin{equation}
    \frac{1}{2^{\alpha-1}}\sum_{\vec\delta\in\{0,\frac{1}{2}\}^{\alpha-1}}(-1)^{4(\vec \varepsilon+\vec{\mu}_1+\vec{\mu}_2)\cdot\vec\delta}=\begin{cases}
        1, & 2(\vec{\mu}_1+\vec{\mu}_2)=2\vec\varepsilon \quad (\mathrm{mod}\ 2)\\
        0, & \mathrm{otherwise}
    \end{cases}
\end{equation}
Since $\Theta[\vec e]=\Theta[\tfrac{1}{2} (2\vec e \ \mathrm{mod} \ 2)]$ for $\vec e\in(\mathbb Z/2)^{2(\alpha-1)}$, after summing over $\vec{\delta}$ in Eq.~\eqref{step tripartite info JW 2 deformation} we can replace $\vec\varepsilon$ by $\vec{\mu}_1+\vec{\mu}_2$ in the characteristic of the theta functions . For the same reason, vectors $2\vec{\mu}_2+\vec{\mu}_1$ and $2\vec{\mu}_1+\vec{\mu}_2$ in the characteristic can be replaced by $\vec{\mu}_1$ and $\vec{\mu}_2$ respectively. Grouping the factors that depend on $\vec\mu_1$ and those that depend on $\vec{\mu}_2$ we recognize
\begin{equation}
 \frac{1}{2^{\alpha-1}}\sum_{\vec\varepsilon,\vec\delta\in\{0,\frac{1}{2}\}^{\alpha-1}}(-1)^{4\vec \varepsilon\cdot\vec\delta}\left(\F[(\mathrm{XXZ})]{\vec\varepsilon}{\vec\delta}x\right)^2=\left( \F[(\mathrm{XXZ})]{\vec 0}{\vec 0}x\right)^2 \; . 
\end{equation}
Therefore, the tripartite information in the Jordan-Wigner 2-deformation is the same as the fermionic one in the corresponding fermion model, which is equal to twice the one in the fermionic XXZ chain, given by Eq.~\eqref{tripartite info fermionic XXZ}. This result is expressed by Eq.~\eqref{tripartite info JW 2}.

\paragraph{Small $x$ expansion} For $\nu=2$ we have shown that the tripartite information is equal to twice the one of the fermionic XXZ chain (see Eq.~\eqref{tripartite info JW 2}). Therefore its small $x$ expansion can be read from the small $x$ expansion of the latter, given by Eqs~\eqref{expansion tripartite information fermionic XXZ} and \eqref{expansion tripartite information fermionic XXZ cases} for R\'enyi index $\alpha=2,3,\ldots$ and by Eqs~\eqref{expansion tripartite information fermionic vN} and \eqref{expansion tripartite information fermionic vN cases} for the von Neumann case. 

For $\nu\geq 3$ the result can be obtained from the small $x$ expansion for different spin structures in the XXZ chain (Eqs~\eqref{expansion spin structure epsilon zero}, \eqref{expansion spin structure epsilon two domain walls} and \eqref{expansion spin structure epsilon more than two domain walls}), by raising them to power $\nu$, in accordance with Eq.~\eqref{eq:I3expansion JWdef}. The result reads
\begin{align}
   & I_{3}^{(\alpha;\nu=3)}(x)=\frac{1}{\alpha-1}\left[-3 s(\alpha;1) \frac{x}{2\alpha^2}+3 s(\alpha;\tfrac{1}{\ETA}) \left(\frac{x}{4\alpha^2}\right)^{\tfrac{1}{\ETA}}+4s(\alpha;3\ETA) \left(\frac{x}{4\alpha^2}\right)^{3\ETA}+o\big(x^{\min (1,\tfrac{1}{\ETA},3\ETA)}\big)\right] \; ,\\
  & I_{3}^{(\alpha;\nu\geq 4)}(x)=\frac{\nu}{\alpha-1}\left[- s(\alpha;1) \frac{x}{2\alpha^2}+ s(\alpha;\tfrac{1}{\ETA}) \left(\frac{x}{4\alpha^2}\right)^{\tfrac{1}{\ETA}}+ s(\alpha;4\ETA) \left(\frac{x}{4\alpha^2}\right)^{4\ETA}+o\big(x^{\min (1,\tfrac{1}{\ETA},4\ETA)}\big)\right]
\end{align}
for the R\'enyi-$\alpha$ tripartite information, and
\begin{align}
   & I_{3}^{(\mathrm{vN};\nu=3)}(x)=-3 s_{\mathrm{vN}}(1) \frac{x}{2}+3 s_{\mathrm{vN}}(\tfrac{1}{\ETA}) \left(\frac{x}{4}\right)^{\tfrac{1}{\ETA}}+4s_{\mathrm{vN}}(3\ETA) \left(\frac{x}{4}\right)^{3\ETA}+o\big(x^{\min (1,\tfrac{1}{\ETA},3\ETA)}\big) \; ,\\
  & I_{3}^{(\mathrm{vN};\nu\geq 4)}(x)=\nu\left[ -s_{\mathrm{vN}}(1) \frac{x}{2}+ s_{\mathrm{vN}}(\tfrac{1}{\ETA}) \left(\frac{x}{4}\right)^{\tfrac{1}{\ETA}}+ s_{\mathrm{vN}}(4\ETA) \left(\frac{x}{4}\right)^{4\ETA}\right]+o\big(x^{\min (1,\tfrac{1}{\ETA},4\ETA)}\big)
\end{align}
for the von Neumann tripartite information, where the function $s_{\mathrm{vN}}$ is defined in Eq.~\eqref{s vN}. We note that, any time the tripartite information is negative, the leading term is linear in $x$, as in the non-interacting case $\ETA=1/2$, also discussed in Ref.~\cite{Fagotti2012New}.

\section{Spin structure from the fusion of models}\label{sec Spin structure from the fusion of models}

{In this section we propose two different procedures for computing the contribution $\mathcal{F}_f\left[\begin{smallmatrix}
    \vec\varepsilon \\ \vec\delta
\end{smallmatrix}\right]$
to the R\'enyi-$\alpha$ tripartite information in the XXZ chain, with the main focus on the torus, i.e. $\alpha=2$. The first procedure is based on computing the tripartite information in JW deformations for different values of $\nu$.  The second procedure is based on computing the tripartite information in two-site shift invariant interleaved models. We mention that our numerical checks for the contributions of different torus spin structures are based on the second procedure, which we find advantageous for numerical implementations.  
}

\paragraph{With one-site shift invariance.}

{The tripartite information of the JW deformations has the form given by Eq.~\eqref{eq:I3expansion JWdef}. If we consider a large enough number of deformations $\nu$, we can generally express $\F[(\mathrm{XXZ})]{\vec\varepsilon}{\vec\delta}x$ in terms of $I_3^{(\alpha;\nu)}(x)$ by solving polynomial equations, assuming only smoothness as a function of $x$. We have performed this procedure for $\alpha=2$ and $\alpha=3$ in Appendix~\ref{appendix from JW deformations to spin structure}. The result for the R-R spin structure for $\alpha=2$ is also given in Eq.~\eqref{from tripartite info in JW def to spin structure}. While a priori effective, the method with JW spin deformations has a rather serious downside: it requires the analysis of models with next-to-next-to-next-to-nearest-neighbor interactions. This can be numerically challenging both because higher range is generally associated with higher numerical complexity and because the typical lengths of a JW $\nu$-deformation is $\nu$ times the typical length of the original model.}

\paragraph{With two-site shift invariance.}
We can ease {the problems inherent to the procedure with JW deformations by considering} the fusion of different models. We can then set an upper bound to the range, for example next-to-nearest-neighbor interactions, and consider a large enough number of combinations of models that allow us to invert the problem. 

Specifically, we consider the interleaved XXZ-XY model 
\begin{multline}\label{interleaved XXZ XY model}
\bs H_{\mathrm{XXZ}\oplus \mathrm{XY}}=\sum_{\ell}(1+\gamma)\bs\sigma_{2\ell}^x\bs\sigma_{2\ell+1}^z\bs\sigma_{2\ell+2}^x+(1-\gamma)\bs\sigma_{2\ell}^y\bs\sigma_{2\ell+1}^z\bs\sigma_{2\ell+2}^y+\bs\sigma_{2\ell-1}^x\bs\sigma_{2\ell}^z\bs\sigma_{2\ell+1}^x+\bs\sigma_{2\ell-1}^y\bs\sigma_{2\ell}^z\bs\sigma_{2\ell+1}^y\\
+2h \bs\sigma_{2\ell}^z+\Delta\bs\sigma_{2\ell-1}^z\bs\sigma_{2\ell+1}^z\, .
\end{multline}
The model is useful because in terms of Jordan-Wigner fermions $\bs c^\dag_\ell=\prod_{j<\ell}\bs \sigma_j^z\, \frac{\bs \sigma_\ell^x+i\bs \sigma_\ell^y}{2}$ it is a sum of the fermionic XXZ chain~\eqref{XXZ fermionic chain Hamiltonian} on the odd sublattice and the fermionic version of the quantum XY chain~\eqref{XY chain Hamiltonian}, i.e. the Kitaev chain, on the even sublattice. For simplicity we work here with the XXZ chain in zero magnetic field, and we denote by $h$ the magnetic field corresponding to the XY chain, but the same considerations would apply to the XXZ with the magnetic field.

Since the model is a sum of fermionic Hamiltonians we have the factorization of the function $\mathcal{F}_f$, defined in Eq.~\eqref{eq:spinstruct}, of the interleaved model into the ones of the two constituting models. Let us denote by $X|_j=X \cap(2\mathbb{Z}+j)$ for $j=0,1$ the restriction of a set $X$ to even and odd sublattice respectively. We have then
\begin{equation}\label{F interleaved model fusion factorization}
    \F[\mathrm{XXZ}\oplus \mathrm{XY}]{\vec\varepsilon}{\vec\delta}{A,B,C}=\F[\mathrm{XXZ}]{\vec\varepsilon}{\vec\delta}{A|_1,B|_1,C|_1}\F[\mathrm{XY}]{\vec\varepsilon}{\vec\delta}{A|_0,B|_0,C|_0} \, ,
\end{equation}
where the superscripts indicate the model in which $\mathcal{F}_f$ is computed, namely, the XXZ spin chain~\eqref{XXZ spin chain Hamiltonian} and the XY spin chain~\eqref{XY chain Hamiltonian}.

The relation~\eqref{F interleaved model fusion factorization} can be exploited to compute $\mathcal{F}_f$ in the original model, i.e. the XXZ spin chain, for any configuration $A,B,C$. For simplicity we focus on the R\'enyi index $\alpha=2$, corresponding to the torus. To compute $\F[\mathrm{XXZ}]{\varepsilon}{\delta}{A,B,C}$ we can consider the tripartite information in the XXZ chain and in the interleaved model corresponding to three different values of $\gamma,h$ close to or at the circle $\gamma^2+h^2=1$. One could be $\gamma=1, h=0$, for which $\F[\mathrm{XY}]{\varepsilon}{\delta}{A,B,C}=(-1)^{4\varepsilon\delta}$ irrespectively of the configuration and the others should be taken not too far from the multicritical point $\gamma=0, h=1$ (and not too close either). This will give us a system of linear equations whose solution gives the quantity of interest, i.e. $\F[\mathrm{XXZ}]{\varepsilon}{\delta}{A,B,C}$ for all torus spin structures $\varepsilon,\delta$. We want to use the parameters $h,\gamma$ close to the circle $h^2+\gamma^2=1$ because only there the dependence of $\mathcal{F}_f$ on the configuration is sufficiently complex to give us distinct enough coefficients of the linear problem (see Sec.~\ref{s:noncrit}).

Let us use the shorthand notation $\mathcal{C}=(A,B,C)$ for the configuration of interest. Let us denote by $\tilde{A},\tilde{B},\tilde{C}$ three adjacent subsystems that are twice the size of $A,B,C$, i.e. $\tilde{A}=\{1,2,\ldots,2|A|\}$, $\tilde{B}=\{2|A|+1,2|A|+2,\ldots,2|A|+2|B|\}$, $\tilde{C}=\{2(|A|+|B|)+1,2(|A|+|B|)+2,\ldots,2(|A|+|B|+|C|)\}$, and let us denote the whole configuration by $\tilde{\mathcal{C}}=(\tilde A, \tilde B, \tilde C)$. The interleaved model satisfies $\F[\mathrm{XXZ}\oplus \mathrm{XY}]{\varepsilon}{\delta}{\tilde{\mathcal{C}}}=\F[\mathrm{XY}]{\varepsilon}{\delta}{\mathcal{C}}\F[\mathrm{XXZ}]{\varepsilon}{\delta}{\mathcal{C}}$. Considering the configuration $\mathcal{C}$ in the XXZ spin chain and the configuration $\tilde{\mathcal{C}}$ in interleaved model for $(\gamma_1,h_1)=(1,0)$ and other two values $(\gamma_j,h_j)$, $j=1,2$, close to the circle $\gamma^2+h^2=1$, we obtain the system of equations
\begin{equation}\label{linear system spin structures interleaved}
\frac{1}{2}\left(\begin{smallmatrix}
    1 & 1 & 1 & -1\\
    1 & 1 & 1 & 1\\
    \F[\mathrm{XY}_2]{0}{0}{\mathcal{C}} &  \F[\mathrm{XY}_2]{0}{1/2}{\mathcal{C}} &  \F[\mathrm{XY}_2]{1/2}{0}{\mathcal{C}} &  -\F[\mathrm{XY}_2]{1/2}{1/2}{\mathcal{C}}\\
     \F[\mathrm{XY}_3]{0}{0}{\mathcal{C}} &  \F[\mathrm{XY}_3]{0}{1/2}{\mathcal{C}} &  \F[\mathrm{XY}_3]{1/2}{0}{\mathcal{C}} &  -\F[\mathrm{XY}_3]{1/2}{1/2}{\mathcal{C}}
\end{smallmatrix}\right)
    \left(\begin{smallmatrix}
        \F[\mathrm{XXZ}]{0}{0}{\mathcal{C}} \\ \F[\mathrm{XXZ}]{0}{1/2}{\mathcal{C}} \\ \F[\mathrm{XXZ}]{1/2}{0}{\mathcal{C}} \\ \F[\mathrm{XXZ}]{1/2}{1/2}{\mathcal{C}}
    \end{smallmatrix}\right)
   =
    \left(\begin{smallmatrix}
        \exp\left[I_3^{\mathrm{XXZ}}(\mathcal{C})\right] \\  \exp\left[I_3^{\mathrm{XXZ}\oplus\mathrm{ XY}_1}(\tilde{\mathcal{C}})\right]  \\ \exp\left[I_3^{\mathrm{XXZ}\oplus\mathrm{ XY}_2}(\tilde{\mathcal{C}})\right] \\ \exp\left[I_3^{\mathrm{XXZ}\oplus\mathrm{ XY}_3}(\tilde{\mathcal{C}})\right] 
    \end{smallmatrix}\right)\, ,
\end{equation}
where by $\mathrm{XY}_j$ we denote the quantum XY chain~\eqref{XY chain Hamiltonian} with $(\gamma,h)=(\gamma_j,h_j)$ and we have also omitted writing the R\'enyi index $\alpha$ in the tripartite information $I_3^{(\alpha)}$. The solution to this linear system of equations unfolds the contribution from each the spin structure in XXZ.

In practice, we compute the tripartite information numerically using tensor network algorithms. In the interleaved model we take the values of $\gamma_j,h_j$ for $j=2,3$ exactly on the circle $\gamma^2+h^2=1$. 
The accuracy of the numerical simulations  is however crucial since a small discrepancy in the parameters results in huge discrepancies in the $\mathcal F$s.
Thus, instead of  computing $\mathcal{F}_f^{\mathrm{XY}}$ directly using analytical formulas \eqref{F XY circle exact}, we ease such sensitivity by renormalising the parameters as follows.
We assume that the state is close to the ground state of the XY model for some effective $h$ and $\gamma$ and compare the predictions for the expectation values of two local observables with their numerical values. For instance, we use the identities
\begin{equation}\label{for renormalization of parameters}
\begin{split}
& \braket{\bs\sigma^z_{2\ell}}=-\frac{1}{\pi}\int\limits_{0}^\pi \frac{h-\cos k}{\sqrt{(\gamma\sin k)^2+(h-\cos k)^2}}dk \, ,\\
&\braket{\bs\sigma^x_{2\ell}\bs\sigma^z_{2\ell+1}\bs\sigma^x_{2\ell+2}-\bs\sigma^y_{2\ell}\bs\sigma^z_{2\ell+1}\bs\sigma^y_{2\ell+2}}=-\frac{2}{\pi}\int\limits_{0}^\pi \frac{\gamma \sin^2 (k)}{\sqrt{(\gamma\sin k)^2+(h-\cos k)^2}}dk \, ,
\end{split}
\end{equation}
which are easily obtained considering that the interleaved model is equivalent to an XY chain when restricted to even sites. We extract the left hand sides of \eqref{for renormalization of parameters} from the numerical simulations and invert the system of integral equations for $h$ and $\gamma$. Finally, we use such effective $\gamma,h$ for computing $\mathcal{F}_f^{\mathrm{XY}}$. 
Appendix~\ref{appendix effective parameters renormalization} demonstrates explicitly the importance of this step by comparing some results obtained with and without it. The procedure for computing $\mathcal{F}_f^{\mathrm{XY}}$ has been developed in Ref.~\cite{Fagotti2010disjoint} and discussed also in Refs~\cite{Maric2023Universality,Maric2023universality2}. For completeness, it is also reported in Appendix~\ref{appendix: Noninteracting systems: free fermion techniques}.

\subsection{Tensor network implementation}\label{sec tensor networks}
In this section we describe the tensor network algorithms that we used to obtain numerical data for the tripartite information of three adjacent subsystems in the spin chains of interest.  
Let $\ket{\psi}$ be the ground state of a given Hamiltonian $H$. After fixing a local basis of the Hilbert space associated to each spin, we can expand the state as follows
\begin{equation}
    \ket{\psi} = 
    \sum_{i_1,...,i_L=1}^d T_{i_1,i_2,...,i_L}\ket{i_1}\ket{i_2}
...\ket{i_L}\, ,
\end{equation}
where $L$ is the chain size, $d$ is the dimension of the local Hilbert space ($d=2$ in our case of spin-$\frac{1}{2}$ chains). 
The ground state minimizes the  expectation value of the energy,
which corresponds to minimising $\braket{\psi|\bs H|\psi}$ over the $d^L$-dimensional space of the coefficients $T_{i_1,i_2,...,i_L}$.
This is a hard numerical problem to solve for growing $L$, hence one generally resorts to variational principles.
Instead of searching for the minimum over all the states of the Hilbert space, one restricts the minimisation to the states that have a particular form, referred to as the \textit{variational ansatz}.
The minimum over this subset of the Hilbert space is obviously an approximation from above to the true minimum.
How well the variational minimum approximates the actual minimum depends on the quality of the variational ansatz.
The most flexible and successful variational ansatz for 1D problems are arguably \textit{matrix product states} (MPS), which are states that can be parametrized as
\begin{equation}
    \ket{\psi} = 
    \sum_{i_1,...,i_L=1}^d
    \sum_{\ell_1,...,\ell_{L-1}=1}^\chi
    (M^{(1)})^{i_1}_{\ell_1}
    (M^{(2)})^{i_2}_{\ell_1,\ell_2}
    (M^{(3)})^{i_3}_{\ell_2,\ell_3}
    ...
    (M_L)^{i_1}_{\ell_{L-1}}
    \ket{i_1}\ket{i_2}
...\ket{i_L}\,,
\end{equation}
where $\chi$ is known as \textit{bond dimension} of the MPS, and $M^{(j)}$ is a tensor with two indices for $i\in\{1,L\}$ and three indices otherwise (see e.g.~\cite{Orus_TNreview}).
For a given bond dimension, our ansatz has $\sim Ld\chi^2$ variational parameters. Therefore the bond dimension controls the ``expressivity'' of the ansatz: a larger bond dimension corresponds to a larger subset of the Hilbert space in the minimization problem (note that, in principle, a bond dimension that scales exponentially in system size allows for an exact representation of any state).

\begin{figure}
    \centering
    \includegraphics[width=\textwidth]{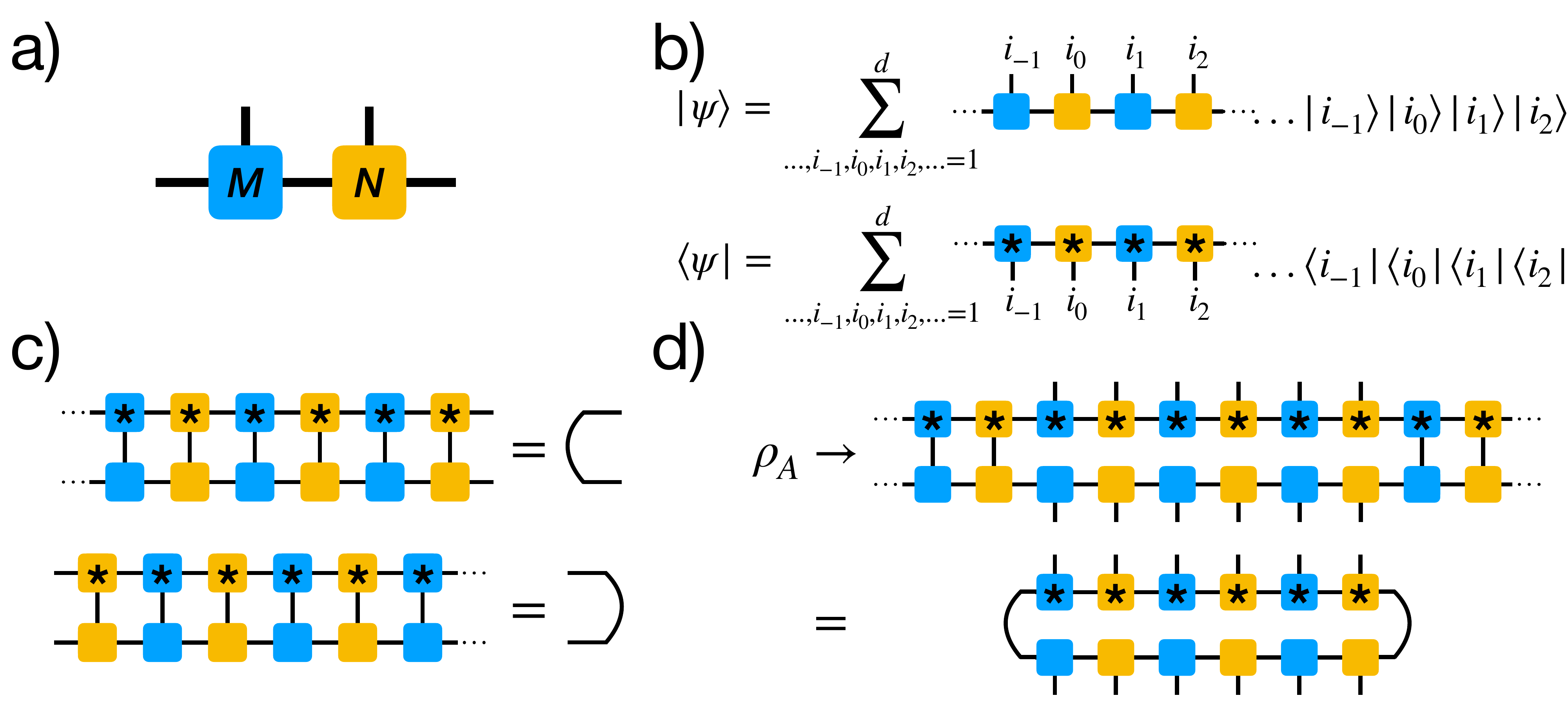}
    \caption{Conventional drawings representing tensor networks. Tensors are represented by rectangles having as many legs as the number of their indices. Since we assume 2-site-shift invariance, the MPS is parametrized by two distinct 3-index tensors. a) The building block of the MPS ansatz, which is denoted by $M$ and $N$ in the main text. b) MPS for the full state, given by the infinite repetition of the building block, and its adjoint, where the asterisk stands for complex conjugation. c) Canonical form of the MPS, allowing one to simplify the contraction of the infinite tails of the MPS. d) Example of a reduced density matrix for a block $A$ containing $6$ spins; for the sake of brevity, we omitted the sum over the basis, analogous to the sums in b).}
    \label{fig:TN_generalities}
\end{figure}

We are interested in working in the thermodynamic limit under the assumption of invariance under shifts of two lattice sites (this is a symmetry common to all  the interleaved models that we considered).
Shift-invariance allows us to encode the full variational ansatz in just two distinct tensors with three indices each, which should be repeated infinitely many times in order to reconstruct the full state.
In particular, following~\cite{iTEBD_first,iTEBD_second}, our MPS variational ansatz will look like
\begin{equation}
\label{eq:variational_ansatz}
    \ket{\psi} = 
    \sum_{...,i_{-1},i_0, i_1, i_2,...=1}^d
    \sum_{...,\ell_{-1},\ell_0, \ell_1, \ell_2,...=1}^\chi
    ...M^{i_{-1}}_{\ell_{-2},\ell_{-1}}
    N^{i_0}_{\ell_{-1},\ell_0}
    M^{i_1}_{\ell_0,\ell_1}
    N^{i_2}_{\ell_{1},\ell_2}
    ...
    \ket{i_{-1}}\ket{i_0}\ket{i_1}\ket{i_2}...\,,
\end{equation}
for a total of $2d\chi^2$ variational parameters, contained in the two tensors $M^i_{\ell,\ell'}$ and $N^i_{\ell,\ell'}$.
To simplify the notations, drawings are usually employed to represent the various quantities occurring in the computation. Tensors are represented by rectangles having as many legs as the number of their indices, while a shared leg between two tensors stands for a sum over the corresponding index.
For example, a drawing representing the building blocks $M$ and $N$ is reported in Fig.~\ref{fig:TN_generalities}a, and the state~\eqref{eq:variational_ansatz} can be represented as in Fig.~\ref{fig:TN_generalities}b.

Note that there is some gauge freedom in the MPS representation of a state: given an invertible matrix $X$, the two tensors $XM^iX^{-1}$ and $XN^iX^{-1}$ describe  the same state as $M^i$ and $N^i$. One can use this freedom in the parametrization to recast the state in a form that is particularly convenient for computations, known as \textit{canonical form of the MPS}. When working in the thermodynamic limit, the canonical form can be defined such that the contraction of the semi-infinite tails of the MPS simplifies to the identity, as shown diagrammatically in Fig.~\ref{fig:TN_generalities}c (see e.g.~\cite{iTEBD_second} for more details about canonical forms).
This allows one to express many important objects in a very compact form; as an example, we report in Fig.~\ref{fig:TN_generalities}d the tensor network representation of a reduced density matrix.

\begin{figure}
    \centering
    \includegraphics[width=\textwidth]{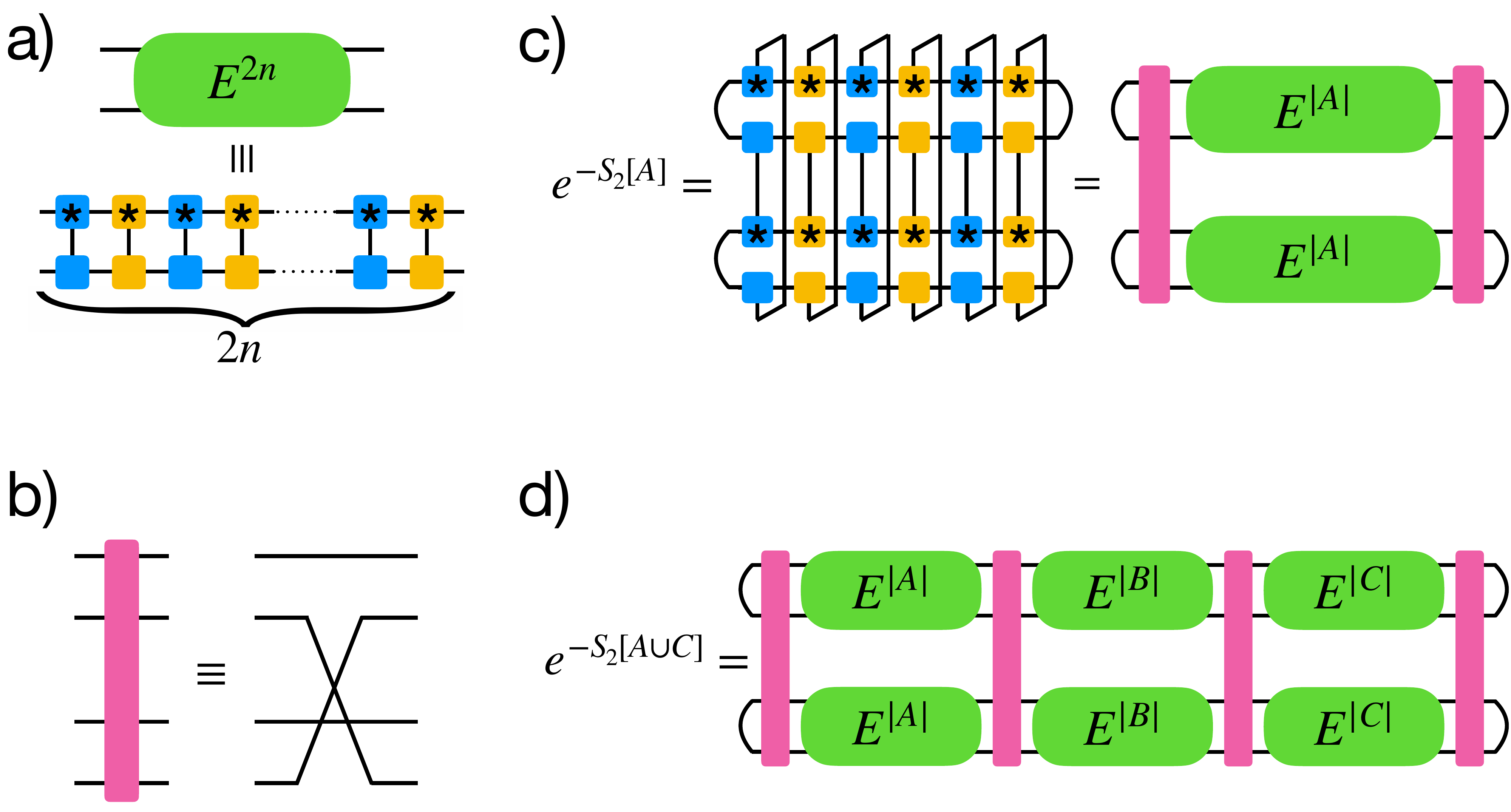}
    \caption{Tensor networks used to compute the Rényi-2 entropy.
    a) Definition of the tensor representing the transfer matrix (the definition can be easily adapted to the case in which the first site is associated to the tensor $N$ instead of the tensor $M$ and/or to the case of an odd number of sites).
    b) Definition of the twist field tensor; it is used, together with the transfer matrix, to get more compact representations of the tensor networks that follow.
    c) Entropy for a subsystem $A$ containing 6 spins. d) Entropy for a subsystem made by two disjoint blocks $A$ and $C$ at distance $|B|$.}
    \label{fig:TN_entropies}
\end{figure}

Once the variational ansatz~\eqref{eq:variational_ansatz} is chosen, the next step is to develop an algorithm to find the parameters that minimize the energy.
We use the variational uniform matrix product states (VUMPS) algorithm, formulated in Ref.~\cite{VUMPSfirst}  to perform variational energy minimization in infinite 1D quantum systems.
Ref.~\cite{VUMPSfirst} shows that the VUMPS algorithm generally outperforms its more-famous predecessors, the infinite Density Matrix Renormalization Group (iDMRG)~\cite{iDMRG} and the infinite Time Evolving Block Decimation (iTEBD)~\cite{iTEBD_first}, both in convergence speed and accuracy of the approximation. It also seems to be a good choice for studying the critical systems we are interested in.
The algorithm comes with two main schemes of convergence to the variational minimum: \textit{sequential} and \textit{parallel}.
The original paper~\cite{VUMPSfirst} showed that these two approaches are typically equivalent in terms of performance and our simulations are based on the sequential algorithm.
In practice, we use the Julia-language implementation of the algorithm available in the work-in-progress package ITensorInfiniteMPS.jl, based on the ITensors.jl library~\cite{ITensor}.
The results that we show in this work have been obtained setting the bond dimension up to $\chi=180$.

Note that the true GS in the (critical) systems that we investigated cannot be represented as an MPS: on the one hand the entanglement of a bipartition in two extensive subsystems grows as the logarithm of the subsystems size, which is divergent in the thermodynamic limit; on the other hand, for any state represented as an MPS with finite bond dimension, the entanglement follows an area law and is finite (see e.g.~\cite{Orus_TNreview}).
Thus, a finite bond dimension introduced an effective length~\cite{effective_length_MPS1,effective_length_MPS2,Pollmann2009Theory}, and only the observables with a range sufficiently smaller than that can be  described by the approximate ground state state well enough. We have checked the accuracy of the approximation by verifying that, at the scales that we consider, the numerical data are not significantly affected by an increase in the bond dimension. 

The final step is to compute the Rényi entropy by contracting the tensor networks shown in Fig.~\ref{fig:TN_entropies}.
The computational cost of the contraction scales as $\chi^5$, which is much better than the scaling of the same computation in an MPS with periodic boundary conditions, shown to be $\chi^9$ in Ref.~\cite{Coser2014OnRenyi}. The reason is that here we can simplify the contraction by employing the canonical form of the MPS, which is not possible with periodic boundary conditions. This is another advantage of working directly in the thermodynamic limit.
Note that the green and pink tensors used in Fig.~\ref{fig:TN_entropies} to give a compact representation of the tensor network of interest can be interpreted as powers of the transfer matrix and twist fields respectively, as was done e.g.~in Ref.~\cite{Coser2014OnRenyi}. We do not discuss them here and we simply use them as a convenient way to represent our tensor network.

\subsection{On the numerical data}\label{sec numerical analysis}

\begin{figure}
    \centering
	\begin{minipage}{0.85\textwidth}
		\centering
		\includegraphics[width=1\textwidth]{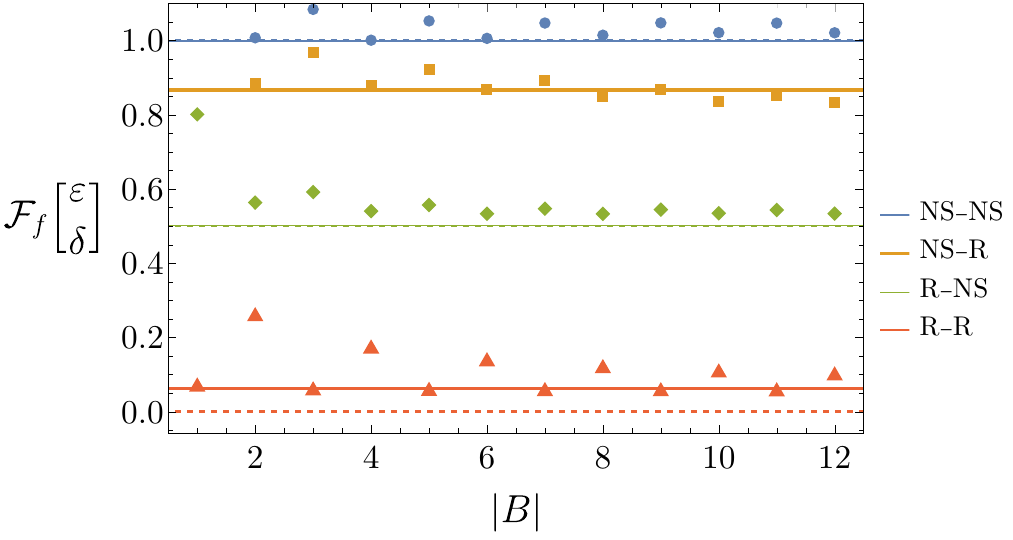}
		\subcaption{Contribution from different spin structures for $\alpha=2$ and $(\Delta,h)=(0.1, 0) $. The three subsystems have equal lengths $|A|=|B|=|C|$, so the cross ratio is $x=1/4$. The dashed lines show the free case ($\Delta=0$) for comparison.}
	\end{minipage}
\\ \vspace{0.8 cm}
	\begin{minipage}{0.85\textwidth}
		\centering
		\includegraphics[width=1\textwidth]{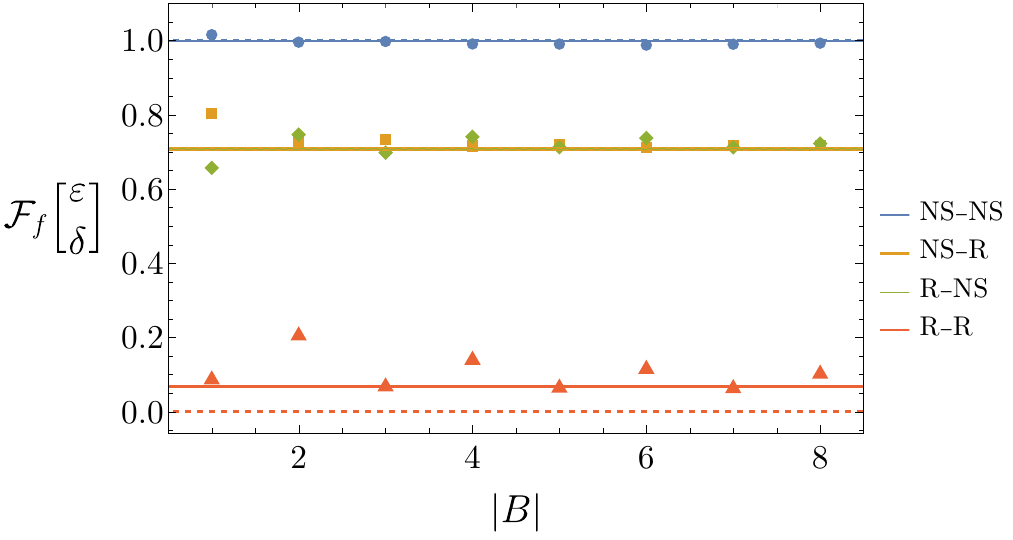}
		\subcaption{Contribution from different spin structures for $\alpha=2$ and $(\Delta,h)=(0.1, 0)$. The subsystem lengths are set to be $(|A|,|B|,|C|)=(2\ell,\ell,3\ell)$ for $\ell=1,2,3,\ldots$, so the cross ratio is $x=1/2$. The lines R-NS and NS-R overlap. The dashed lines show the free case for comparison.}
	\end{minipage}
     
        \caption{The same as in Fig.~\ref{fig:fixed_x}, but for $(\Delta,h)=(0.1, 0)$. The factor $(-1)^{4\varepsilon\delta}$ is now omitted since the prediction for the R-R term is positive. }
    \label{fig:fixed_x_Delta01}
\end{figure}

\begin{figure}
		\centering
		\includegraphics[width=0.95\textwidth]{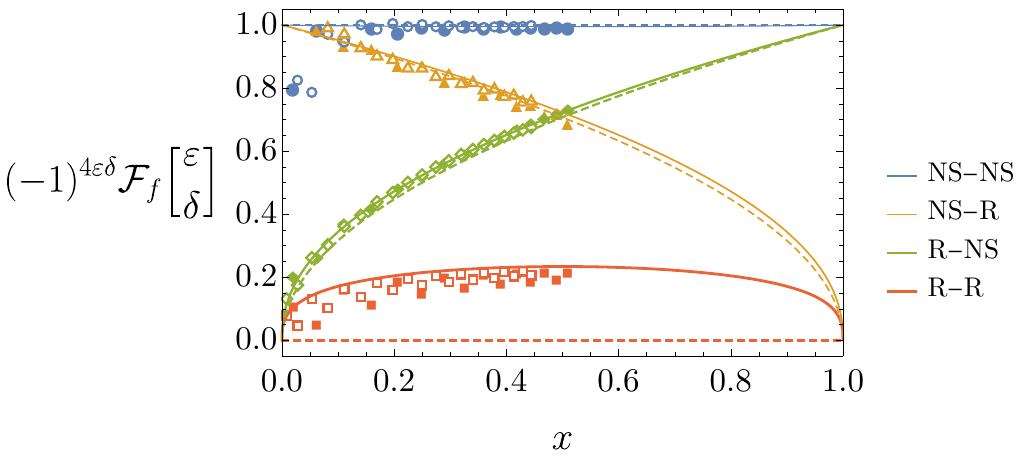}

        \caption{Contributions of different spin structures $(\varepsilon,\delta)$ to the R\'enyi-$2$ tripartite information of three finite adjacent subsystems $A,B,C$ in the infinite XXZ spin chain with $\Delta=-0.3$.  Function $\F{\varepsilon}{\delta}{A,B,C}$ (see Eq.~\eqref{eq:I3spinstr} and Eq.~\eqref{eq:spinstruct}) is shown versus the cross ratio $x=|A||C|/[(|A|+|B|)(|B|+|C|)]$ for all four torus spin structures. The lengths $|A|=|C|$ are varied, while the length of $|B|$ is set to  $6$ (filled markers) or $10$ (open markers). The points are obtained using the method of fusion of models, incorporating the  data  from the tensor network simulations. The solid lines are the  predictions~\eqref{eq:FfXXZ}, whereas dashed lines are the corresponding contributions in the free case ($\Delta=0$). }
    \label{fig:fixed B}
\end{figure}

\begin{figure}
		\centering
		\includegraphics[width=0.75\textwidth]{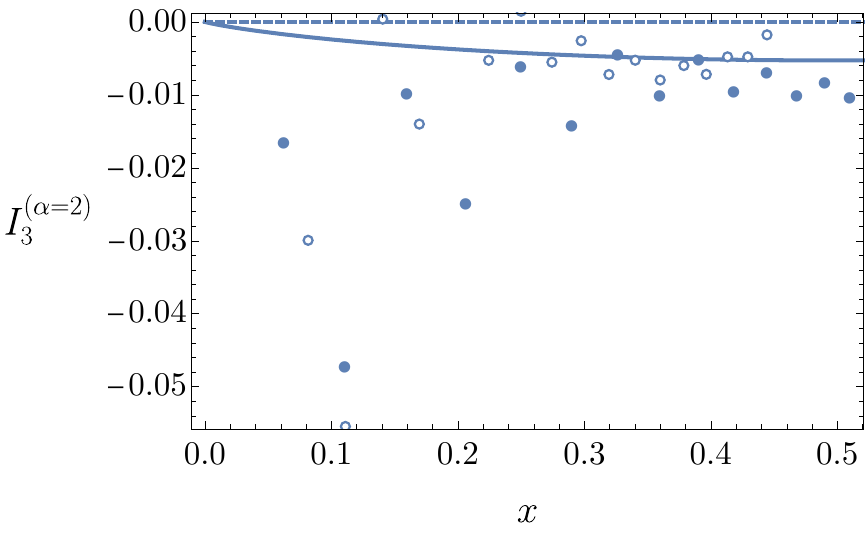}

        \caption{The R\'enyi-$2$ tripartite information in the infinite fermionic XXZ chain~\eqref{fig:fermionic} with $\Delta=-0.3$ for finite adjacent subsystems $A,B,C$. The lengths $|A|=|C|$ are varied, while the length of $|B|$ is set to $6$ (filled markers) or $10$ (open markers). The points are obtained using the method of fusion of models, incorporating the  data  from the tensor network simulations. The solid line is the  prediction~\eqref{tripartite info fermionic XXZ},  whereas the dashed line is the R\'enyi-$2$ tripartite information in the fermionic XX chain ($\Delta=0$), which is zero for any $x$. }
    \label{fig:fixed B fermionic}
\end{figure}

We have used the described procedure with the interleaved model~\eqref{interleaved XXZ XY model} to compute the functions $\F[]{\varepsilon}{\delta}{A,B,C}$ in the ground state of the XXZ spin chain~\eqref{XXZ spin chain Hamiltonian} for all spin structures $\varepsilon,\delta$ of the torus, for different configurations $(A,B,C)$ and different values of $\Delta$. We discuss here the data for $(\Delta,h)=(-0.3,0)$ and $(\Delta,h)=(0.1,0)$, for which we reached a satisfactory numerical convergence and are exemplar of the general behaviour observed. 

Figures~\ref{fig:fixed_x} and~\ref{fig:fixed_x_Delta01} show the data at fixed cross ratio~\eqref{eq:x4} equal to either $x=1/4$ or $x=1/2$. 
In Fig.~\ref{fig:fixed B}, instead,  the length of the central subsystem $B$ is fixed to two different values and the other lengths are set to the same value, varying over a range that is numerically accessible with high enough accuracy. Finally, Fig.~\ref{fig:fixed B fermionic} shows a comparison between prediction and numerical data for the fermionic R\'enyi-$2$ tripartite information. 

In all the cases, the data are in good visual agreement with the predictions~\eqref{eq:FfXXZ}. Note that the larger disagreement for smaller $x$ in Fig.~\ref{fig:fixed B} and Fig.~\ref{fig:fixed B fermionic} should be expected, in that, for fixed $|B|$, a smaller cross ratio corresponds to smaller $|A|=|C|$ and, therefore, to larger finite-size effects. Despite the overall good visual agreement, we point out that, in order to check the agreement quantitatively, one would need to work out the asymptotics of the finite-size corrections, which could depend on the particular spin structure and would not necessarily match the leading corrections expected in the tripartite information. We have not addressed this issue, therefore we leave a real quantitative check to future investigations. 
Even if our analysis is more qualitative than quantitative, it provides anyway the first numerical check of its kind in an interacting model. It also  provides an indirect check of the R\'enyi-$2$ tripartite information in the ground state of the Jordan-Wigner-$\nu$ deformations~\eqref{eq:JWdef} for any $\nu$.

\section{Insights into the tripartite information after global quenches}\label{sec quench}
One of the commonest protocols of global quench starts with preparing the system in the ground state of a local, translationally invariant Hamiltonian, which is then suddenly perturbed, for example, by changing a global magnetic field. In generic situations the system is expected to locally thermalise, in the sense that the expectation values of local observables will relax to thermal values. An exception to that behaviour, which has attracted much interest, is when the system has infinitely many (pseudo)local conservation laws, which generally means that the system is integrable. The stationary state can then be characterised by a so-called generalised Gibbs ensemble (GGE)~\cite{Rigol2007Relaxation,Doyon2017Thermalization,Vidmar2016Generalized,Essler2016Quench}. A GGE shares many properties in common with thermal states, for example, subsystems are generally described by reduced density matrices with an extensive entropy. It can however also exhibit exceptional features. Specifically, static correlations in thermal states of systems with local interactions decay exponentially with the distances~\cite{Bluhm2022exponentialdecayof}; in a generalised Gibbs ensemble of a system with local interactions, that is not always the case. Ref.~\cite{Maric2022Universality} considered some settings in which one should expect also algebraically decaying correlations  at late times after a global quench in an integrable model. It was observed that such  exceptional property leaves clear marks in the entanglement entropies of subsystems. 
In particular, Refs~\cite{Maric2022Universality,Maric2023Universality,Maric2023universality2} studied the tripartite information after global quenches in noninteracting spin chains and identified contributions that are generally killed by the quench and others that are instead only mildly affected by nonequilibrium time evolution. 
As a result, the tripartite information of large adjacent subsystems is not trivialised by the limit of infinite time and, remarkably, remains a function of the cross ratio such as in conformal critical ground states. 
It was also pointed out that the final result has a universal property: when the tripartite information survives the quench, the limit $x\rightarrow 1^-$ turns out to be nonzero and, in the large class of noninteracting models investigated in Ref.~\cite{Maric2023Universality}, it approaches a universal $-\log 2$. 

The analogous problem in the presence of interactions deserves a separate investigation. Here we simply compute what would be the late time tripartite information in 1D lattice models if the observations of Ref.~\cite{Maric2022Universality} about the quench  having the main effect of killing the expectation values of operators including fermionic strings between the blocks held true in more general models. Namely, we do not expect this property to depend on the boson radius in the underlying field theory. This has the effect of killing all terms~\eqref{eq:spinstruct} in the tripartite information with $s_k\neq 1$ for some $k$. These are the terms with $\vec\varepsilon\neq 0$. We will also assume the quench to be small, so that we can presumably ignore the modifications in the remaining contributions.

If we keep only the terms that we expect to be nonzero we find
\begin{equation}
     I_3^{(\alpha)}(x)=\frac{1}{\alpha-1}\log \left( \frac{1}{2^{\alpha-1}} \sum\limits_{\vec \delta} \mathcal{F}_f \left[
    \begin{smallmatrix}
  \vec 0 \\ \vec \delta    
    \end{smallmatrix}\right](x)\right)=\frac{1}{\alpha-1}\log \frac{\Theta(\vec 0|\tfrac{1}{\ETA} \Omega)\Theta(\vec 0|4\ETA \Omega)}{[\Theta(\vec 0 | \Omega)] ^2}\, ,
\end{equation}
which can also be written as
\begin{equation}\label{tripartite quench in terms of ground state}
 I_3^{(\alpha)}(x)=I_{3,\mathrm{GS}}^{(\alpha)}(x)+\frac{1}{\alpha-1}\log\frac{\Theta(\vec 0 | 4\ETA\Omega)}{\Theta(\vec 0 | \ETA\Omega)} \, ,
\end{equation}
where $I_{3,\mathrm{GS}}^{(\alpha)}$ is the tripartite information in the ground state of the XXZ chain, given by Eq.~\eqref{tripartite information compactified boson}.

\paragraph{The large $x$ expansion: residual tripartite information.}

The large $x$ expansion for the tripartite information in the stationary state can be obtained from the corresponding expansion of two terms in Eq.~\eqref{tripartite quench in terms of ground state}. The large $x$ expansion of the first term, describing the ground state, follows directly from the crossing symmetry $I_{3,\mathrm{GS}}^{(\alpha)}(x)=I_{3,\mathrm{GS}}^{(\alpha)}(1-x)$ and the small $x$ expansion, given by Eq.~\eqref{expansion XXZ small x}. It remains to deal with the large $x$ expansion of the second term in \eqref{tripartite quench in terms of ground state}.

We start from the large $x$ expansion of $\beta_{k/\alpha}(x)$ appearing in the expression \eqref{period matrix def} for the period matrix. The expansion follows from the relation $\beta_{k/\alpha}(x)=1/\beta_{k/\alpha}(1-x)$ and the small $x$ expansion of $\beta_{k/\alpha}(x)$, reported in Refs~\cite{Calabrese2009Entanglement1,Calabrese2011Entanglement} and in Eq.~\eqref{beta small x expansion} in the appendix. Namely, we have for $x\to 1$
\begin{equation}
\begin{split}\label{def f euler gamma}
   &\beta_{\frac{k}{\alpha}}(x)=-\frac{\pi}{\sin\left(\pi\frac{k}{\alpha}\right)}\left[\log(1-x)+f_{\frac{k}{\alpha}}+\sum_{\ell=1}^\infty p_\ell\left(\frac{k}{\alpha}\right)(1-x)^\ell) \right]^{-1} , \\ & f_{\frac{k}{\alpha}}\equiv 2\gamma_E+\psi\left(\frac{k}{\alpha}\right)+\psi\left(1-\frac{k}{\alpha}\right),
   \end{split}
\end{equation}
where $\gamma_E$ is the Euler gamma, $\psi(z)\equiv\Gamma'(z)/\Gamma(z)$ is the polygamma function and $p_l(z)$ is a polynomial of degree 2$l$, whose explicit expression is not needed. Factoring out the leading term $\log(1-x)$ and exploiting the formula for the geometric series we obtain further
\begin{equation}
    \beta_{\frac{k}{\alpha}}(x)=\frac{\pi}{\sin\left(\pi\frac{k}{\alpha}\right)} \frac{1}{|\log(1-x)|} \left[1-\frac{f_{\frac{k}{\alpha}}}{|\log(1-x)|}+O\left(\frac{1-x}{|\log(1-x)|}\right)\right], \quad \textrm{as } x\to 1 \; .
\end{equation}
This gives the large $x$ expansion of the period matrix~\eqref{period matrix def}
\begin{equation}\label{period matrix large x expansion}
    \Omega=\frac{2\pi i}{\alpha} \frac{\alpha\mathbb{I}-\vec u \vec u^\mathrm{T}}{|\log(1-x)|} \left[\mathbb{I}-\frac{\mathbb{I}+\vec u \vec u^\mathrm{T}}{\alpha|\log(1-x)|}B+O\left(\frac{1-x}{|\log(1-x)|}\right) \right],
\end{equation}
where $\vec u=(1,1,\ldots,1)^\mathrm{T}$ is a vector with $\alpha -1$ columns and all elements equal to $1$, satisfying $(\alpha\mathbb I-\vec u \vec u^\mathrm{T})^{-1}=(\mathbb{I}+\vec u \vec u^\mathrm{T})/\alpha$, and we have defined the matrix
\begin{equation}\label{def matrix B large x expansion}
    B_{\ell,n}=  \sum_{k=1}^{\alpha-1} f_{\frac{k}{\alpha}} \cos\left[2\pi\frac{k}{\alpha}(\ell-n)\right] \; .
\end{equation}
We can see that $\log(1-x)$ in Eq.~\eqref{period matrix large x expansion} appears in the denominator, making it complicated to conclude about the behavior of the theta function \eqref{theta function with characteristic def}. We can exploit the fact that it appears in the numerator in the inverse matrix. The inverse, that follows from Neumann series formula, is given by
\begin{equation}\label{inverse period matrix large x expansion}
 \left(-i \Omega \right)^{-1}=\frac{1}{2\pi}|\log(1-x)|\left(\mathbb{I}+\vec u \vec u^\mathrm{T}\right)+\frac{1}{2\pi\alpha} \left(\mathbb{I}+\vec u \vec u^\mathrm{T}\right)B\left(\mathbb{I}+\vec u \vec u^\mathrm{T}\right) +O(1-x) , \quad \textrm{as } x \to 1 \; .
\end{equation}

To express the theta function in terms of $\Omega^{-1}$ we exploit the formula
\begin{equation}\label{formula discrete gaussians}
    \sum_{\vec m\in\mathbb Z^{\alpha-1}}e^{-\vec m \cdot M \vec m}=\sqrt{\det(\pi M^{-1})}\sum_{\vec n\in\mathbb Z^{\alpha-1}}e^{-\pi^2\vec n \cdot M^{-1} \vec n}
\end{equation}
for $(\alpha-1) \times (\alpha-1)$ positive definite orthogonal matrices $M$. For $\alpha=2$ the derivation of the formula is an application of the Poisson summation formula, while for higher $\alpha$ it is a straightforward generalization. Using formula~\eqref{formula discrete gaussians} we get
\begin{equation}
    \frac{\Theta(0|4\ETA \Omega)}{\Theta(0|\ETA\Omega)}=\frac{1}{2^{\alpha-1}}\frac{\sum\limits_{\vec n \in \mathbb{Z}^{\alpha-1}} \exp\left[-\frac{1}{4\ETA}\pi\vec n \cdot (-i\Omega)^{-1} \vec n\right]}{\sum\limits_{\vec n \in \mathbb{Z}^{\alpha-1}} \exp\left[-\frac{1}{\ETA}\pi\vec n \cdot (-i\Omega)^{-1} \vec n\right]}
\end{equation}
and plugging the expansion~\eqref{inverse period matrix large x expansion} we get further
\begin{equation}\label{step fraction large x expansion}
    \frac{\Theta(0|4\ETA \Omega)}{\Theta(0|\ETA\Omega)}=\frac{1}{2^{\alpha-1}}\frac{\sum\limits_{\vec n \in \mathbb{Z}^{\alpha-1}} (1-x)^{\frac{1}{8\ETA}\vec n \cdot (\mathbb{I}+\vec u \vec u^\mathrm{T}) \vec n} \exp\left\{-\frac{1}{8\ETA\alpha}\vec n \cdot \left[ \left(\mathbb{I}+\vec u \vec u^\mathrm{T}\right)B\left(\mathbb{I}+\vec u \vec u^\mathrm{T}\right) +O(1-x)\right]\vec n \right\}}{\sum\limits_{\vec n \in \mathbb{Z}^{\alpha-1}} (1-x)^{\frac{1}{2\ETA}\vec n \cdot (\mathbb{I}+\vec u \vec u^\mathrm{T}) \vec n} \exp\left\{-\frac{1}{2\ETA\alpha}\vec n \cdot \left[ \left(\mathbb{I}+\vec u \vec u^\mathrm{T}\right)B\left(\mathbb{I}+\vec u \vec u^\mathrm{T}\right) +O(1-x)\right]\vec n \right\}} \; .
\end{equation}

Both in the numerator and the denominator in Eq.~\eqref{step fraction large x expansion} the leading term in the expansion in $1-x$ as $x\to 1$ is a constant term obtained for $\vec n =\vec 0$. The first subleading term is given by those $\vec n \neq \vec 0$ that minimize
\begin{equation}\label{step large x expansion to minimize}
    \vec n \cdot (\mathbb{I}+\vec u \vec u^\mathrm{T}) \vec n= \sum_{j=1}^{\alpha-1}n_j^2 +\left(\sum_{j=1}^{\alpha-1}n_j\right)^2 \; .
\end{equation}
By inspection, it is easy to see that this expression is minimized by $\vec n =\pm \vec e_j$ for any $j$ and by $\vec n= \vec e_j - \vec e_{j'} $ for any $j\neq j'$, where we have denoted $(\vec e_j)_\ell=\delta_{j\ell}$. In these cases the expression in Eq.~\eqref{step large x expansion to minimize} is equal to $2$. If three $n_j$'s are nonzero we cannot get anything smaller than $3$, so we can conclude that the previous solutions complete the totality of the minima. Therefore, we find
\begin{equation}\label{step ratio theta functions to simplify the coefficient}
\begin{split}
       &\frac{\Theta(0|4\ETA \Omega)}{\Theta(0|\ETA\Omega)} \\ &=\frac{1}{2^{\alpha-1}} \left\{1+(1-x)^{\frac{1}{4\ETA}}\left[2\sum_{j=1}^{\alpha-1} e^{-\frac{1}{8\ETA \alpha}(\vec e_j+\vec u)\cdot B(\vec e_j +\vec u)} + \sum_{j\neq j' =1}^{\alpha-1} e^{-\frac{1}{8\ETA \alpha}(\vec e_j-\vec e_{j'})\cdot B(\vec e_j -\vec e_{j'} )} \right] +o\left((1-x)^{\frac{1}{4\ETA}} \right) \right\}
\end{split}
\end{equation}
as $x\to 1$.

The coefficient of the subleading term in Eq.~\eqref{step ratio theta functions to simplify the coefficient} can be simplified. From the definition~\eqref{def matrix B large x expansion} of matrix $B$ we get
\begin{equation}
   (\vec e_j+\vec u)\cdot B(\vec e_j +\vec u)= 4\sum_{k=1}^{\alpha-1} f_{\frac{k}{\alpha}} \sin^2\left(\frac{\pi k j}{\alpha}\right) , \qquad (\vec e_j-\vec e_{j'})\cdot B(\vec e_j -\vec e_{j'} )= 4\sum_{k=1}^{\alpha-1} f_{\frac{k}{\alpha}} \sin^2\left(\frac{\pi k (j-j')}{\alpha}\right) \; .
\end{equation}
Now, similarly to what was done in Ref.~\cite{Calabrese2011Entanglement} for the small $x$ expansion, using the definition of $f(k/\alpha)$ in Eq.~\eqref{def f euler gamma} and the integral representation of the polygamma function
\begin{equation}
    \psi\left(\frac{k}{\alpha}\right)+\gamma_E= \int_{0}^\infty \frac{e^{-t}-e^{-\frac{k}{\alpha}t}}{1-e^{-t}} dt \; ,
\end{equation}
exchanging the order of sum and integral and then performing the latter, after simple algebra we find
\begin{equation}
    \sum_{k=1}^{\alpha-1} f_{\frac{k}{\alpha}} \sin^2\left(\frac{\pi k j}{\alpha}\right)=-\alpha \log \alpha-\alpha \log \left[2\sin\left(\frac{\pi j}{\alpha}\right)\right] , \qquad j=1,2,\ldots,\alpha-1 \; .
\end{equation}
It follows
\begin{equation}\label{step large x expansion ratio theta functions}
       \frac{\Theta(0|4\ETA \Omega)}{\Theta(0|\ETA\Omega)} =\frac{1}{2^{\alpha-1}} \left[1+s(\alpha; \tfrac{1}{4\ETA})\left(\frac{1-x}{4\alpha^2}\right)^{\frac{1}{4\ETA}} +o\left((1-x)^{\frac{1}{4\ETA}} \right) \right],
\end{equation}
where the function $s$ is defined in Eq.~\eqref{s CCT}.

Finally, the large $x$ expansion of the tripartite information follows from Eqs~\eqref{tripartite quench in terms of ground state},~\eqref{expansion XXZ small x} and ~\eqref{step large x expansion ratio theta functions},
\begin{equation}
         I_3^{(\alpha)}(x)=-\log 2 \ + \frac{s(\alpha;\ETA)}{\alpha-1}\left(\frac{1-x}{4\alpha^2}\right)^{\ETA}+\frac{s(\alpha;\tfrac{1}{4\ETA})}{\alpha-1}\left(\frac{1-x}{4\alpha^2}\right)^{\frac{1}{4\ETA}}+o\left((1-x)^{\min(\ETA, \tfrac{1}{4\ETA})}\right), \quad \textrm{as } x\to 1 .
\end{equation}
The large $x$ expansion of the von Neumann tripartite information follows immediately, using formula~\eqref{s vN step}. The result is given in Eq.~\eqref{tripartite expansion quench vN}.

\paragraph{A physical argument.}
Our predictions are based on the conjecture that the contributions from spin structures associated with operators that have a fermionic string in the central block are killed by the global quench. In addition, in the limit of small quench the other contributions are not affected at all by the quench and keep their ground state values. 
If that picture is correct, we can interpret the residual tripartite information as a measure of the fraction of correlations surviving the quench. Imagine indeed to focus on the contribution from a correlation corresponding to a product of spins. For it to be nonzero, the number of $\bs \sigma^{x,y}$ in the first and third block should be even (otherwise it would be represented with a string of JW fermions in the block in between). 
On the other hand, in the ground state of the model, if $|B|\ll |A|,|C|$, $\mathrm{tr}[\rho_{A\cup C}^\alpha]\gg \mathrm{tr}[\rho_{(A\setminus A_R)\cup (C\setminus C_L)}^\alpha] $, whenever $A_R\in A$ and  $C_L\in C$, both adjacent to $B$, have extent comparable with $A$ and $C$ (the mutual information diverges in the limit of zero distance). Since $\mathrm{tr}[\rho_{A\cup C}^\alpha]$ and $ \mathrm{tr}[\rho_{(A\setminus A_R)\cup (C\setminus C_L)}^\alpha] $ share the correlations in which there are no spins in $A_R\cup C_L$, it is reasonable to assume that the typical correlation contributing to $\mathrm{tr}[\rho_{A\cup C}^\alpha]$ should feature spin operators with a distance from the edges of the central block that is negligible with respect to the lengths of $A$ and $C$.   
We now argue that the value of a typical  correlation in the ground state should not be affected by a small displacement (by a length comparable with $|B|$) of one of its spins. Let the latter be the operator $\bs\sigma^{x,y}$ closest to the edge with $B$. Moving it to the other block is not expected to change its contribution to the ground state entropy, but it changes the parity of the number of spins in the blocks. Thus, such a small displacement maps a correlation that is relevant to the quench tripartite information to an irrelevant one, both of them giving the same contribution in the ground state of the model. This suggests that half of the correlations contributing to the ground state tripartite information are killed by the global quench, reducing the space of relevant correlations by $\frac{1}{2}$. The logarithm of such ratio  matches indeed our predictions for the residual tripartite information.

\section{Conclusions}
We revisited the R\'enyi-$\alpha$ entropies of two disjoint blocks in the ground state of the XXZ quantum spin chain, focusing on the universal contribution that is captured by the tripartite information of the corresponding three adjacent blocks.  
We have pointed out the connection between the tripartite information and  the partition functions of the underlying massless Thirring quantum field theory with different spin structures.  
This has allowed us to uncover the behaviour of the tripartite information in other models, e.g., in  the fermionic XXZ chain. For small cross ratio $x$, using the methods of Ref.~\cite{Calabrese2011Entanglement}, we computed the limit $\alpha\to 1^+$ of the analytic continuation, obtaining in turn also the small $x$ expansion of the von Neumann tripartite information.

    The results for the fermionic XXZ chain generalise the works  of Refs~\cite{Casini2005,Casini2009reduced_density,Casini2009Entanglement} on the entanglement entropy of disjoint blocks in non-interacting hopping chains, which are described by the free massless Dirac fermion. While for the latter the tripartite information is zero for any $\alpha$, the tripartite information in the fermionic XXZ chain, described by the massless Thirring QFT with NS boundary conditions on all cycles, exhibits a more interesting behavior. In particular,  it can be both positive and negative.
    
Very recently Ref.~\cite{Fujimura2023} studied the R\'enyi-$2$ entanglement entropy of two disjoint intervals in the massless Thirring model, with the main focus on the NS-NS spin structure of the torus. This problem is equivalent to the one of the fermionic XXZ chain, studied in this work. They have obtained the results by expressing the fermionic partition function in terms of bosonic ones, in accordance with the duality discussed in Ref.~\cite{Karch2019}. Our result~\eqref{tripartite info fermionic XXZ} for $\alpha=2$, while manifestly in a different form, is in agreement with their result, and, moreover, the numerical checks we have performed provide also a check for their formulas.

The identification of the contributions from different spin structures in the XXZ spin chain has also allowed us to predict the behaviour of the tripartite information in a family of models that we called ``Jordan-Wigner-$\nu$ deformations'', the XXZ spin chain corresponding to the special case $\nu=1$. Those are simple models with interactions beyond the two-body ones, incorporating $\nu+1$ adjacent spins. Their low-energy properties are described by a sum of massless Thirring models for $\nu$ different species of fermions with correlated spin structures. The corresponding central charge is equal to $\nu$. For $\nu=2$ we have obtained the curious result that the tripartite information is exactly twice the one of the fermionic XXZ chain. This provides an additional relation between the entanglement properties of the fermionic XXZ model and those of a spin model. Furthermore, we have shown that, for large enough $\nu$, the tripartite information in Jordan-Wigner-$\nu$ deformations can become negative, generalizing  the free case addressed in Ref.~\cite{Fagotti2012New}.

Finally, the identification of the  operators contributing to each spin structure has allowed us to speculate about the stationary behavior of the tripartite information at late times after a small quantum quench of the anisotropy in the gapless XXZ spin chain. The resulting predictions  depend on the anisotropy and reduce to exact formulas  analytically computed in Refs~\cite{Maric2022Universality,Maric2023universality2,Maric2023Universality} for the free case $\Delta=0$.
We also extracted the limit $x\to 1^-$, which corresponds to configurations in which the central subsystem is much smaller than the others. We found that the tripartite information becomes equal to $-\log 2$ irrespectively of the model parameters. Refs~\cite{Maric2022Universality,Maric2023universality2,Maric2023Universality} uncovered the same universal value in a variety of non-interacting quantum spin chains and termed it ``residual tripartite information''. The results of our work suggest that such universality is likely to extend to interacting integrable models.

There are several open problems that our work brings to light, particularly in the direction of quantum quenches. 
First, our conjecture about the asymptotic behaviour of the tripartite information after a small quench in an interacting integrable model should be confirmed and generalised to larger quenches by an exact calculation of time evolution after the quench. 
Second, Ref.~\cite{Hayden2013Holographic} argued that a negative tripartite information should suggest a domination of quantum entanglement over classical correlations. Since we have confirmed that the residual tripartite information after global quenches is negative, it is natural to wonder whether that is measuring a quantum or classical feature (or both of them).  To that aim it might be useful to study some other quantity in which the distinction between classical and quantum correlations is sharper, for example the entanglement negativity~\cite{Vidal2002,Plenio2005,Calabrese2012negativityPRL}. We also wonder whether the residual tripartite information might be connected to some unusual property of the entanglement Hamiltonian~\cite{Dalmonte2022} of disjoint subsystems.

\acknowledgments

VM is grateful to Shunsuke Furukawa, Tomotaka Kuwahara, Vincent Pasquier, Tadashi Takayanagi and Tomonori Ugajin for stimulating discussions on topics related to this work.

\paragraph{Funding information}
This work was supported by the European Research Council under the Starting Grant No. 805252 LoCoMacro.
SB acknowledges the financial support from PEPR-Q (QubitAF project).

\appendix

\section{Spin structures and gapped spin chains}\label{appendix spin structures and gapped spin chains}

In this appendix we discuss how to show the formulas given in Sec.~\ref{s:noncrit}.

\paragraph{Disordered phases.}
Let us consider the fermionic pseudo-RDMs~\eqref{eq:pseudoRDM} underlying the expansion in the spin structures. After separating the contribution between even and odd observables, clustering allows us for an immediate simplification 
\begin{align}
\rho_{AC}^{+,s'}\sim& \rho_A^{(e)}\otimes \rho_C^{(e)}+s'\frac{1}{2^{|A|+|C|}}\sum_{\bs O_A^{(o)},\bs O_C^{(o)}}
\braket{\bs O_A^{(o)} \bs T(v_1)\bs T(u_2-1) \bs O_C^{(o)}} O_A O_C\label{eq:rho-sp} \\
\rho_{AC}^{-,s'}\sim&\frac{1}{2^{|A|+|C|}}\sum_{\bs O_A^{(e)},\bs O_C^{(e)}}
\braket{\bs O_A^{(e)} \bs T(v_1)\bs T(u_2-1)\bs O^{(e)}_C} O_A O_C+s' \rho_A^{(o)}\otimes \rho_C^{(o)}\nonumber\\
&\qquad\qquad\qquad \overset{sp.flip}{=}\frac{1}{2^{|A|+|C|}}\sum_{\bs O_A^{(e)},\bs O_C^{(e)}}
\braket{\bs O_A^{(e)} \bs T(v_1)\bs T(u_2-1)\bs O^{(e)}_C} O_A O_C
\end{align}
where the last identity follows from spin-flip symmetry, we introduced the semilocal string $\bs T(u)=(-1)^{\sum_{\ell\leq u}\bs c^\dag_\ell \bs c_\ell}$, and we used the notations for the boundaries of the intervals $A=[u_1,v_1]$ and $C=[u_2,v_2]$.
These expressions can be simplified using clustering for semilocal operators\footnote{Clustering for semilocal operators can be understood as follows.
In the product state with all spins aligned in the $z$ direction, the second term of \eqref{eq:rho-sp} vanishes and the first term matches $\rho_A^{(e)}\otimes \rho_C^{(e)}$. We then observe that any spin-flip symmetric state with a finite correlation length is mapped to that product state by a spin-flip symmetric quasilocal unitary transformation. Rather than transforming the product state, we can apply the inverse transformation to the operators. Such a transformation would preserve the parity of the operators and map the string $(-1)^{\bs F_B}$ into itself, up to even unitary  operators $\bs Q_{\partial_L B}\bs Q_{\partial_R B}$ quasilocalised around the boundaries $\partial_L B$ and $\partial_R B$ of $B$: $(-1)^{\bs F_B}\rightarrow \bs Q_{\partial_L B}\bs Q_{\partial_R B}(-1)^{\bs F_B}\bs Q^{-1}_{\partial_R B}\bs Q^{-1}_{\partial_L B}$. Since the string acts as $1$ on the product state, this allows us to use clustering also in this case.}, which gives
\begin{align}
\rho_{AC}^{+,s'}\sim& \rho_A^{(e)}\otimes \rho_C^{(e)}\\
\rho_{AC}^{-,s'}\sim&\left(\frac{1}{2^{|A|}}\sum_{\bs O_A^{(e)}}
\braket{\bs O_A^{(e)} \bs T(v_1)} O_A\right)\otimes\left( \frac{1}{2^{|C|}}\sum_{\bs O_C^{(e)}}
\braket{\bs T(u_2-1)\bs O_C^{(e)} } O_C\right)
\end{align}
We can manipulate  the pseudo-RDMs in each block as follows
\begin{equation}\label{eq:pseudo-manip}
\sum_{\bs O_A^{(e)}}
\braket{\bs O_A^{(e)} \bs T(v_1)} \bs O_A=\sum_{\bs O_A^{(e)}}
\braket{\bs O_A^{(e)} \bs T(x)} \bs O_A\bs T(x)\bs T(v_1)=\bs R_A^{(e)} \bs U_x^\dag \bs T(x)\bs T(v_1)\equiv \bs T(v_1) \bs T(x)\bs U_x\bs R_A^{(e)}
\end{equation}
where $\bs U_x$ is a unitary transformation quasilocalised around $x$, which we assumed to be deep in the bulk of $A$, such that
$
\bs T(x)\ket{\Psi}=\bs U_x\ket{\Psi}
$. The last identity in \eqref{eq:pseudo-manip} comes from applying $\bs T(x)$ to the bra instead of the ket. Note that, even if the result is supposed to be independent of $x$, we cannot simply replace   $\bs U_x^\dag \bs T(x)$ by the identity. On the other hand we have
$
(\sum_{\bs O_A^{(e)}}
\braket{\bs O_A^{(e)} \bs T(v_1)} O_A)^2=[\rho_A^{(e)}]^2
$
which implies that the pseudo-RDM is equal to $\rho_A^{(e)}$ up to changes of signs in the eigenvalues.  Since each term of the 
spin structure of the tripartite information consists of an even number of $\rho^{-,s'}_{AC}$, this difference is irrelevant, so we can overlook that subtlety and assume that all pseudo-RDMs are identical, which gives \eqref{eq:noncrit}. 

\paragraph{Ordered phases.}
Using clustering and the vanishing of the expectation values of strings of the form  $(-1)^{\bs F_X}$ in the limit of large $|X|$ in $\ket{\psi}$ we find
\begin{align}
&\rho_{AC}^{+,s'}\sim \frac{1}{2^{|A|+|C|}}\sum_{\bs O_A^{(e)},\bs O_C^{(e)}}\braket{\psi|\bs O_A^{(e)}\bs O_C^{(e)}|\psi}O_A O_C\rightarrow \rho_{A;\ket{\psi}}^{+}\otimes \rho_{C;\ket{\psi}}^{+}\\
&\rho_{AC}^{-,s'}\sim s'\frac{1}{2^{|A|+|C|}}\sum_{\bs O_A^{(o)},\bs O_C^{(o)}}\braket{\psi|\bs O_A^{(o)}\bs O_C^{(o)}|\psi}O_A O_C\rightarrow s' \rho_{A;\ket{\psi}}^{-}\otimes \rho_{C;\ket{\psi}}^{-}
\end{align}
Thus, each term of the spin structure is proportional to the trace of a product of $\rho_{A;\ket{\psi}}^{\pm}$  and $\rho_{C;\ket{\psi}}^{\pm}$.  
We note
\begin{equation}
\rho_{A;\ket{\psi}}^{\pm}=\frac{\rho_{A;\ket{\psi}}\pm (-1)^{\bs F_A}\rho_{A;\ket{\psi}}(-1)^{\bs F_A}}{2}=\frac{\rho_{A;\ket{\psi}}\pm \rho_{A;(-1)^{\bs F}\ket{\psi}}}{2}\, .
\end{equation}
Since $\rho_{A;\ket{\psi}}$ satisfies the area law, it has only a finite number of eigenvalues that are significantly different from zero. Let us call them $\ket{\psi_i,A}$ with eigenvalues $p_i$ such that $p_i\geq p_{i+1}$.  By assumption we have
\begin{equation}
|\sum_i p_i\braket{\psi_i,A|(-1)^{\bs F_A}|\psi_i,A}|=|\mathrm{tr}[(-1)^{\bs F_A}\rho_{A;\ket{\psi}}]|\leq e^{-\kappa |A|}
\end{equation}
for some $\kappa>0$ and it is reasonable to expect the inequality to hold term by term, \emph{i.e.},   $|\braket{\psi_i|(-1)^{\bs F_A}|\psi_i}|\leq e^{-\kappa |A|}$. This implies
\begin{equation}
\mathrm{tr}[\rho_{A;\ket{\psi}}^{s_1}\cdots\rho_{A;\ket{\psi}}^{s_n}]\sim \frac{1}{2^{n-1}}\mathrm{tr}[\rho_{A;\ket{\psi}}^{n}]\, .
\end{equation}
Thus, in the spin structure the only information that matters is how many $\rho_{AC}^{-,-}$ are present in a particular term, indeed each of them carries a factor $-1$. This is exactly \eqref{eq:Fcat}, which is therefore expected to hold  in general within this class of states.

\section{Reproducing the tripartite information of the XXZ chain}\label{appendix reproducing CCT}

In this appendix we show how summing the massless Thirring model over all spin structures reproduces the tripartite information of the XXZ chain~\eqref{tripartite info fermionic XXZ}, obtained from the consideration of the free compactified boson~\cite{Calabrese2009Entanglement1}. We also derive some useful identities. The tripartite information of the XXZ chain is obtained by summing over all spin structures, as given by Eq.~\eqref{eq:I3spinstr},
\begin{equation}\label{step tripartite info XXZ}
     I_3^{(\alpha)}(x)=\frac{1}{\alpha-1}\log \left( \frac{1}{2^{\alpha-1}} \frac{\sum\limits_{\vec e}\Theta\left[
    \begin{matrix}
  \vec E \\ \vec F    
    \end{matrix}\right]\left(\vec 0|\tilde \Omega\right)}{\left[\Theta\left(\vec 0 | \Omega\right)\right]^2 }\right) \; .
\end{equation}
We show that \eqref{step tripartite info XXZ} reproduces the formula~\eqref{tripartite information compactified boson} obtained from the considerations of the free compactified boson in Ref.~\cite{Calabrese2009Entanglement1}.

We remind the reader that the Riemannn-Siegel theta function with characteristic is defined as follows
\begin{equation}
\Theta\left[
    \begin{matrix}
  \vec E \\ \vec F    
    \end{matrix}\right]\left(\vec 0|\tilde \Omega\right)=\sum_{\vec M\in \mathbb Z^{2(\alpha-1)}}e^{i\pi  (\vec M+\vec E)^t \tilde \Omega (\vec M+\vec E)+2\pi i \vec M\cdot \vec F}\, .
\end{equation}
Let us then split the sum over $\vec M$ into two sums
\begin{equation}
\sum_{\vec M}\equiv \sum_{\vec \mu\in \{0,\frac{1}{2}\}^{\alpha-1}}\sum_{\vec m,\vec m'\in\mathbb Z^{\alpha-1}}\qquad \text{with}\quad \vec M= \begin{pmatrix}
    1 \\ 1
\end{pmatrix}\otimes (\vec m+\vec\mu)+\begin{pmatrix}
    1 \\ -1
\end{pmatrix}\otimes (\vec m'+\vec \mu)
\end{equation}
We then find
\begin{equation}
\sum_{\vec \mu\in \{0,\frac{1}{2}\}^{\alpha-1}}(-1)^{4\vec \mu\cdot\vec\delta}\sum_{\vec m,\vec m'\in\mathbb Z^{\alpha-1}}\exp\left[i\pi  4\ETA(\vec m'+\vec \varepsilon+\vec\mu)^t  \Omega (\vec m'+\vec \varepsilon+\vec \mu)+i\pi \tfrac{1}{\ETA}(\vec m+\vec \mu)^t \Omega(\vec m+\vec \mu)\right]\, ,
\end{equation}
which can be compactly written as
\begin{equation}
\Theta\left[
    \begin{matrix}
  \vec E \\ \vec F    
    \end{matrix}\right]\left(\vec 0|\tilde \Omega\right)=\sum_{\vec \mu\in \{0,\frac{1}{2}\}^{\alpha-1}}(-1)^{4\vec \mu\cdot\vec\delta}\Theta\left[
    \begin{matrix}
  \vec \mu \\ \vec 0    
    \end{matrix}\right]\left(\vec 0|\tfrac{1}{\ETA} \Omega\right)\Theta\left[
    \begin{matrix}
  \vec \varepsilon+\vec \mu \\ \vec 0    
    \end{matrix}\right]\left(\vec 0|4\ETA\Omega\right)\, .
\end{equation}
Incidentally, we point out the following identity:
\begin{equation}
\Theta\left[
    \begin{matrix}
  \vec E \\ \vec F    
    \end{matrix}\right]\left(\vec 0|\tilde \Omega\right)\xrightarrow{\ETA\rightarrow \frac{1}{4\ETA}}(-1)^{4\vec\varepsilon\cdot\vec\delta} \Theta\left[
    \begin{matrix}
  \vec E \\ \vec F    
    \end{matrix}\right]\left(\vec 0|\tilde \Omega\right) \; ,
\end{equation}
obtained by shifting the summation index as $\vec\mu \to \vec\varepsilon+\vec\mu$. This also gives the identity~\eqref{eq:FfXXZ}.

Therefore we have
\begin{equation}\label{eq:I3XXZ}
     I_3^{(\alpha)}(x)=\frac{1}{\alpha-1}\log \left( \frac{1}{2^{\alpha-1}} \frac{\sum\limits_{\vec \varepsilon,\vec\delta,\vec \mu}(-1)^{4\vec \mu\cdot\vec\delta}\Theta\left[
    \begin{smallmatrix}
  \vec \mu \\ \vec 0    
    \end{smallmatrix}\right]\left(\vec 0|\tfrac{1}{\ETA} \Omega\right)\Theta\left[
    \begin{smallmatrix}
  \vec \varepsilon+\vec \mu \\ \vec 0    
    \end{smallmatrix}\right]\left(\vec 0|4\ETA\Omega\right)}{\left|\Theta\left(\vec 0 | \Omega\right)\right| ^2}\right)
\end{equation}
Let us work out the sums. 
If we sum over $\vec \delta$ we get
\begin{equation}
\sum_{\vec\delta\in \{0,\frac{1}{2}\}^{\alpha-1}}\Theta\left[
    \begin{matrix}
  \vec E \\ \vec F    
    \end{matrix}\right]\left(\vec 0|\tilde \Omega\right)=2^{\alpha-1}\Theta\left(\vec 0|\tfrac{1}{\ETA} \Omega\right)\Theta\left[
    \begin{matrix}
  \vec \varepsilon \\ \vec 0    
    \end{matrix}\right]\left(\vec 0|4\ETA\Omega\right)\, .
\end{equation}
If we sum also over $\vec\varepsilon$ we find 
\begin{multline}
\sum_{\vec\delta,\vec\varepsilon\in \{0,\frac{1}{2}\}^{\alpha-1}}\Theta\left[
    \begin{matrix}
  \vec E \\ \vec F    
    \end{matrix}\right]\left(\vec 0|\tilde \Omega\right)=2^{\alpha-1}\Theta\left(\vec 0|\tfrac{1}{\ETA} \Omega\right)\sum_{\vec\varepsilon\in \{0,\frac{1}{2}\}^{\alpha-1}} \sum_{\vec m'\in\mathbb Z^{\alpha-1}}\exp\left[i4\pi \ETA(\vec m'+\vec \varepsilon)^t  \Omega (\vec m'+\vec \varepsilon)\right]=\\
    2^{\alpha-1}\Theta\left(\vec 0|\tfrac{1}{\ETA} \Omega\right)\sum_{\vec\varepsilon\in \{0,\frac{1}{2}\}^{\alpha-1}} \sum_{\vec m'\in\mathbb Z^{\alpha-1}}\exp\left[i\pi  \ETA(2\vec m'+2\vec \varepsilon)^t  \Omega (2\vec m'+2\vec \varepsilon)\right]=
    2^{\alpha-1}\Theta\left(\vec 0|\tfrac{1}{\ETA} \Omega\right) \Theta\left(\vec 0|\ETA\Omega\right)\, ,
\end{multline}
which is consistent with the result~\eqref{tripartite information compactified boson}.

\section{Noninteracting systems: free fermion techniques}\label{appendix: Noninteracting systems: free fermion techniques}

In non-interacting spin chains the computation of the entanglement entropies is based on the matrix of fermionic two-point correlations and the Wick theorem. The two-point correlation matrix reads $\Gamma_{2\ell+i, 2n+j}=\delta_{\ell n}\delta_{i j}-\braket{\vec{\bs a}_\ell\vec{\bs a}_n^\dagger}_{i j}$, where we have introduced the Majorana fermions $\bs a_{2\ell-1}=\bs c_\ell+\bs c_\ell^\dagger=\bs \sigma_\ell^x\prod_{j<\ell}\bs\sigma_j^z, \ \bs a_{2\ell}=i(\bs c_\ell-\bs c_\ell^\dagger)=\bs \sigma_\ell^y\prod_{j<\ell}\bs\sigma_j^z$ and the shorthand vector notation $\vec{\bs a_\ell}=(\bs{a}_{2\ell-1}, \ \bs{a}_{2\ell})^{\mathrm{T}}$. The Majorana fermions are self-adjoint and satisfy the anticommutation relations $\{\bs a_\ell, \bs a_n\}=2\delta_{\ell n}$. We refer the reader to Ref.~\cite{Maric2023Universality} (Table 1) for a brief review how to compute the correlation matrix in different settings. In particular, in the ground state of the quantum XY chain
\begin{equation}
    H=\sum_{\ell} (1+\gamma)\bs\sigma_\ell^x \bs\sigma_{\ell+1}^x +(1-\gamma)\bs\sigma_\ell^y \bs\sigma_{\ell+1}^y +2h\bs \sigma_\ell^z
\end{equation}
we have
\begin{equation}
\begin{split}
&\Gamma_{2j+1,2\ell+2}=-\Gamma_{2\ell+2,2j+1}=\int\limits_{-\pi}^\pi\frac{-\gamma\sin k+i(h-\cos k)}{|-\gamma\sin k+i(h-\cos k)|}e^{ik(j-\ell)}\frac{dk}{2\pi} \, , \\ &\Gamma_{2j+1,2\ell+1}=\Gamma_{2j+2,2\ell+2}=0 \; .
\end{split}
\end{equation}

In the case of a single block $A$, the entropy of spins is equal to the entropy of the corresponding block of Jordan-Wigner fermions \cite{Jin2004,Vidal2003,Peschel2003}.  It can be written in terms of the correlation matrix $\Gamma_A=(\Gamma_{j,\ell})_{j,\ell=1}^{2|A|}$,
\be\label{eq:Ssingleblock}
S_\alpha(A)=\tfrac{1}{1-\alpha}\tfrac{1}{2}\tr \Bigl[\log\Bigl(\bigl(\tfrac{\mathrm I+\Gamma_A}{2}\bigr)^\alpha+\bigl(\tfrac{\mathrm I-\Gamma_A}{2}\bigr)^\alpha\Bigr)\Bigr]\, .
\ee
Such a simplicity is related to the property that the reduced density matrix~\eqref{reduced density matrix fermions smaller Hilbert space} is a fermionic Gaussian, i.e. an operator of the form $\rho(\Gamma)\propto \exp{\left(\sum_{j,\ell }  a_j^\dagger W_{j,l} a_\ell /4\right)}$, where $W$ is an antisymmetric matrix. Clearly, the moments appearing in Eq.~\eqref{eq:spinstruct} can be then computed using $\tr\rho_A^\alpha=\exp[(1-\alpha)S_\alpha(A)]$.

In the case of disjoint blocks the non-locality of the Jordan-Wigner transformation complicates the correspondence. A procedure for computing the R\'enyi entropies of two disjoint blocks on the lattice that takes into account these non-locality effects has been proposed in Ref.~\cite{Fagotti2010disjoint}. The main ingredient of this procedure is to notice that each pseudo-RDM~\eqref{eq:pseudoRDM} is a fermionic Gaussian. Then the procedure is reduced to computing the trace of products of different Gaussians.

The pseudo-RDMs $\rho_{AC}^{s,s'}$ can be expressed in terms of four correlation matrices: $\Gamma_1\equiv\Gamma^{+,+}\equiv \Gamma_{A\cup C, A\cup C}$, $\Gamma_2\equiv\Gamma^{+,-}\equiv \mathrm P \Gamma_1 \mathrm P$, $\Gamma_3\equiv\Gamma^{-,+}\equiv \Gamma_1-\Gamma_{A\cup C,B}\Gamma_{B,B}^{-1}\Gamma_{B,A\cup C}$ and $\Gamma_4\equiv\Gamma^{-,-}\equiv P\Gamma_3 P$, where $\Gamma_{A',A''}$ for some subsystems $A',A''$ is the correlation matrix with the indices running in $A',A''$ respectively,
\begin{equation}
\begin{split}
    & \left(\Gamma_{A',A''}\right)_{2(\ell-1)+i,2(n-1)+j}=\delta_{A'(\ell),A''(n)}\delta_{ij}-\braket{\vec{\bs a }_{A'(\ell)} \vec{\bs a}^\dagger _{A''(n)}}_{ij} \; ,\\  \quad & \ell\in\{1,2,\ldots,|A'|\},\ n\in\{1,2,\ldots,|A''|\}, \  i,j\in\{1,2\}, \\ & A'(1)<A'(2)<\ldots <A'(|A'|) \textrm{ are the elements of $A'$ in ascending order (analogously for $A''$)},
\end{split}
\end{equation}
and
\be\label{eq:operator P definition}
\mathrm P_{\ell n}\equiv \delta_{\ell n}\begin{cases}
	-1& 1\leq \ell\leq 2|A|\\
	1&\text{otherwise}
\end{cases}
\ee
introduces a minus sign for each fermion in block $A$. Specifically, we have
\begin{equation}
\mathrm{tr}\left[\prod\limits_{k=1}^\alpha\rho_{AC}^{s_k,s_k'}\right]=|\det \Gamma_{B,B}|^{\frac{N^{-+}+N^{--}}{2}}\{\Gamma^{s_1,s_1'},\Gamma^{s_2,s_2'},\ldots, \Gamma^{s_\alpha,s_\alpha'}\} \, ,
\end{equation}
where $N^{-+}$ and $N^{--}$ is the number of matrices $\Gamma^{-+}$ and $\Gamma^{--}$, respectively, appearing in the expression.

Each term can be evaluated using the  recursive formula derived in Ref.~\cite{Fagotti2010disjoint}, which expresses a product of two normalised  Gaussians as a Gaussian
\begin{equation}\label{eq:product of two gaussians}
	\rho(\Gamma)\rho(\Gamma')=\{\Gamma,\Gamma'\}\rho\left(\Gamma\times \Gamma'\right), \quad \Gamma\times \Gamma'\equiv \mathrm I-(\mathrm I-\Gamma')(\mathrm I+\Gamma\Gamma')^{-1}(\mathrm I-\Gamma)\, ,
\end{equation}
together with a formula for the trace of a product of normalised Gaussians
\begin{equation}\label{eq:trace of a product of two gaussians determinant formula}
	\{\Gamma,\Gamma'\}^2=\mathrm{tr}[\rho(\Gamma)\rho(\Gamma')]^2=\det \bigg(\frac{\mathrm I+\Gamma \Gamma'}{2}\bigg) \; .
\end{equation}
Alternatively,  \eqref{eq:product of two gaussians} could be replaced by the formula for the trace of a product of an arbitrary number of Gaussians reported in Ref.~\cite{Klich2014}. Note that \eqref{eq:trace of a product of two gaussians determinant formula} fixes the value of $\{\Gamma,\Gamma'\}$ only up to a sign; in general such a sign ambiguity can be lifted by computing the product of eigenvalues with halved degeneracy \cite{Fagotti2010disjoint}:
\begin{equation}
    \{\Gamma,\Gamma'\}=\prod\limits_{\{\lambda\}/2}\lambda \, ,
\end{equation}
where the product is over all eigenvalues $\lambda$ of the matrix $(I+\Gamma\Gamma')/2$ with degeneracy, that is always even, reduced by half.

\section{About the effective parameters in the interleaved model}\label{appendix effective parameters renormalization}
\begin{figure}[t]
    \centering
    \includegraphics[width=0.85\textwidth]{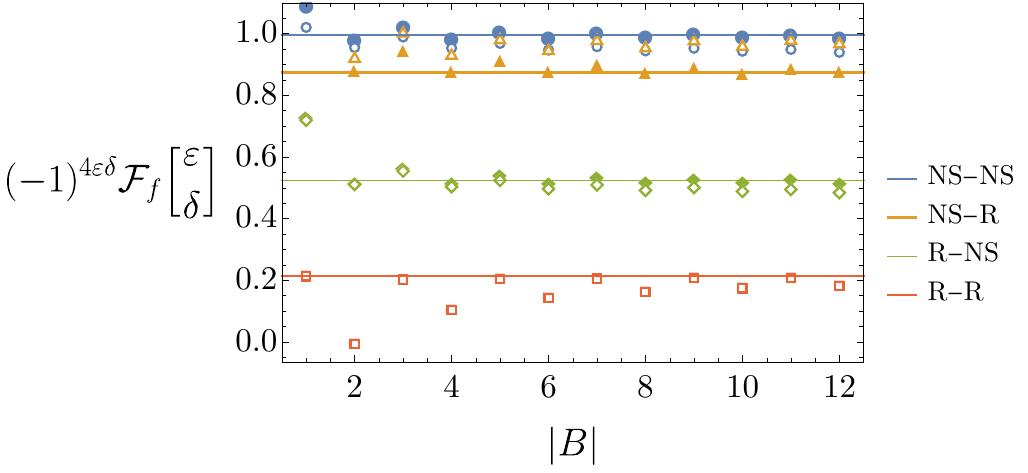}
    \caption{The same as in Fig.~\ref{fig:fixed_x}a, where open point markers correspond to data obtained  using the parameters $\gamma,h$ of the interleaved model~\eqref{interleaved XXZ XY model} in Eq.~\eqref{linear system spin structures interleaved}, whereas filled point markers correspond to renormalised data. The latter data are obtained by choosing effective parameters in the XY subchain so as to match the numerical expectation values of two local observables. The R-R is not affected by the change, but the other spin structures are very sensitive to it and  only the renormalised data show good agreement with the predictions. See the discussion around Eq.~\eqref{for renormalization of parameters}.}
    \label{fig:unrenormalized}
\end{figure}

Here we comment a detail about the implementation of our numerical procedure. Around eq.~\eqref{for renormalization of parameters} we have discussed the importance of using the effective parameters $\gamma,h$ in Eq.~\ref{linear system spin structures interleaved}, obtained by considering local observables in our tensor network simulations, instead of the ones we initially put in the Hamiltonian. In Fig.~\ref{fig:unrenormalized} we explicitly compare the results of Fig.~\ref{fig:fixed_x}a, obtained with the effective parameters, with the ones obtained using the initial ones. There we can see the importance of using the effective values.

\section{From the tripartite information of Jordan-Wigner spin deformations to XXZ spin structure}\label{appendix from JW deformations to spin structure}

{The tripartite information of the JW deformation has the form
\begin{equation}
e^{(\alpha-1)[I_3^{(\alpha;\nu)}(x)+\log 2]}=\sum_{n=1}^{M_\alpha} g_{\alpha,n} \gamma_{\alpha,n}^\nu-\sum_{n=M_\alpha+1}^{M_\alpha+\tilde M_\alpha} \tilde g_{\alpha,n} \gamma_{\alpha,n}^\nu\, ,
\end{equation}
where each $\gamma_{\alpha,n}$ denotes the contribution of a different spin structure $\F[(\mathrm{XXZ})]{\vec\varepsilon}{\vec\delta}{x}$, $M_\alpha$ and $\tilde M_\alpha$  are integers denoting the limits of the sum and $g_{\alpha,n}$ and $\tilde g_{\alpha,n}$ are positive integers that represent the degeneracy of the terms.}
We can generally express $\{\gamma_{\alpha,n}\}$    in terms of $\{e^{(\alpha-1)[I_3^{(\alpha;\nu)}(x)+\log 2]}\}$ if we consider a large enough number of deformations. 
If $g_{\alpha,n}=1$ and $\tilde g_{\alpha,n}=0$, we can readily express $\{\gamma_{\alpha,n}\}$ as the solutions to the polynomial
\begin{equation}\label{eq:poly}
x^{M_\alpha}+\sum_{n=1}^{M_\alpha}(-1)^n e_n x^{M_\alpha-n}=0\, ,
\end{equation}
where $e_n$ are the elementary symmetric polynomials, which can be expressed in terms of the power-sum symmetric polynomials through the Newton's identities. Specifically, we have
$$
e_n=\frac{1}{n!}\det \left|\begin{matrix}
e^{(\alpha-1)[I_3^{(\alpha;1)}(x)+\log 2]}&1&0&\dots&\\
e^{(\alpha-1)[I_3^{(\alpha;2)}(x)+\log 2]}&e^{(\alpha-1)[I_3^{(\alpha;1)}(x)+\log 2]}&2&0\dots\\
\vdots&&\ddots&\ddots\\
e^{(\alpha-1)[I_3^{(\alpha;n-1)}(x)+\log 2]}&e^{(\alpha-1)[I_3^{(\alpha;n-2)}(x)+\log 2]}&\dots&e^{(\alpha-1)[I_3^{(\alpha;1)}(x)+\log 2]}&n-1\\
e^{(\alpha-1)[I_3^{(\alpha;n)}(x)+\log 2]}&e^{(\alpha-1)[I_3^{(\alpha;n-1)}(x)+\log 2]}&\dots&e^{(\alpha-1)[I_3^{(\alpha;2)}(x)+\log 2]}&e^{(\alpha-1)[I_3^{(\alpha;1)}(x)+\log 2]}
\end{matrix}\right|\, .
$$
Since each tripartite information is symmetric under $x\rightarrow 1-x$, also the roots of this polynomial are, a priori, symmetric. The symmetry is however broken if we require the solutions to be smooth in $x\in (0,1)$. Those will be the functions that should be identified with $\{\gamma_{\alpha,n}\}$.

In order to solve the  inverse problem for given $\alpha$, we will have to relax the constraints on $g_{\alpha,n}$ and $\tilde g_{\alpha,n}$ that we imposed for the sake of exploration. Note however that the simplified solution that we exhibited above will be useful to derive the exact expressions. 

\paragraph{$2$-sheeted Riemann (torus, $g=1$). }
In the $2$-sheeted Riemann surface, $M_2=3$, $\tilde M_2=1$, and the coefficients are all equal to $1$.  
To solve this problem, it is convenient to treat $\gamma_{2,4}$ as it were a correction ``renormalizing'' the tripartite information (this abuse of terminology comes from the observation that, like any other term multiplying the coefficients $\tilde g_{\alpha, n}$ in the Riemann surfaces with genus $g$, $\gamma_{2,4}$ is zero in the noninteracting case~\cite{Coser2016Spin}). 
Let us then define
\begin{equation}\label{eq:Itilde}
e^{\tilde I_3^{(2;j)}(x)}=e^{I_3^{(2;j)}(x)}+\tfrac{\gamma_{2,4}^j(x)}{2}
\end{equation}
We then have
\begin{equation}
\begin{aligned}
e_1=&2e^{\tilde I_3^{(2;1)}(x)}=2e^{ I_3^{(2;1)}(x)}+\gamma_{2,4}(x)\\
e_2=&2e^{2\tilde I_3^{(2;1)}(x)}-e^{\tilde I_3^{(2;2)}(x)}=2e^{2 I_3^{(2;1)}(x)}-e^{ I_3^{(2;2)}(x)}+2e^{I_3^{(2;1)}(x)}\gamma_{2,4}(x)\\
e_3=&\tfrac{4e^{3\tilde I_3^{(2;1)}(x)}+2e^{\tilde I_3^{(2;3)}(x)}-6e^{\tilde I_3^{(2;2)}(x)}e^{\tilde I_3^{(2;1)}(x)} }{3}=\tfrac{4e^{3 I_3^{(2;1)}(x)}+2e^{ I_3^{(2;3)}(x)}-6e^{ I_3^{(2;2)}(x)}e^{ I_3^{(2;1)}(x)} }{3}\\
&\qquad\qquad\qquad\qquad\qquad\qquad\qquad\qquad\qquad\qquad+(2e^{2 I_3^{(2;1)}(x)}-e^{ I_3^{(2;2)}(x)})\gamma_{2,4}(x)\, .
\end{aligned}
\end{equation}
Before discussing the roots of the corresponding polynomial---\eqref{eq:poly}, let us determine $\gamma_{2,4}(x)$. To that aim, it is convenient to rewrite \eqref{eq:poly} with the unknown replaced by $\gamma_{2,j}(x)$, with $j\in\{1,2,3\}$, multiply it by $\gamma_{2,j}(x)$ and finally sum over $j\in\{1,2,3\}$. This gives a linear equation for $\gamma_{2,4}(x)$, which is solved by 
\begin{equation}\label{eq:gamma4}
\gamma_{2,4}(x)=-\frac{4e^{4 I_3^{(2;1)}(x)}-12 e^{2I_3^{(2;1)}(x)}e^{I_3^{(2;2)}(x)}+3e^{2I_3^{(2;2)}(x)}+8e^{ I_3^{(2;1)}(x)}e^{ I_3^{(2;3)}(x)}-3e^{ I_3^{(2;4)}(x)}}{8e^{3I_3^{(2;1)}(x)}-12e^{ I_3^{(2;1)}(x)}e^{ I_3^{(2;2)}(x)}+4 e^{ I_3^{(2;3)}(x)}}\, .
\end{equation}
We point out that $\gamma_{2,4}(x)$ exhibits the symmetry $x\leftrightarrow 1-x$ and vanishes at $x=0$ (and $x=1$). We can now solve the cubic polynomial in terms of $\tilde I_3^{(2,j)}(x)$ (they can be eventually expressed in terms of $I_3^{(2,j)}(x)$ by using \eqref{eq:gamma4}). 
The roots of \eqref{eq:poly} read
$$
x_k=\frac{1}{3}\Bigl(e_1-e^{\frac{2\pi i k}{3}}C-e^{-\frac{2\pi i k}{3}}\frac{\Delta_0}{C}\Bigr)\qquad k\in\{-1,0,1\}
$$
where
\begin{equation}
\begin{aligned}
C=&\sqrt[3]{\frac{\Delta_1+\sqrt{\Delta_1^2-4\Delta_0^3}}{2}}\\
\Delta_0=&e_1^2-3e_2=3e^{\tilde I_3^{(2;2)}(x)}-2e^{2\tilde I_3^{(2;1)}(x)}\\
\Delta_1=&-2e_1^3+9 e_1 e_2-27 e_3=-16e^{3\tilde I_3^{(2;1)}(x)}+36e^{\tilde I_3^{(2;1)}(x)}e^{\tilde I_3^{(2;2)}(x)}-18e^{\tilde I_3^{(2;3)}(x)} 
\end{aligned}
\end{equation}
For $x=0$ we can choose 
$$
e_1=2\qquad \Delta_0=1\qquad \Delta_1=2\qquad 
C=1
$$
which gives
$$
x_k(0)=\frac{4}{3}\sin^2(\pi k /3)\, .
$$
It is not accidental that at $x=0$ all roots are either $0$ or $1$. This can be seen as a consequence of clustering for all Jordan-Wigner deformations. Importantly, the symmetry $x\leftrightarrow 1-x$ implies the same set of roots at $x=1$. 

As we commented before, the roots that we obtained using the standard solution of a third order equation are not exactly our unknowns, as they're generally not smooth. In fact, the ratios of partition functions should be identified with some root for given $x$, but the specific root could depend on $x$. They can jump from one root to the other only if there is a multiple root, i.e., if the discriminant vanishes
$$
\Delta_1^2=4\Delta_0^3\, .
$$
We've already seen that this happens at $x=0$ (and hence, at $x=1$). 
Since we expect real roots, we are going to assume $\Delta_1^2(x)\leq  4\Delta_0^3(x)$ and $\Delta_0(x)\geq 0$ (the latter follows from the former). Thus we have
$$
C=\sqrt[3]{\frac{\Delta_1+i\sqrt{4\Delta_0^3-\Delta_1^2}}{2}}=\sqrt{\Delta_0}e^{\frac{i}{3}\arctan\frac{\sqrt{4\Delta_0^3-\Delta_1^2}}{\Delta_1}+\frac{\pi}{3}\theta_H(-\Delta_1)}\, ,
$$
where the $\frac{\pi}{3}$ shift in the argument has been added to keep $C$ smooth even when  $\Delta_1$ changes sign. The three solutions then read
\begin{equation}
x_k=\frac{2}{3}\Bigl(e^{\tilde I_3^{(2;1)}(x)}-\sqrt{\Delta_0}\cos\Bigl(\tfrac{2\pi k}{3}+\tfrac{1}{3}\arctan\tfrac{\sqrt{4\Delta_0^3-\Delta_1^2}}{\Delta_1}+\tfrac{\pi}{3}\theta_H(-\Delta_1)\Bigr)\Bigr)
\end{equation}
Since $\Delta_1^2(x)\leq  4\Delta_0^3(x)$, when we expand $4\Delta_0^3(x)-\Delta_1^2(x)$ about its zeroes $z_j$ we get (assuming $\Delta_1(z_j)\neq 0$)
$$
x_k(x)\xrightarrow{x\sim z_j}\tfrac{2}{3}\Bigl(e^{\tilde I_3^{(2;1)}(x)}-\sqrt{\Delta_0}\cos\Bigl(\tfrac{2\pi k}{3}\mathrm{sgn}(x-z_j)+\tfrac{1}{3}\tfrac{(x-z_j)\sqrt{\partial_x^2(4\Delta_0^3-\Delta_1^2)|_{x=z_j}}}{\sqrt{2}\Delta_1(z_j)}+\tfrac{\pi}{3}\theta_H(-\Delta_1)\mathrm{sgn}(x-z_j)\Bigr)\Bigr)
$$
which shows that  $k=1$ matches $k=-1$ when $\Delta_1(z_j)>0$ and $k=0$ does $k=-1$ when $\Delta_1(z_j)<0$. As long as the tripartite information is smooth around $z_j$, the jump will result in a smooth function. 
If there's an even number of zeroes of $4\Delta_0^3-\Delta_1^2$ with  $\Delta_1>0$, the solution corresponding to $k=1$ at $x=0$  corresponds to $k=1$ at $x=1$. 

Let us study the solutions depending on the number of zeroes of $4\Delta_0^3-\Delta_1^2$ in $x\in(0,1)$, boundary excluded. If there's no zero, the roots are in a one-to-one correspondence with the partition functions. If there's a zero (hence it should be at $x=\frac{1}{2}$), there 
are two possibilities. If $\Delta_1>0$ at the zero, then $k=1$ interchanges with $k=-1$ at $x=\frac{1}{2}$, which we could indicate as $k=(1,-1)$ for one solution and $k=(1,-1)$ for the other, whereas the root with $k=0$ remains unchanged, i.e., $k=(0,0)$. This situation cannot really occur if we consider that the fermionic tripartite information is supposed to satisfy clustering (which means, the corresponding partition function should be $1$ at $x=0$) and to be symmetric under $x\leftrightarrow 1-x$. The other possibility is $\Delta_1(1/2)<0$, in which case the root $k=1$ remains unchanged, $k=(1,1)$, and can be identified with the fermionic entropy, whereas $k=0$ and $k=-1$ interchange at $x=\frac{1}{2}$, i.e., the other solutions are $k=(0,-1)$ and $k=(-1,0)$. This situation occurs, in particular, in the noninteracting case. If $4\Delta_0^3-\Delta_1^2$ has two zeroes (clearly, symmetric around $x=\frac{1}{2}$), by symmetry $\Delta_1$ has the same sign at both zeroes. If $\Delta_1>0$, the smooth solutions are $k=(1,-1,1)$, $k=(-1,1,-1)$, and $k=(0,0,0)$. If $\Delta_1<0$, they are $k=(1,1,1)$, $k=(-1,0,-1)$, and $k=(0,-1,0)$. This means that also the smooth solutions are symmetric under $x\rightarrow 1-x$. If there are three zeroes, one is necessarily at $x=1/2$. There are four possibilities for the sign of $\Delta_1$. If it is always positive we have $k=(1,-1,1,-1)$, $k=(-1,1,-1,1)$, and $k=(0,0,0,0)$, which cannot occur for the same reason why we couldn't have a single zero with $\Delta_1>0$. If $\Delta_1<0$ at each zero we have $k=(1,1,1,1)$, $k=(0,-1,0,-1)$, and $k=(-1,0,-1,0)$. If $\Delta_1(1/2)>0$ but $\Delta_1<0$ at the other zeroes, then $k=(1,1,-1,0)$, $k=(0,-1,1,1)$, and $k=(-1,0,0,-1)$, where the latter corresponds to the fermionic entropy.  If $\Delta_1(1/2)<0$ but $\Delta_1>0$ at the other zeroes, then $k=(1,-1,0,0)$, $k=(0,0,-1,1)$, and $k=(-1,1,1,-1)$, where, again, the latter corresponds to the fermionic entropy. 
\begin{table}
$$
\begin{array}{c||c|c|c|c|c|c}
\mathrm{sgn}(\Delta_{1})&-&++&--&+-+&-+-&---\\
\hline
\gamma_{2,1}&(1,1)&(\pm 1,\mp 1,\pm 1)&( 1,1, 1)\vee (-1,0,-1)&(-1,1,1,-1)&(-1,0,0,-1)&(1,1,1,1)\\
\gamma_{2,2}&(-1,0)&(\mp 1,\pm 1,\mp 1)& (-1,0,-1)\vee ( 1,1, 1)&(1,-1,0,0)&(1,1,-1,0)&(-1,0,-1,0)\\
\gamma_{2,3}&(0,-1)&(0,0,0)&(0,-1,0)&(0,0,-1,1)&(0,-1,1,1)&(0,-1,0,-1)\\
\hline
\text{RTI}&-\log 2&0&0&-\log 2&-\log 2&-\log 2
\end{array}
$$
\caption{The smooth roots written in terms of the standard ones depending on the sign of $\mathrm{sgn}(\Delta_{1})$ at the zeroes of $4\Delta_0^3-\Delta_1^2$, for the smallest number of zeroes. In some cases our qualitative considerations are not sufficient to unequivocally identify all terms. 
The last line reports the expected residual tripartite information after a small global quench (see Section~\ref{sec quench}). }
\end{table}
The reader can easily generalise this discussion to potential cases with more zeroes. 

As a matter of fact, this qualitative analysis was more pedagogical than necessary, indeed, using the explicit result for the spin structure of XXZ, we find that, independently of $\Delta$, XXZ falls into the same category of the noninteracting case, i.e., $4\Delta_0^3-\Delta_1^2$ has a single zero at $x=\frac{1}{2}$ and $\mathrm{sgn}(\Delta_{1}(\frac{1}{2}))<0$. 
We can therefore write 
\begin{equation}
\gamma_{2,1}(x)=\frac{2}{3}\Bigl(e^{\tilde I_3^{(2;1)}(x)}-\sqrt{\Delta_0(x)}\cos\Bigl(\frac{2\pi}{3}+\frac{1}{3}\arctan\frac{\sqrt{4\Delta_0^3(x)-\Delta_1^2(x)}}{\Delta_1(x)}+\frac{\pi}{3}\theta_H(-\Delta_1(x))\Bigr)\Bigr)\, .
\end{equation}
\begin{equation}
\gamma_{2,2}(x)=\frac{2}{3}\left[e^{\tilde I_3^{(2;1)}(x)}-\sqrt{\Delta_0(x)}\cos\Bigl(\tfrac{\pi}{3}\mathrm{sgn}(x-\tfrac{1}{2})+\tfrac{1}{3}\arctan\tfrac{\sqrt{4\Delta_0^3(x)-\Delta_1^2(x)}}{\Delta_1(x)}-\tfrac{\pi}{3}\theta_H(\Delta_1(x))\Bigr)\right]
\end{equation}
\begin{equation}
\gamma_{2,3}(x)=\frac{2}{3}\left[e^{\tilde I_3^{(2;1)}(x)}-\sqrt{\Delta_0(x)}\cos\Bigl(\tfrac{\pi}{3}\mathrm{sgn}(\tfrac{1}{2}-x)+\tfrac{1}{3}\arctan\tfrac{\sqrt{4\Delta_0^3(x)-\Delta_1^2(x)}}{\Delta_1(x)}-\tfrac{\pi}{3}\theta_H(\Delta_1(x))\Bigr)\right]
\end{equation}
which, together with \eqref{eq:gamma4} and \eqref{eq:Itilde}, allows us to access the spin structure in the torus through the R\'enyi-2 tripartite information of four local spin-chain models. A similar procedure can be applied also to the Riemann surfaces with higher genus, but the inversion becomes more complicated and the number of models to consider increases.
In particular, the $3$-sheeted Riemann surface corresponding to the R\'enyi-$3$ tripartite information is worked out in Appendix~\ref{a:Inverse3}.

\paragraph{$3$-sheeted Riemann surface ($g=2$)}\label{a:Inverse3}
For the R\'enyi-$3$ tripartite information, if we exploit some symmetries of the partition functions, we obtain the system
\begin{equation}
\gamma_{3,1}^j+3\gamma_{3,2}^j+3\gamma_{3,3}^j+3\gamma_{3,4}^j-6\gamma_{3,5}^j=4e^{2I_3^{(3;j)}(x)}\, ,
\end{equation}
where $\gamma_{3,j}$ are the ratios of partition functions associated with given boundary conditions and $I_3^{(3;j)}(x)$ is the R\'enyi-$3$ tripartite information in the JW $j$-deformation of XXZ. 
Although we suspect that the tripartite information of 5 deformations should be enough to invert that system of equations, we propose a less refined approach that involves more models but that can be easily generalised to surfaces with higher genus. 

We start by giving up the information about the degeneracy of the terms, splitting them in the corresponding number of unknowns, e.g., $3\gamma_{3,2}^j\rightarrow \gamma_{3,2_1}^j+\gamma_{3,2_2}^j+\gamma_{3,2_3}^j$. We then use a similar trick to that employed for the torus. Specifically, we write
\begin{equation}\label{Aeq:Inverse3split}
\gamma_{3,1}^j+\sum_{k=1}^3(\gamma_{3,2_k}^j+\gamma_{3,3_k}^j+\gamma_{3,4_k}^j)=4e^{2I_3^{(3,j)}(x)}+\sum_{k=1}^6\gamma_{3,5_k}^j\, ,
\end{equation}
where we have relaxed the symmetry constraints $\gamma_{3,y_k}=\gamma_{3,y}$. Let us now work out the elementary symmetric polynomial associated with the unknowns on the left hand side of the equation
\begin{multline}
e_n[\{\gamma_{3,1},\gamma_{3,2_\bullet},\gamma_{3,3_\bullet},\gamma_{3,4_\bullet}\}]=\\
\frac{4^n}{n!}\det \left|\begin{matrix}
e^{2I_3^{(3;1)}}+\tfrac{\sum_k\gamma_{3,5_k}}{4}&\tfrac{1}{4}&0&\dots&\\
e^{2I_3^{(3;2)}}+\tfrac{\sum_k\gamma_{3,5_k}^2}{4}&e^{2I_3^{(3;1)}}+\tfrac{\sum_k\gamma_{3,5_k}}{4}&\frac{2}{4}&0\dots\\
\vdots&&\ddots&\ddots\\
e^{2I_3^{(3;n-1)}}+\tfrac{\sum_k\gamma_{3,5_k}^{n-1}}{4}&e^{2I_3^{(3;n-2)}}+\tfrac{\sum_k\gamma_{3,5_k}^{n-2}}{4}&\dots&e^{2I_3^{(3;1)}}+\tfrac{\sum_k\gamma_{3,5_k}}{4}&\frac{n-1}{4}\\
e^{2I_3^{(3;n)}}+\tfrac{\sum_k\gamma_{3,5_k}^{n}}{4}&e^{2I_3^{(3;n-1)}}+\tfrac{\sum_k\gamma_{3,5_k}^{n-1}}{4}&\dots&e^{2I_3^{(3;2)}}+\tfrac{\sum_k\gamma_{3,5_k}^2}{4}&e^{2I_3^{(3;1)}}+\tfrac{\sum_k\gamma_{3,5_k}}{4}
\end{matrix}\right|\, .
\end{multline}
We construct the associated polynomial $x^{10}+\sum_{n=1}^{10} (-1)^n e_n x^{10-n}$ and multiply it by $x^m$. 
We then evaluate the polynomial at its roots and sum over the roots, using \eqref{Aeq:Inverse3split}. The result is
\begin{multline}
4e^{2I_3^{(3;10+m)}}+\sum_{k=1}^6\gamma_{3,5_k}^{10+m}+\\
\sum_{n=1}^{10} (-1)^n e_n[\{\gamma_{3,1},\gamma_{3,2_\bullet},\gamma_{3,3_\bullet},\gamma_{3,4_\bullet}\}] (4e^{2I_3^{(3;10+m-n)}}+\sum_{k=1}^6\gamma_{3,5_k}^{10+m-n})=0 \, .
\end{multline}
We restrict ourselves to $m\in \{1,2,3,4,5,6\}$; this turns out to be a linear system of equations for the elementary symmetric polynomials associated with the variables $\gamma_{3,5_k}$ and, in turn, can be inverted with standard means. We can then express $e_n[\{\gamma_{3,5_1},\dots,\gamma_{3,5_6}\}]$ as functions of tripartite informations as follows
$$
e_n[\{\gamma_{3,5_1},\dots,\gamma_{3,5_6}\}]=\mathcal E_n[I_3^{(3;1)},\hdots,I_3^{(3;16)}]\qquad n\in \{1,2,3,4,5,6\}\, .
$$
Imposing $\gamma_{3,5_k}=\gamma_{3,5}$ for every $k$ gives immediately $\gamma_{3,5}$, for example using 
$$
\gamma_{3,5}=\frac{1}{6}\mathcal E_1[I_3^{(3;1)},\hdots,I_3^{(3;16)}]\, .
$$
Since we are free to use any other $\mathcal E_n$, as a by-product we obtain  $5$ identities satisfied by $\{I_3^{(3;1)},\cdots,I_3^{(3;16)}\}$. Similarly to the case $g=1$, we are finally reduced to 
\begin{equation}
\gamma_{3,1}^j+\sum_{k=1}^3(\gamma_{3,2_k}^j+\gamma_{3,3_k}^j+\gamma_{3,4_k}^j)=4e^{2\tilde I_3^{(3,j)}(x)}\, ,
\end{equation}
where 
$$
4e^{2\tilde I_3^{(3,j)}(x)}=4e^{2 I_3^{(3,j)}(x)}+\sum_{k=1}^6 \gamma_{3,5_k}^j
$$
are now known functions of the tripartite information. 
We now split variable $\gamma_{3,1}$ in 5 unknowns that will be only eventually forced to match:
\begin{equation}\label{Aeq:systemsplit}
\sum_{k=1}^3(\gamma_{3,1_k}^j+\gamma_{3,2_k}^j+\gamma_{3,3_k}^j+\gamma_{3,4_k}^j)=4e^{2\tilde I_3^{(3,j)}(x)}+\gamma_{3,1_4}^j+\gamma_{3,1_5}^j\, .
\end{equation}
In this way we have increased the number of variables to $14$, but we will momentarily treat $\gamma_{3,1_4}$ and $\gamma_{3,1_5}$ as if they were known functions. As before, we construct the associated elementary symmetric polynomials and perform the same manipulations. We end up with 
$$
4e^{2\tilde I_3^{(3;12+m)}}+\sum_{k=4}^5\gamma_{3,1_k}^{12+m}+\sum_{n=1}^{10} (-1)^n \tilde e_n[\{\gamma_{3,1_\bullet},\gamma_{3,2_\bullet},\gamma_{3,3_\bullet},\gamma_{3,4_\bullet}\}] (4e^{2\tilde I_3^{(3;12+m-n)}}+\sum_{k=4}^5\gamma_{3,1_k}^{12+m-n})=0 
$$
where $\sim$ in $\tilde e_n$ is used to emphasise that $I_3^{(3,j)}$ is now replaced by $\tilde I_3^{(3,j)}$. In this case we can restrict to $m\in\{1,2\}$. As before, this is a linear system of equations for the elementary symmetric polynomials associated with the variables $\gamma_{3,1_4}$ and $\gamma_{3,1_5}$, which can be readily inverted. We end up with
$$
\tilde e_n[\{\gamma_{3,1_4},\gamma_{3,1_5}\}]=\tilde{\mathcal E}_n[\tilde I_3^{(3;1)},\hdots,\tilde I_3^{(3;14)}]\qquad n\in\{1,2\}\, .
$$
Imposing $\gamma_{3,1_k}=\gamma_{3,1}$ for every $k$ immediately gives $\gamma_1$, for example using 
$$
\gamma_{3,1}=\frac{1}{2}\tilde{\mathcal E}_1[\tilde I_3^{(3;1)},\hdots,\tilde I_3^{(3;14)}]\, .
$$
Since we could also use $\mathcal E_2$, we also get one identity satisfied by $\{\tilde I_3^{(3;1)},\cdots,\tilde I_3^{(3;14)}\}$. Now that we know $\gamma_{3,1}$, we can reduce the system of equations~\eqref{Aeq:systemsplit} into one of the same form as for the torus:
\begin{equation}
\gamma_{3,2}^j+\gamma_{3,3}^j+\gamma_{3,4}^j=\frac{4}{3}e^{2\tilde I_3^{(3,j)}(x)}-\frac{1}{3}\gamma_{3,1}^j(x)\, .
\end{equation}
This can be solved as we did for the torus, and the entire discussion about the roots holds true. 

This procedure has allowed us to invert the problem using $16$ deformed models, with a by-product of $6$ potentially independent identities satisfied by $\{ I_3^{(3;1)},\hdots, I_3^{(3;16)}\}$. In the best case scenario those identities could be used to express $\{ I_3^{(3;11)},\hdots, I_3^{(3;16)}\}$ in terms of $\{ I_3^{(3;1)},\hdots, I_3^{(3;10)}\}$, realising a more refined inversion based on  just $10$ deformed models.

\section{Small $x$ expansion}\label{appendix small x expansion}
In this section we derive the results~\eqref{expansion spin structure epsilon zero}, \eqref{expansion spin structure epsilon two domain walls} and \eqref{expansion spin structure epsilon more than two domain walls}. The results are obtained by studying the small $x$ expansion of
\begin{equation}
    \Theta\left[
    \begin{matrix}
  \vec \mu \\ \vec 0    
    \end{matrix}\right]\left(\vec 0|\kappa \Omega\right)
\end{equation}
for $\vec\mu\in \{0,1/2\}^{\alpha-1}$ and a generic parameter $\kappa\in(0,\infty)$, and considering different contributions in Eq.~\eqref{eq:FfXXZ} .

The small $x$ expansion relies heavily on the methods developed in Refs~\cite{Calabrese2009Entanglement1,Calabrese2011Entanglement}, which start from the expansion of $\beta_{k/\alpha}$, appearing in the period matrix \eqref{period matrix def}, for $x\to 0$,
\begin{equation}\label{beta small x expansion}
    \beta_{\frac{k}{\alpha}}(x)=-\frac{\sin(\pi\frac{k}{\alpha})}{\pi}\left(\log x + f_{\frac{k}{\alpha}}+\sum\limits_{l=1}^\infty p_l\left(\frac{k}{\alpha}\right)x^l\right), \quad f_{\frac{k}{\alpha}}\equiv 2\gamma_E+\psi\left(\frac{k}{\alpha}\right)+\psi\left(1-\frac{k}{\alpha}\right),
\end{equation}
where $\gamma_E$ is the Euler gamma, $\psi(z)\equiv\Gamma'(z)/\Gamma(z)$ is the polygamma function and $p_l(z)$ is a polynomial of degree 2$l$, whose explicit expression is not needed. As in Ref.~\cite{Calabrese2011Entanglement}, we plug this expansion in the theta function, getting
\begin{equation}\label{step theta function expansion}
    \Theta\left[
    \begin{matrix}
  \vec \mu \\ \vec 0    
    \end{matrix}\right]\left(\vec 0|\kappa \Omega\right)=\sum\limits_{\vec{m}\in \mathbb{Z}^{\alpha-1}}x^{\kappa a_{\vec{m}+\vec{\mu}}} e^{\kappa b_{\vec{m}+\vec{\mu}}} (1+O(x)) \; ,
\end{equation}
where the two vector-dependent coefficients are defined by
\begin{align}
    & \label{appendix coefficient a def} a_{\vec{v}}\equiv\frac{2}{\alpha} \sum\limits_{k=1}^{\alpha-1} \sin^2 \left(\pi\frac{k}{\alpha}\right) \vec{v} \cdot  C_{k/\alpha} \vec{v} \; ,\\
  & \label{appendix coefficient b def} b_{\vec{v}}\equiv\frac{2}{\alpha} \sum\limits_{k=1}^{\alpha-1} \sin^2 \left(\pi\frac{k}{\alpha}\right) f_{k/\alpha} \vec{v} \cdot  C_{k/\alpha} \vec{v} \; ,
\end{align}
and the matrix $C_{k/\alpha}$ has elements
\begin{equation}
    \left(C_{\frac{k}{\alpha}}\right)_{\ell,n}\equiv \cos\left[2\pi\frac{k}{\alpha}(\ell-n)\right] \; .
\end{equation}
Furthermore, for any vector $\vec{v}$ the coefficient $a_{\vec{v}}$ is equal to~\cite{Calabrese2011Entanglement}
\begin{equation}
    a_{\vec{v}}=\sum_{\ell=1}^{\alpha-1}v_\ell^2- \sum_{\ell=1}^{\alpha-2} v_\ell v_{\ell+1}=\frac{v_\ell^2+v_{\alpha-1}^2}{2}+\frac{1}{2} \sum_{\ell=1}^{\alpha-2}(v_{\ell}-v_{\ell+1})^2 \; .
\end{equation}
For $\vec{v}\in \mathbb{Z}^{\alpha-1}$ the first equality makes it apparent that $a_{\vec{v}}\in\mathbb{Z}$, while from the second equality we see $a_{\vec{v}}\geq 0$. It follows $a_{\vec{m}+\vec{\mu}}= a_{2(\vec{m}+\vec{\mu})}/4\in \mathbb N_0/4$, where the first equality follows from the definition \eqref{appendix coefficient a def}.

The leading term in the expansion \eqref{step theta function expansion} is obtained by collecting all $\vec{m}\in\mathbb{Z}^{\alpha-1}$ that minimize the coefficients $a_{\vec{m}+\vec{\mu}}$. 
By inspection, $a_{\vec{m}+\vec{\mu}}=0$ is only possible for $\vec\mu =0$ (for $\vec m=0$) and in that case the leading term is a constant term. The case $\vec \mu=0$ has already been covered in Ref.~\cite{Calabrese2011Entanglement}, which derived the first two subleading terms. Focusing just on the first subleading term, the result reads
\begin{equation}\label{appendix expansion CCT}
 \Theta(0|\kappa\Omega)=1+s(\alpha;\kappa)x^{\kappa}+o(x^{\kappa}) \; ,
\end{equation}
where $s(\kappa;\alpha)$ is defined in Eq.~\eqref{s CCT}.

For $\vec \mu\neq 0$ we have $a_{\vec{m}+\vec{\mu}}\neq 0$ for all $m\in\mathbb Z^{\alpha-1}$. Thus the leading term in \eqref{step theta function expansion} will have a non-zero power. The smallest possible non-zero value of $a_{\vec{m}+\vec{\mu}}$ is $1/4$, achieved when $\vec \mu$ contains two ``domain walls'',
\begin{equation}\label{step mu two domain walls}
        \vec{\mu}^\mathrm{T}=\frac{1}{2}(\underbrace{0,0,\ldots,0}_{j_1}, \underbrace{1,1,\ldots,1}_{j_2-j_1},0,0,\ldots,0) \; , \quad 0\leq j_1< j_2\leq \alpha-1 \; ,
\end{equation}
in combination with $\vec m =\vec 0$ and $\Vec m=-2\vec\mu$, which give $\vec m+\vec\mu=\pm \vec\mu$ respectively. Thus for $\vec\mu$ of the type \eqref{step mu two domain walls}
\begin{equation}
      \Theta\left[
    \begin{matrix}
  \vec \mu \\ \vec 0    
    \end{matrix}\right]\left(\vec 0|\kappa \Omega\right)=x^{\kappa /4}\left( e^{\kappa b_{\vec{\mu}}}+e^{\kappa b_{-\vec{\mu}}}\right)+o(x^{\kappa/4}) \; .
\end{equation}
Now we resort to another formula derived in Ref.~\cite{Calabrese2011Entanglement}, about the coefficients $b_{\pm\vec\mu}$ for $\vec \mu$ given by \eqref{step mu two domain walls},
\begin{equation}
    b_{\pm\vec \mu}=-\frac{1}{2}\log\left[2\alpha\sin\left(\pi\frac{j_2-j_1}{\alpha}\right)\right] \; .
\end{equation}
Using this formula we obtain
\begin{equation}\label{appendix expansion two domain walls mu}
       \Theta\left[
    \begin{matrix}
  \vec \mu \\ \vec 0    
    \end{matrix}\right]\left(\vec 0|\kappa \Omega\right)=\frac{2}{\left[\sin\left(\pi\frac{j_2-j_1}{\alpha}\right)\right]^{\kappa/2}}\left(\frac{x}{4\alpha^2}\right)^{\kappa/4}+o(x^{\kappa/4})
\end{equation}
for $\vec\mu$ given by \eqref{step mu two domain walls}.

When $\vec\mu\neq 0$ and when $\vec \mu$ is not of the type \eqref{step mu two domain walls}, there are more than two ``domain walls'' in $\vec{\mu}$, i.e. four or more. In this case $a_{\vec m+\vec \mu}\geq 1/2$ for any $\vec m \in \mathbb Z^{\alpha-1}$ and, therefore,
\begin{equation}\label{appendix expansion more than two domain walls mu}
      \Theta\left[
    \begin{matrix}
  \vec \mu \\ \vec 0    
    \end{matrix}\right]\left(\vec 0|\kappa \Omega\right)=O(x^{\kappa/2}) \; .
\end{equation}

To derive the result \eqref{expansion spin structure epsilon zero}, in \eqref{eq:FfXXZ} we separate the contributions with respect to the number of domain walls in $\vec \mu$,
\begin{equation}
\begin{split}
    &\mathcal{F}_f  \left[\begin{matrix}
  \vec 0 \\ \vec \delta    
    \end{matrix}\right]=\frac{1}{\left[\Theta(0|\Omega)\right]^2}\Bigg(\Theta(0|\tfrac{1}{\ETA}\Omega)\Theta\left(0|4\ETA\Omega\right)\\ & +\sum_{\vec \mu\textrm{ with two domain walls}}(-1)^{4\vec \mu\cdot\vec\delta}\Theta\left[
    \begin{matrix}
  \vec \mu \\ \vec 0    
    \end{matrix}\right]\left(\vec 0|\tfrac{1}{\ETA}\Omega\right)\Theta\left[
    \begin{matrix}
  \vec \mu \\ \vec 0   \end{matrix}\right]\left(\vec 0|4\ETA\Omega\right)+
  \\ & 
  +\sum_{\vec \mu\textrm{ with more than two domain walls}}(-1)^{4\vec \mu\cdot\vec\delta}\Theta\left[
    \begin{matrix}
  \vec \mu \\ \vec 0    
    \end{matrix}\right]\left(\vec 0|\tfrac{1}{\ETA}\Omega\right)\Theta\left[
    \begin{matrix}
  \vec \mu \\ \vec 0   \end{matrix}\right]\left(\vec 0|4\ETA\Omega\right)
  \Bigg)
    \end{split}
\end{equation}
Then we apply the expansion~\eqref{appendix expansion CCT} to the denominator and the first term in the brackets, the expansion~\eqref{appendix expansion two domain walls mu} to the second term and the bound~\eqref{appendix expansion more than two domain walls mu} to the third term. Collecting the $\ETA$-dependent leading terms we get the result~\eqref{expansion spin structure epsilon zero}.

To derive the result \eqref{expansion spin structure epsilon two domain walls}, we rewrite \eqref{eq:FfXXZ} as
\begin{equation}\label{appendix step expansion two domain walls}
\begin{split}
    \mathcal{F}_f  \left[\begin{matrix}
  \vec \varepsilon \\ \vec \delta    
    \end{matrix}\right]=&
    \frac{1}{\left[\Theta(0|\Omega)\right]^2}
  \Bigg(  
    \Theta\left[\begin{matrix}
  \vec \varepsilon \\ \vec 0    
    \end{matrix}\right](0|\tfrac{1}{\ETA}\Omega)\Theta\left(0|4\ETA\Omega\right)+(-1)^{4\vec\varepsilon\cdot\vec\delta}\Theta(0|\tfrac{1}{\ETA}\Omega)\Theta\left[\begin{matrix}
  \vec \varepsilon \\ \vec 0    
    \end{matrix}\right]\left(0|4\ETA\Omega\right)\\
   & +\sum_{\vec \mu \neq \vec 0,\vec\varepsilon}(-1)^{4\vec \mu\cdot\vec\delta}\Theta\left[
    \begin{matrix}
  \vec\varepsilon+\vec \mu \\ \vec 0    
    \end{matrix}\right]\left(\vec 0|\tfrac{1}{\ETA} \Omega\right)\Theta\left[
    \begin{matrix}
  \vec \mu \\ \vec 0   \end{matrix}\right]\left(\vec 0|4\ETA \Omega\right)
  \Bigg)
    \end{split}
\end{equation}
Here we have also used that the theta function is invariant under the shift of a component of the characteristic by an integer, implying $ \Theta\left[
    \begin{matrix}
  2\vec \varepsilon\\ \vec 0    
    \end{matrix}\right]= \Theta\left[
    \begin{matrix}
  \vec 0\\ \vec 0    
    \end{matrix}\right]$. Now, the expansion of the denominator in \eqref{appendix step expansion two domain walls} is given by \eqref{appendix expansion CCT}. The first term in the brackets is a product of two theta functions, one with zero characteristic and one with $\vec\varepsilon$ given by the right hand side of \eqref{step mu two domain walls}. The leading term in the theta function with zero characteristics is the constant term in \eqref{appendix expansion CCT}, while the leading term for the second theta function is described by \eqref{appendix expansion two domain walls mu}. The second term in the brackets is treated in an analogous way. The third term in the brackets is subleading because both theta functions in the product have a non-zero characteristic. Summing all these contributions gives the result~\eqref{expansion spin structure epsilon two domain walls}.

To derive the bound \eqref{expansion spin structure epsilon more than two domain walls} we recognize that when $\vec\varepsilon$ has more than two domain walls then each term
\begin{equation}
    \Theta\left[
    \begin{matrix}
  \vec\varepsilon+\vec \mu \\ \vec 0    
    \end{matrix}\right]\left(\vec 0|\tfrac{1}{\ETA} \Omega\right)\Theta\left[
    \begin{matrix}
  \vec \mu \\ \vec 0   \end{matrix}\right]\left(\vec 0|4\ETA \Omega\right)
\end{equation}
in \eqref{eq:FfXXZ} has the following property. Either both characteristics, $\vec\mu$ and $\vec\varepsilon+\mu$, have two domain walls (e.g. $\vec\varepsilon^\mathrm{T}=1/2(1,0,1),\ \vec\mu^\mathrm{T}=1/2(0,1,0)$ for $\alpha=4$), or one of the two characteristics, $\vec\mu$ or $\vec\varepsilon+\mu$, has more than two domain walls (e.g. $\vec\varepsilon^\mathrm{T}=1/2(1,0,1)$ with $\vec\mu=0$ or $\vec\mu=\vec\varepsilon$ for $\alpha=4$). In the first case we apply the expansion~\eqref{appendix expansion two domain walls mu} and in the second case the bound~\eqref{appendix expansion more than two domain walls mu}, giving us the result~\eqref{expansion spin structure epsilon more than two domain walls}




 \bibliographystyle{SciPost_bibstyle.bst}
 \bibliography{references.bib}
\end{document}